\documentclass[fleqn,usenatbib]{mnras}

\usepackage[T1]{fontenc}

\DeclareRobustCommand{\VAN}[3]{#2}
\let\VANthebibliography\thebibliography
\def\thebibliography{\DeclareRobustCommand{\VAN}[3]{##3}\VANthebibliography}

\usepackage{graphicx}	
\usepackage{amsmath}	
\usepackage{amssymb}	
\usepackage{comment}
\usepackage{hyperref}

\usepackage{newtxtext,newtxmath}

\title[A kinematic study of quenching]{What drives galaxy quenching? A deep connection between galaxy kinematics and quenching in the local Universe}

\author[S Brownson et al.]{Simcha Brownson$^{1,2}$\thanks{E-mail: sbb33@cam.ac.uk}, Asa F. L. Bluck$^{1,2,3}$\thanks{E-mail: asa.bluck@mrao.cam.ac.uk}, Roberto Maiolino$^{1,2}$ and Gareth C. Jones$^{1,2}$
\\ 
$^{1}$Kavli Institute for Cosmology, University of Cambridge, Madingley Road, Cambridge CB3 0HA, UK\\
$^{2}$Cavendish Laboratory Astrophysics Group, University of Cambridge, 19 J. J. Thomson Ave., Cambridge CB3 0HE, UK\\
$^{3}$Florida International University, Department of Physics, 11200 SW 8th Street, Miami, FL, USA\\
}
\date{Accepted XXX. Received YYY; in original form ZZZ}

\pubyear{2022}

\begin{document}
\label{firstpage}
\pagerange{\pageref{firstpage}--\pageref{lastpage}}
\maketitle

\begin{abstract}
We develop a 2D inclined rotating disc model, which we apply to the stellar velocity maps of 1862 galaxies taken from the MaNGA survey (SDSS public Data Release 15). We use a random forest classifier to identify the kinematic parameters that are most connected to galaxy quenching. We find that kinematic parameters that relate predominantly  to the disc (such as the mean rotational velocity) and parameters that characterise whether a galaxy is rotation- or dispersion-dominated (such as the ratio of rotational velocity to velocity dispersion) are not fundamentally linked to the quenching of star formation. Instead, we find overwhelmingly that it is the absolute level of velocity dispersion (a property that relates primarily to a galaxy's bulge/spheroidal component) that is most important for separating star forming and quenched galaxies. Furthermore, a partial correlation analysis shows that many commonly discussed correlations between galaxy properties and quenching are spurious, and that the fundamental correlation is between quenching and  velocity dispersion. In particular, we find that at fixed velocity dispersion, there is only a very weak dependence of quenching on the disc properties, whereby more discy galaxies are slightly more likely to be forming stars. By invoking the tight relationship between black hole mass and velocity dispersion, and noting that black hole mass traces the total energy released by AGN, we argue that these data support a scenario in which quenching occurs by preventive feedback from AGN. The kinematic measurements from this work are publicly available.
\end{abstract}




\section{Introduction}\label{Introudction}
Many galaxy properties are bimodally distributed (e.g. \citealt{Strateva2001, Baldry2004, Brinchmann2004, Driver2006, Cameron2009, Wuyts2011}). This bimodality can be summarised in terms of two broad fundamental galaxy features: 1) star formation activity and 2) morphological and kinematic structure. In terms of star formation activity, observations of the local Universe reveal `star forming' blue galaxies that have relatively large specific star formation rates ($\rm sSFR=SFR/\mathit{M_\star}$), as well as `quenched' red galaxies that have suppressed sSFR.  In terms of morphological and kinematic structure, there exists both `rotation-dominated' galaxies that have small bulge-to-total mass ratios ($B/T$) and large ordered to disordered kinematic  ratios ($V/\sigma$); as well as `dispersion-dominated' galaxies that have large $B/T$ and small $V/\sigma$.

There appears to be a deep connection between these two bimodalities (e.g. \citealt{CameronDriver2009,  Gadotti2009, Cappellari2011a, Bell2012, Lang2014, Omand2014, Bluck2014, Bluck2016}). This is succinctly expressed by the `morphology-colour' relation, which claims that star-forming galaxies are generally rotation-dominated or `disc-dominated', and quenched galaxies are generally dispersion-dominated or `bulge-dominated'. Despite the wealth of observational support for a morphology-colour relation, understanding the physical mechanisms responsible for its existence remains an important outstanding question in the field of galaxy evolution. 

The goal of this work is to understand why galaxies quench. Given the observed morphology-colour relation, it is reasonable, and tempting, to look for mechanisms that are simultaneously capable of quenching galaxies $and$ transforming them from being rotation- to dispersion-dominated. For example, galaxy mergers provide a plausible pathway for triggering the morphological transition, and simultaneously feeding the growth of the central supermassive black hole, which could quench the galaxy through feedback from an active galactic nucleus (AGN, e.g. \citealt{DiMatteo2005, Springel2005b, Croton2006, Bower2008, Hopkins2008, Maiolino2012}). For the sake of completeness, we note that mergers could also quench galaxies through alternative pathways. These include elevated star formation and supernovae feedback (e.g. \citealt{Cole2000, Henriques2019}), halo growth and virial shock heating (e.g. \citealt{Dekel2006, Woo2013}), and increased kinematic stabilisation of the galaxy disc by the galaxy bulge in the morphological quenching scenario (e.g. \citealt{Martig2009, Gensior2020}). 

We must be clear about what we mean by galaxy quenching before introducing common quenching mechanisms. Indeed, there are two common definitions of galaxy quenching. First, there is the trigger event that initially shuts off star formation within a galaxy, causing a galaxy to depart the star forming main sequence. Second, there is the maintenance mode which keeps a galaxy quenched and does not allow star formation to rejuvenate over many billions of years of cosmic history. 

In this work we are primarily concerned with the second definition. To appreciate why, it is important to recognise that galaxies are not closed box systems, and that the vast majority of baryons in massive systems reside in hot ($T\gtrsim10^7\,\mathrm{K}$) gas halos \citep{Lin2003, Fabian2006, McNamara2007}. These baryons are expected to cool via thermal bremsstrahlung emission on timescales shorter than $1\,\mathrm{Gyr}$,  form a cooling flow, and thereby trigger dramatic late-time star formation within the galaxy \citep{Fabian2012}. In other words, naively one would expect star formation to rejuvenate within once-quenched galaxies. This theoretical expectation is inconsistent with the observed  suppression of star formation, and  hence ultimately stellar mass to halo mass ratios in massive galaxies (e.g. \citealt{Baldry2006, Peng2010, Moster2010}). It is also inconsistent with the observation that 90 per cent of baryonic matter remains unprocessed through stars \citep{Fukugita2004, Shull2012}. These inconsistencies remain the key theoretical challenge of galaxy quenching, commonly referred to as the `cooling catastrophe' (e.g. \citealt{Binney1995, Ruszkowski2002}).

Solutions to the cooling catastrophe invoke an additional heating mechanism to offset the cooling. Three proposed mechanisms are as follows: 1) heating from supernovae feedback (e.g. \citealt{Cole2000, Henriques2019}), which offers a natural explanation for the strong correlation between stellar mass (which is a tracer of previous star formation and number of supernovae) and quenching, in the `mass-quenching' paradigm (e.g. \citealt{Baldry2006, Peng2010, Peng2012}); 2) virial shock heating (e.g. \citealt{Dekel2006, Woo2013}), which is supported by the observation that quenching is more closely related to halo mass than stellar mass \citep{Woo2013, Bluck2016}; and 3) heating from AGN feedback both in the high Eddington ratio `quasar mode' (e.g. \citealt{DiMatteo2005, Hopkins2008, Maiolino2012, Bischetti2019}) and the low Eddington ratio `preventative mode' (e.g. \citealt{Croton2006, Bower2008, Fabian2006, Sijacki2007, Zinger2020}), which is supported by the observation that quenching is most strongly related to parameters that probe the mass of the black hole (e.g. \citealt{Wake2012, Bluck2016,Terrazas2016, Terrazas2017, Bluck2020a, Bluck2020b}).

On the structural side, previous studies have used a range of morphological parameters derived from photometric/spectroscopic observations to constrain different quenching mechanisms (e.g. \citealt{Peng2010, Peng2012, Wuyts2011, Bell2012, Bluck2016, Bluck2021}). This approach faces a number of critical challenges. Firstly, morphological parameters are waveband dependent (e.g. \citealt{Bluck2019}).  Spectral energy distribution (SED) fitting is often used to overcome this waveband dependence, but this approach is highly dependent on key assumptions, such as the adopted initial mass function (IMF), the simple stellar population (SSP) library, and the star formation history (e.g. \citealt{Conroy2013, Lower2020}). Secondly, photometric/spectroscopic measurements are not sensitive to all phases of matter in the galaxy (i.e. stellar, gas,  dust and dark matter). Moreover, it is only economically feasible to observe the gas in all its phases for small galaxy samples (e.g. \citealt{Saintonge2016, Piotrowska2020, Brownson2020,Lin2020}). Finally, and most importantly, galaxy morphology only indirectly traces the fundamental  structure of galaxies. For example, low $B/T$ disc-dominated galaxies are often assumed to be rotation-dominated. In reality $B/T$ is a mere proxy of galaxy kinematics that simply quantifies the light or mass associated with a disc or bulge structure, not revealing the kinematics and hence true dynamics of the system.

Kinematic studies of galaxies, on the other hand, directly probe the motion of gas/stars and are sensitive to the fundamental physics of gas/stellar orbits. Indeed, the kinematics of any single component (stellar, gas or dark matter) traces the total mass budget and is therefore a probe of the galaxy's gravitational potential in virialised systems. Moreover, kinematic measurements  provide a more accurate quantification of galaxy structure, which could be used to refine the morphology-colour relation. For example, the dimensionless spin parameter, $\lambda$, which is a proxy of the angular momentum, has been particularly effective at identifying and cleanly separating rotation-dominated (or `fast rotator') and dispersion-dominated (`slow rotator') galaxies (e.g. \citealt{Emsellem2007, Emsellem2011, Fogarty2015, Cappellari2016, Graham2018, Wang2020}). Moreover, \citet{Cappellari2011b} show that two thirds of face-on fast rotator early-type galaxies are wrongly classified as photometric spheroids.  

These advantages of galaxy kinematics motivate an update of previous morphological studies of galaxy quenching with new kinematic studies of galaxy quenching.  We note that there is already evidence for galaxy kinematics being more predictive of quenching than morphology. In particular, the best morphological predictor of galaxy quenching is the mass of the bulge \citep{Bluck2014, Lang2014}, but its kinematic counterpart, the velocity dispersion, is even more effective at separating the star forming and quenched populations \citep{Wake2012, Teimoorinia2016, Bluck2016, Bluck2020a, Bluck2020b, Bluck2021}. The natural extension of these works is a full kinematic study  that replaces all morphological galaxy properties with their kinematic counterparts.

The advent of large integral field unit (IFU) galaxy surveys is providing astronomers with invaluable spatially resolved spectroscopic information of galaxies \citep{Cappellari2011a, Sanchez2012, Bundy2014, Cappellari2016}. The Mapping Nearby Galaxies at Apache Point Observatory survey (MaNGA) is the largest survey of this kind to date \citep{Bundy2014}.  MaNGA can be used to estimate many  galaxy properties on  kpc scales, but in this paper we focus on its estimates of the line-of-sight velocity and line-of-sight velocity dispersion, which can be used to model galaxy kinematics.

In this work we develop our own 2D kinematic code (i.e. separately modelling integrated light [moment-0], line-of-sight velocity [moment-1] and line-of-sight velocity dispersion [moment-2]) which models fast rotators as inclined rotating discs. We choose to develop a 2D code (i.e. rather than a 3D code) since it is suitable for modelling stellar kinematics, which is essential for the study of quenched galaxies that generally do not have strong emission lines. However, our 2D code incorporates the most important features of the latest 3D fitting codes. In particular, it carefully incorporates the effect of beam smearing both to model the moment-1 maps and to correct the observed moment-2 maps for the observed velocity dispersion artificially induced by differential disc rotation (e.g. \citealt{Bosma1978, Begeman1987, Lelli2010, Teodoro2015}).

We use the model to update common morphological parameters and derive kinematic estimates that are fundamentally connected to the physics of stellar orbits, such as the mean specific angular momentum  and the mean specific kinetic energy. We exploit the size of the MaNGA survey to achieve these estimates for 1862 galaxies, which is the largest homogeneous kinematic sample to date. We validate our kinematic estimates against traditional kinematic scaling relations such as the Tully-Fisher \citep{Fisher1977} and Faber-Jackson scaling relations \citep{Faber1976}, as well as against their traditional morphological counterparts. We then exploit rigorous statistical techniques, combining a  random forest analysis  with a partial correlation analysis, to identify the kinematic parameters which are most fundamentally effective at separating star forming and quenched galaxies. We thereby place powerful new constraints on  theoretically motivated quenching mechanisms. 

This paper is structured as follows. In Section \ref{Data}, we introduce the data used in this work. In Section \ref{Galaxykinematics}, we describe and validate our 2D kinematic model. In Section \ref{Results}, we perform a statistical analysis to identify the kinematic parameters that are important for quenching. In Section \ref{Discussion - How do galaxies quench?}, we interpret our results and discuss the importance of the kinematic parameters in the context of different quenching mechanisms. Finally, in Section \ref{Summary}, we summarise our key findings. We also include various appendices which show additional examples of our kinematic modelling and test the stability of our results. We assume a  $\rm \Lambda$CDM cosmology  throughout this paper, with $\rm H_{0}$ = 70  $\rm {km} \ \rm{ s}^{-1}$ $\rm Mpc{}^{-1}$, $\rm {\Omega}_{M}$ = 0.3 and $\rm {\Omega}_{\Lambda}$ = 0.7.


\section{Data}\label{Data}

\subsection{MaNGA, DAP and Pipe3D}
MaNGA is an IFU galaxy survey targeting 10,000 galaxies in the redshift range $0.01<z<0.15$ \citep{Bundy2014,Yan2016}\footnote{\href{https://www.sdss.org/dr15/manga/}{https://www.sdss.org/dr15/manga/}}. We briefly review the survey selection criteria, and refer the interested reader to \citet{Wake2017} for a full discussion. MaNGA survey galaxies are drawn from the SDSS legacy parent sample and are chosen to have a flat number density distribution in stellar mass with $\rm log(M_{\star}/M_{\odot})>9.0$. The survey consists of the following two samples: the primary sample, which observes galaxies out to $\rm 1.5\,R_{e}$ and contains two thirds of MaNGA galaxies, and the secondary sample, which observes galaxies out to $\rm 2.5\,R_{e}$ and contains the remaining one third of MaNGA galaxies. In this work, we utilise the  publicly available data release 15 \citep{Aguado2019}, which contains $\rm \sim4500$ galaxies. This is the largest spatially resolved spectroscopic sample of local galaxies, which offers an unprecedented opportunity to conduct a statistical study of galaxy kinematics and quenching. 

The MaNGA IFU system is mounted on the SDSS 2.5m telescope at the Apache Point Observatory \citep{Gunn2006} and contains 17 IFUs on a single plate. The IFUs vary in size, but they each contain spectroscopic fibres arranged in a hexagonal configuration. The field of view (FOV) diameter of the IFUs depends on the IFU size, and ranges from 12 arcsec for IFUs composed of 19 fibres, to 32 arcsec for IFUs composed of 127 fibres. This range of diameters enables the MaNGA survey to map a large sample of galaxies, with a wide range of sizes and redshifts, out to at least $\rm 1.5\,R_{e}$. The fibres are fed to the Baryon Oscillation Spectroscopic Survey (BOSS) spectrographs, which span the wavelength range $\rm 3600-10000$\AA~with an average spectral resolution of $\rm \sim 2000$. Reduced data cubes have 0.5 arcsec spaxels (spectroscopic pixels) and a spatial resolution of $\rm 2.5~arcsec$ \citep{Law2016, Yan2016}. The MaNGA survey thus provides spatially resolved information about the stellar and gas properties of local galaxies.

In this paper, we use two publicly available MaNGA catalogues: the data analysis pipeline (\textsc{dap}) v2.2.1 \citep{Westfall2019, Belfiore2019}\footnote{\href{https://www.sdss.org/dr15/manga/manga-analysis-pipeline/}{https://www.sdss.org/dr15/manga/manga-analysis-pipeline/}} and \textsc{pipe3d} v2.4.3 \citep{Sanchez2016}\footnote{\href{https://www.sdss.org/dr15/manga/manga-data/manga-pipe3d-value-added-catalog/}{https://www.sdss.org/dr15/manga/manga-data/manga-pipe3d-value-added-catalog/}}. Both catalogues provide spatially resolved estimates of the line-of-sight (LOS) stellar velocity and velocity dispersion, but we choose the \textsc{dap} as our primary source because it employs a binning scheme that is more appropriate for this work. More specifically, the \textsc{dap} uses the Voronoi-binning algorithm  to achieve a signal-to-noise threshold required for spectral fitting \citep{Westfall2019}. This algorithm enforces a roundness criterion which prevents spatial bins, commonly referred to as voxels, from becoming elongated \citep{Cappellari2003}. \textsc{pipe3d} similarly bins spaxels to achieve a signal-to-noise threshold, but it deliberately omits the roundness criterion, so that voxels are elongated along isophotes, preserving the shape of the underlying galaxy \citep{Sanchez2016}. In this work, we model the LOS stellar velocity of rotating galaxies, which is a strong function of azimuthal angle in the galaxy plane (see Section \ref{2D Kinematic Model}). Binning along isophotes blurs the azimuthal structure of the stellar velocity, making it more difficult to accurately model the kinematics. Indeed, our kinematic model described in the next section regularly fails to fit the \textsc{pipe3d} kinematics of galaxies that are clearly rotating and have \textsc{dap} kinematics that are consistent with the model. The \textsc{dap} is therefore more suited to this kinematic study.

A number of the \textsc{dap} maps have point spread function (PSF) sized `holes' towards the galaxy centre that lack estimates of the stellar velocity dispersion. These central regions, where the stellar velocity dispersion peaks, are critical for this work, so we fill the holes with estimates taken from \textsc{pipe3d}. We have checked that the \textsc{dap} and \textsc{pipe3d} are consistent by comparing their estimates of the average velocity dispersion measured within $\rm 1\,R_{e}$ for all galaxies in our sample. The two estimates are well correlated ($\rm \rho_{Pearson}=0.91$), and a linear fit comparing the velocity dispersions measured by \textsc{dap} against \textsc{pipe3d} (gradient $\rm m=1.08\pm0.01$, intercept  $\rm c=25.1\pm2.2\, km~s^{-1}$ and scatter about the relation $\rm RMSE=33.9\, km~s^{-1}$) is broadly consistent with the 1-1 relation. We note that $\sim$95 per cent of galaxies do not contain a hole, and when present, the holes represent only a small fraction of the spaxels in a single map. We are therefore confident that our use of \textsc{pipe3d} in these regions does not introduce a bias or affect our key results.

We also use \textsc{pipe3d} for its estimates of the stellar mass surface density, $\Sigma_{\star}$, which is not provided in the \textsc{dap} data release. \textsc{pipe3d} assumes a Salpeter  \citep{Salpeter1955} IMF so we convert the estimates to the Chabrier \citep{Chabrier2003} IMF assumed in this work by using the standard conversion $\rm log\Sigma_{\star}^{Salpeter} = log\Sigma_{\star}^{Chabrier} -0.22$. 

\subsection{SDSS ancillary data}
MaNGA galaxies are drawn from the SDSS parent sample, so they have a wealth of ancillary data. We use the following data in this work. First, we use the NSA-Sloan catalogue \citep{Blanton2011} for its estimates of S\'ersic index ($\rm n_{S\acute{e}rsic}$),  photometric axis ratio ($ (b/a)_\mathrm{phot}$), photometric position angle ($\rm PA_{phot}$), and global stellar mass ($M_{\star}$) for a \citet{Chabrier2003} IMF. The stellar masses are found via SED fitting to SDSS photometry. Second, we use the MPA-JHU catalogue for its estimates of the global star formation rate (SFR). These are calculated using emission lines where possible, and via the strength of the 4000\AA~ break (D4000) otherwise \citep{Brinchmann2004}. We convert the SFR estimates to the \citet{Chabrier2003} IMF. Finally, we identify and exclude galaxies with a spectroscopic companion closer than 100$\rm \,kpc$ in projection and 500$\rm \, km~s^{-1}$ in the LOS \citep{Patton2016}, post-merger galaxies \citep{Thorp2019}, as well as galaxies with a companion in the IFU and galaxies that are visually interacting with a companion that does not have spectroscopic information (private communication from M. D. Thorp). We restrict our focus to isolated galaxies in this work, since interactions disturb galaxy kinematics and contradict the assumption of virialisation (which is important for many of our subsequent analyses).

We require that the MaNGA galaxies in this work are present in each of the above catalogues with good measurements of the relevant parameters.  Additionally, we restrict our focus to  galaxies with $\rm log(M_{\star}/M_{\odot})>9.8$. This cut is chosen for two reasons: first, our focus is on intrinsic galaxy quenching which dominates at high mass, rather than environmental quenching which dominates at low mass; second, the  kinematics of low mass galaxies are more difficult to model since they have lower stellar velocity and velocity dispersion, as prescribed by the virial theorem. Moreover, the stellar continuum in low mass galaxies is much weaker and therefore the stellar kinematics are much more difficult to trace. These cuts, together with our removal of mergers and close encounters, return a sample of 2637 galaxies, which is the largest homogeneous sample used in a joint study of kinematics and quenching. 1862 of these have data of sufficient quality for effective kinematic modelling as we show in the next section.


\section{Galaxy kinematics}\label{Galaxykinematics}
The \textsc{dap} catalogue contains estimates of the spatially resolved LOS stellar velocity (moment-1) and LOS stellar velocity dispersion (moment-2). 

In this section, we use these estimates to derive and validate a set of physically motivated global kinematic parameters, before assessing their relevance for predicting galaxy quenching in Section \ref{Results}. 

\subsection{2D Kinematic Model}\label{2D Kinematic Model}
In this sub-section, we  fit the moment-1 maps taken from the \textsc{dap} with a 2D idealised inclined rotating disc model, which assumes that galaxies are rotators and that their stellar orbits are axisymmetric. We briefly discuss the following key advantages of the inclined disc model for our work:  it is effective at modelling rotators and is able to identify non-rotators; it is the most simple model capable of describing galaxy rotation;  it can be used to model stellar kinematics; and  it is consistent with previous kinematic and photometric measurements.

We are not suggesting that all galaxies display kinematics that can be modelled as inclined rotating discs.  We simply \textit{attempt} to fit all galaxies with the inclined rotating disc model, understanding fully that it will fail for galaxies that are not rotating. Hence, the spirit of our approach is to ask the following question: which galaxies are kinematically consistent with the inclined disc model? 

There exists alternative, more complex models with many more model parameters that are commonly used to describe galaxy kinematics, such as those that account for the finite thickness of the kinematic disc (e.g. \citealt{vanderHulst1992,Kranjnovic2006, Davis2013, Sellwood2015, Teodoro2015, Neeleman2021}). The advantage of the inclined rotating disc model is that it is the simplest conceivable model which accounts for the dominant observational effect (i.e. beam smearing) and is capable of describing galaxy rotation, with the fewest free parameters. Despite this simplicity, we will demonstrate that it achieves a successful fit for the vast majority of our galaxy sample, and hence we select the inclined rotating disc model on the basis of Occam's razor.

The inclined rotating disc model is a 2D kinematic model. We cannot use 3D fitting codes that model data cubes containing gas emission lines (e.g. \citealt{Davis2013, Teodoro2015, Neeleman2021}), since quenched galaxies do not have strong emission lines. We are therefore forced to use  stellar kinematic maps to model the kinematic properties of both star forming and quenched galaxies in this quenching study. To our knowledge, there does not yet exist a 3D kinematic fitting code that analyses data cubes containing the stellar continuum and absorption features, and simultaneously fits a simple stellar population library as well as a kinematic model. Developing such a code is beyond the scope of this work, so we adopt the next best option, which is a 2D kinematic model that accounts for the most significant challenge in kinematic modelling - namely, the effect of beam smearing  (e.g. \citealt{Bosma1978, Begeman1987, Lelli2010, Teodoro2015}).

Previous works have successfully modelled the moment-1 maps of IFU data with inclined rotating disc models. Indeed, \citet{Barrera-Ballesteros2018} use an even simpler inclined rotating disc model with fewer free parameters than that introduced in the next section. More specifically, they do not fit the inclination of the disc (they simply adopt the photometric inclination) and they do not correct for the dominant observational effect of beam smearing. Nonetheless, they are able to derive estimates consistent with the Tully-Fisher relationship \citep{Fisher1977}. In sub-section \ref{Tests: Kinematic scaling relations and connection to morphology}, we similarly cross-validate our kinematic model against alternative kinematic estimates and photometric measurements, as well as against well established scaling relations.

\subsubsection{Inclined rotating disc model}

We briefly orient the reader with the inclined rotating disc model geometry. The disc is assumed to be infinitesimally thin, but we note that to first order, the finite thickness of a disc (or its non-zero velocity dispersion) does not influence its moment-1 map. The disc is circular by construction, but its projection in the sky plane is an ellipse, with semi-major axis ($a$) semi-minor axis ($b$) and ellipticity ($\epsilon=1-b/a$). The geometry of this projection is determined by the disc's inclination angle ($\mathrm{inc}=\arccos(b/a)$) and position angle ($\rm PA$), which we define as the angle between the north-south axis and the semi-major axis, increasing anticlockwise. Face-on discs have $\rm inc=0^{\circ}$ and appear circular in the sky plane, whilst edge-on discs have $\rm inc=90^{\circ}$ and appear as infinitesimally thin straight lines oriented along the semi-major axis.

It is easiest to describe inclined disc rotation in the plane of the disc $(x_\mathrm{disc}, y_\mathrm{disc})$, where $y_\mathrm{disc}$ is the disc-plane coordinate along the major axis and $x_\mathrm{disc}$ is the disc-plane coordinate along the minor axis. For convenience we introduce the sky-plane coordinate system $(x_{\mathrm{sky}}, y_{\mathrm{sky}})$, which has the same orientation as the familiar $\rm (R.A., Dec)$ coordinate system. The coordinates $(x_\mathrm{disc}, y_\mathrm{disc})$ and $(x_{\mathrm{sky}}, y_{\mathrm{sky}})$ are related using the 2D transformation matrix:
\begin{equation}
\label{EQN:Skytogalframe}
\begin{pmatrix}
x_\mathrm{disc} \\
y_\mathrm{disc}
\end{pmatrix} 
= 
\begin{pmatrix}
\cos(\mathrm{PA})/\cos(\mathrm{inc}) & -\sin(\mathrm{PA})/\cos(\mathrm{inc}) \\
\sin(\mathrm{PA}) & \cos(\mathrm{PA})
\end{pmatrix} 
\begin{pmatrix}
x_\mathrm{sky} \\
y_\mathrm{sky}
\end{pmatrix} .
\end{equation}
The $1/\cos(\mathrm{inc})$ term in the transformation matrix deprojects the coordinates from the face-on sky-plane to the inclined disc-plane, and the remaining terms correspond to the standard rotation matrix in 2D. 

The disc is axisymmetric by design, so we adopt the familiar plane polar coordinate system $(r, \theta)$: 
\begin{equation}
\label{EQN:r}
    r 
    =
    \sqrt{({x_\mathrm{disc}-x_{\mathrm{disc},c}})^2+({y_\mathrm{disc}-{y_{\mathrm{disc},c}}})^2} \\
\end{equation}
\begin{equation}
\label{EQN:theta}
    \theta
    =
    \arctan \left(\frac{x_\mathrm{disc}-x_{\mathrm{disc},c}}{y_\mathrm{disc}-y_{\mathrm{disc},c}}\right)
\end{equation}
where $ (x_{\mathrm{disc},c}, y_{\mathrm{disc},c})$ are the coordinates of the disc centre in the disc plane. The key insight from equations \ref{EQN:Skytogalframe}, \ref{EQN:r} and \ref{EQN:theta} is that the $(r, \theta)$ coordinates are strong functions of inc and PA. The radial and azimuthal variance of the disc properties can therefore be used to determine the disc geometry.

We now use the inclined disc geometry to describe galaxy rotation. We seek a rotation curve in which the circular speed of stars increases linearly with galactocentric distance out from the galaxy centre and plateaus at larger radii, which is consistent with the first-order behaviour of orbits distributed on spatial scales probed by the MaNGA survey (e.g. \citealt{Puech2008, Anderson2013}). The hyperbolic tangent ($\tanh$) function is one example of a function that increases and subsequently flattens. We note that there are other functions, such as $\arctan$, that exhibit similar behaviour, but the exact parameterisation is not important. We merely require a function that captures the behaviour of the rotation which, as we will show, the $\tanh$ model achieves.

MaNGA is sensitive only to the LOS component of the circular velocity vector. The LOS component of a hyperbolic rotation curve is given by
\begin{equation}
\label{EQN:LOSVelocity}
    V_\mathrm{LOS}(r,\theta) = V_\mathrm{sys} + V_\mathrm{max} \times \tanh\left(\frac{r}{r_c}\right) \, \cos(\theta) \, \sin(\mathrm{inc}) 
\end{equation}
where $V_\mathrm{sys}$ is the systemic velocity, $V_\mathrm{max}$ is the amplitude of the rotation curve, and $r_c$ is a kinematic lengthscale describing the steepness of the rotation curve. This equation has a relatively simple form, but it is important to stress that the $ r~\mathrm{and}~\theta$ coordinates are functions of four observed geometric parameters: the coordinates of the disc centre (two coordinates), the inclination angle, and the position angle. The LOS velocity is thus a complex non-linear function whose radial and azimuthal structure is mathematically related to, and can therefore be used to determine, the disc geometry.

Equation \ref{EQN:LOSVelocity} describes the true LOS stellar velocity structure of an inclined rotating disc, referred to as the intrinsic LOS velocity hereafter.  However, we must account for two observational effects introduced by MaNGA and the \textsc{dap} before the model can be used to effectively analyse real galaxy data. 

The first observational effect is beam smearing, caused by MaNGA's modest, 2.5 arcsec spatial resolution. MaNGA does not measure a galaxy's true surface brightness, $I(x_\mathrm{sky}, y_\mathrm{sky})$, also referred to as the moment-0 map. Instead, it measures  the convolution of $I(x_\mathrm{sky}, y_\mathrm{sky})$ and the PSF. The \textsc{dap}'s moment-1 maps are derived from this PSF convolved brightness, hence they correspond to the typical intrinsic LOS stellar velocity within PSF sized regions, rather than within individual spaxels. We model this effect as follows:
\begin{equation}
\label{EQN:ModSpaxelLOSVelocityMaps}
    V^\mathrm{model, spaxel}_\mathrm{LOS}(x_\mathrm{sky}, y_\mathrm{sky}) = \frac{(V_\mathrm{LOS}(x_\mathrm{sky}, y_\mathrm{sky})I(x_\mathrm{sky}, y_\mathrm{sky}))* \mathrm{PSF}}{I(x_\mathrm{sky}, y_\mathrm{sky})*\mathrm{PSF}} 
\end{equation}
where $\rm *$ represents a convolution. We now use the $(x_{\mathrm{sky}}, y_{\mathrm{sky}})$ coordinate system since we are discussing observed, rather than intrinsic, properties. We take empirical, reconstructed PSF models from the \textsc{drp} (v2.4.3 \citealt{Law2016}). To describe equation \ref{EQN:ModSpaxelLOSVelocityMaps} in words, the model LOS velocity in spaxel X is given by the average LOS velocity of all other spaxels in the map, with each spaxel weighted by its surface brightness and the amplitude of the PSF. The PSF thus blurs the LOS velocity structure, such that neighbouring spaxels have similar values. 

One could treat the surface brightness adopted during PSF convolution as a free parameter in the model, and simultaneously fit the moment-0 and moment-1 maps. We choose not to adopt this approach since beam smearing is a second order effect.  Instead we adopt two reasonable prior measurements of the moment-0 maps: model photometric $r$-band S\'ersic profiles taken from the NSA catalogue \citep{Blanton2011}, and $g$-band flux maps taken from the \textsc{dap}. We generally favour the S\'ersic models since they estimate a galaxy's intrinsic light profile, whereas the \textsc{dap} flux maps are PSF convolved and Voronoi binned. Nonetheless, as we discuss later in this section, we do adopt the \textsc{dap} moment-0 maps for a number of fits, which exhibit clear problems in the moment-1 models.

The second observational effect is Voronoi binning, first discussed in Section \ref{Data}, which causes all spaxels within a voxel to share the same estimates, including that of the LOS stellar velocity, $V^\mathrm{obs,voxel}_\mathrm{LOS}$. Throughout this paper, the superscript `obs' refers to observed data, in this case taken from the \textsc{dap}. Voronoi binning thus sacrifices spatial resolution for increased sensitivity. In order to account for the impact of the data being Voronoi binned, we Voronoi bin the model by calculating the light-weighted average of $V^\mathrm{model,spaxel}_\mathrm{LOS}$ within each voxel. Voxels composed of only a single spaxel thus have ${V^\mathrm{model, voxel}_\mathrm{LOS}=V^\mathrm{model, spaxel}_\mathrm{LOS}}$. The Voronoi binned model, $\rm V^{model, voxel}_\mathrm{LOS}$, is now in a form consistent with the \textsc{dap} ($V^\mathrm{obs, voxel}_\mathrm{LOS}$), so we can compare $\rm V^{model, voxel}_\mathrm{LOS}$ and $V^\mathrm{obs, voxel}_\mathrm{LOS}$ to determine the best-fitting values of the seven inclined rotating disc model parameter. These are the disc centre $\rm (x_{sky,c}, y_{sky,c})$, inclination ($\rm inc$), position angle ($\rm PA$), maximum rotation velocity ($V_\mathrm{max})$,  systemic velocity ($V_\mathrm{sys}$), and kinematic lengthscale ($r_c$). 

\subsubsection{Kinematic fitting}
The model we have introduced is highly non-linear, so we use a non-linear least squares minimisation python package, \textsc{lmfit}  \citep{Newville2014}\footnote{\href{https://doi.org/10.5281/zenodo.11813}{https://doi.org/10.5281/zenodo.11813}}, to fit the \textsc{dap} moment-1 maps. We adopt the trust region reflective algorithm to minimise the $\rm \chi^{2}$ statistic. We improve the time efficiency of fitting by bounding the parameters within reasonable limits and using photometric parameters from the NSA catalogue to motivate an initial guess of their kinematic counterparts. In particular, we use the photometric position angle, $\rm PA_{phot}$, and the photometric major and minor axis lengths for our initial guess of $\rm PA$ and $\rm inc$ (via $\arccos((b/a)_\mathrm{phot})$), respectively, and we use our alternative, simplistic kinematic model for an initial guess of $V_\mathrm{max}$ (see Section \ref{Simplistic kinematic model}). 

We have tested the model performance on mock galaxy data. In particular, we have investigated the dependence of our model accuracy on data
quality by taking a mock galaxy with known kinematic parameters
and creating 10000 realisations of the observed data, where we vary
the number of PSF beams along the major axis, the number of voxels along the major axis, and the inclination. These tests demonstrate the model's ability to recover $V_\mathrm{max}$ of true inclined rotating discs, with accuracy better than 25 per cent, provided the data passes the following `data quality cuts': more than five PSF beams along the kinematic major axis; more than 25 voxels along the kinematic major axis; and $\rm inc\in[25,80]^{\circ}$. The typical performance, however, is far better than this 25 per cent upper limit. Indeed, the model accuracy for a simulated galaxy with the average data quality of the galaxies in our sample (which is $\sim 9$ PSF beams along the major axis, $\sim 55$ voxels along the major axis, and an inclination of $\sim55^{\circ}$) is $\sim 3$ per cent. 

Data passing these cuts can meaningfully be tested for their consistency with the inclined rotating disc model. The model should recover the kinematics of a genuine inclined rotating disc in this regime, so a failed fit would evidence true inconsistency with the model. It is more difficult to interpret the fits of data failing the data quality cuts. The model is unable to accurately recover the kinematics of a genuine inclined rotating disc  in this regime, so a failed fit cannot be uniquely attributed to a lack of disc rotation, and could equally be the result of inadequate data quality. 

MaNGA galaxies with \textsc{dap} maps failing the data quality cuts must therefore be removed from the sample. The kinematic properties are unknown prior to a successful fit, so we rely on the photometric properties as a crude proxy, removing all galaxies with less than five PSF beams and/or 25 voxels along the photometric major axis, as well as all galaxies with  $\arccos((b/a)_\mathrm{phot})>80^{\circ}$ or $\arccos((b/a)_\mathrm{phot})<25^{\circ}$. These constitute our final cuts, reducing the sample size by $\sim$30 per cent, and leaving a kinematic sample of 1862 galaxies that can meaningfully be tested for their consistency with the inclined rotating disc model.

A galaxy's observed photometric axis ratio, $ (b/a)_\mathrm{phot}$, depends on both its inclination angle and intrinsic axial ratio (see \citealt{Cappellari2016}). At fixed inclination angle, a galaxy with a bulge and/or a finite, non-zero thickness will appear more round (i.e. have larger $ (b/a)_\mathrm{phot}$) than a galaxy that is perfectly thin. Hence, $\arccos((b/a)_\mathrm{phot})$ is really a lower limit of the true inclination angle, with equality only in the case of an infinitesimally thin disc. The $\arccos((b/a)_\mathrm{phot})<25^{\circ}$ cut therefore removes a greater fraction of spheroids than discs. We have tested the effect of this bias by repeating our analysis with $\arccos((b/a)_\mathrm{phot})<25^{\circ}$ spheroids kept in the sample, on the grounds that their large $ (b/a)_\mathrm{phot}$ is more likely a consequence of their large intrinsic axial ratios than their being genuinely face-on. We confirm that the key results of this paper are stable to this test. 

\subsubsection{Assessment of quality of fits}\label{Assessment of quality of fits}
\begin{figure*}
    \includegraphics[width=\textwidth]{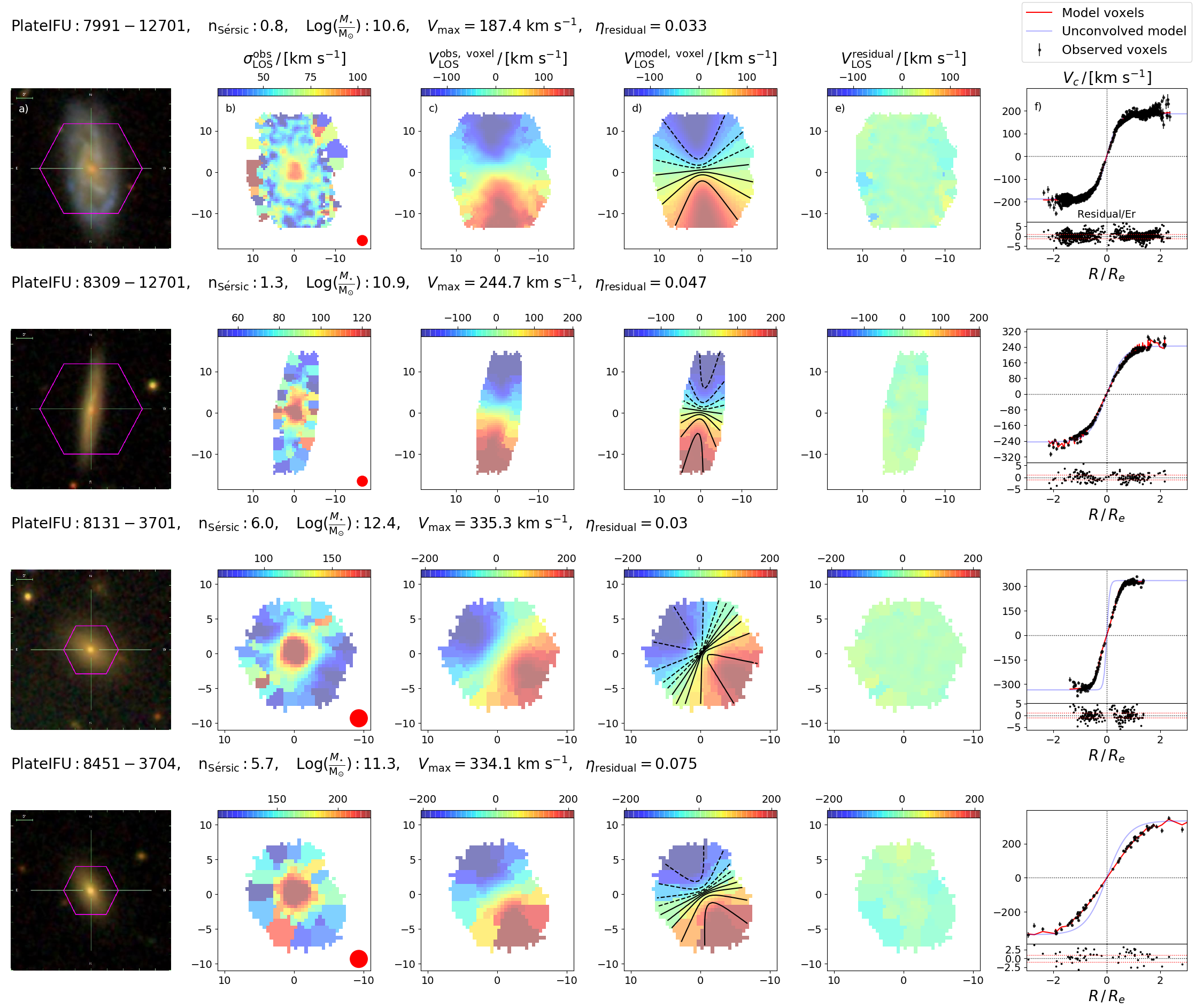}
    \caption{Four examples of successful kinematic modelling, showing low S\'ersic index galaxies in the top two rows and high S\'ersic index galaxies in the bottom two rows.  \textit{In each row, from left to right, panel a:} The SDSS $g,r,i$ composite image with the MaNGA hexagonal FoV overlaid in magenta. \textit{Panel b:} The observed (LOS) stellar velocity dispersion map taken from the DAP. The red ellipse represents the MaNGA PSF. \textit{Panel c:} The observed moment-1 map taken from the DAP. \textit{Panel d:} The model moment-1 map, shown as observed (i.e. in sky coordinates, and with PSF convolution and voxel binning). The black iso-velocity contours represent the estimated intrinsic LOS velocity - i.e. without convolution by the PSF and binning into voxels. \textit{Panel e:} The model residuals, defined as the difference between the data and the model.  \textit{Panel f, upper:} The data (black points) and model (solid red line) now shown in the PV plane - i.e. circular stellar velocity versus galactocentric distance from the kinematic centre. The estimated intrinsic circular stellar velocity (i.e. without convolution) is shown in blue. \textit{Panel f, lower:} The residuals ($\rm Data-Model$, as before) in the PV plane,  normalised by the uncertainties on the observed LOS stellar velocity estimates from the DAP. The dashed red lines indicate $\rm \pm 1~\sigma$ deviations. For each galaxy, we report the MaNGA PlateIFU, $\rm n_{S\acute{e}rsic}$, stellar mass, $V_\mathrm{max}$, which has a typical error of $\sim3$ per cent,  and $\eta_\mathrm{residual}$, which is small (less than 0.2) for  these well fit galaxies. These examples demonstrate the inclined rotating disc model's success in fitting the kinematics of a range of galaxy types, both in terms of S\'ersic index  and $M_{\star}$.}
        \label{fig:GoodFits_2Dfit_pvdiagram}
\end{figure*} 
We attempt to fit all 1862 galaxies that pass the data quality cuts, and adopt a two stage approach for assessing the quality of each fit. The first stage imposes a set of quantitative cuts. We require all of the parameters to be estimated within their limits (i.e. not to have reached their  bounds) and their uncertainties to be well determined (i.e. not NaN). We also define the following statistic to quantify the success of a fit: 
\begin{equation}
\label{EQN:Residual}
    \rm
    \eta_{residual} =  \frac{\sum|\left(V^{model, voxel}_\mathrm{LOS}-V^{obs, voxel}_\mathrm{LOS}\right)\,/\,\delta V^{obs, voxel}_\mathrm{LOS}|} {\sum|V^{obs, voxel}_\mathrm{LOS}\,/\,\delta V^{obs, voxel}_\mathrm{LOS}|}
\end{equation}
In words, $\eta_\mathrm{residual}$ is the weighted average absolute deviation between the data and model, normalised by the weighted average value of the data, such that low values of $\eta_\mathrm{residual}$ are associated with good fits. The weighting is given by the inverse of $\rm \delta V^{obs, voxel}_\mathrm{LOS}$, which imposes a greater penalty on discrepancies between the data and the model in regions where the data are measured with high confidence. 

We compare the residual statistic with our visual assessment of the fits (see next paragraph) and identify $\eta_\mathrm{residual}=0.15$ as the value above which a fit is more likely to be visually classified as `failed' than `passed'. The $\eta_\mathrm{residual}$ statisitc is not rigorous, however, and we place a greater emphasis on the visual classification. We therefore choose a slightly larger value of $\eta_\mathrm{residual}=0.2$ for the quantitative cut, which is the value above which fits are  more than twice as likely to be visually classified as `failed' than as `passed'. We stress that we have visually examined every fit to ensure  that failed fits are identified. The $\eta_\mathrm{residual}$ cut is only included as an extra layer of quality assurance and has a minimal effect on our sample, failing just $\rm \sim5$ per cent of the fits that we visually classified as `passed'.

We visually inspect the fits in the second stage of quality assurance. We describe this process with reference to Fig. \ref{fig:GoodFits_2Dfit_pvdiagram}, which shows four example galaxies whose kinematics are well fit by the inclined rotating disc model. For each fit, we show six panels: the SDSS $g,r,i$ composite image; the moment-2 map from the \textsc{dap} ($\sigma_\mathrm{LOS}^\mathrm{obs}$); the moment-1 map from the \textsc{dap} ($\rm V_\mathrm{LOS}^{obs, voxel}$); the best fitting inclined rotating disc moment-1 model ($\rm V_\mathrm{LOS}^{model, voxel}$); the residuals map ($\rm V_\mathrm{LOS}^{obs, voxel}-V_\mathrm{LOS}^{model, voxel}$); and the position-velocity (PV) diagram, which plots the circular velocity (rather than LOS) as a function of galactocentric radius. All four of these well fit galaxies exhibit the following features: the $\rm V^{obs, voxel}_\mathrm{LOS}$ map clearly shows ordered rotation, with redshifted stars on one side of the kinematic centre and blueshifted stars on the other; the $\rm V^{model, voxel}_\mathrm{LOS}$ and $\rm V^{obs, voxel}_\mathrm{LOS}$ maps are visually consistent; and the residual map shows no evidence of excess structure missed by our kinematic model. We verify the quality of the fit in the PV diagram, where all four galaxies show rotation profiles that are consistent with the smooth `S'-shape typical of rotation-dominated systems (see equation \ref{EQN:LOSVelocity}). Furthermore, we show the residuals normalised by $\rm \delta V^{obs, voxel}_\mathrm{LOS}$ in the lower panel and confirm that they lack radial structure and are consistent with random noise. The three lead authors independently assessed the quality of 250 fits against these visual requirements  (rating them `pass' or `fail') and unanimously agreed on the verdict in over 90 per cent of cases. The lead author subsequently reviewed the remaining $\sim 1600$ fits.

Fig. \ref{fig:GoodFits_2Dfit_pvdiagram} demonstrates the model's success over a wide range of galaxy types (see Appendix \ref{Example Fits} for more examples). In particular, the model is able to fit both low S\'ersic index photometric discs (as expected) as well as high S\'ersic index photometric spheroids in some cases. This result is striking given the simplicity of our kinematic model. We especially highlight the low residuals and note that this is likely because we are using stellar (rather than gas) velocity maps, which are less affected by non-virialised motions such as inflows and outflows. In other words, the stellar systems are highly relaxed. Of course, the model fails for galaxies that are not rotating, but its success in describing fast rotator photometric spheroids validates our methodology of attempting to fit the kinematics of all galaxies (i.e. not only photometric discs).  We further validate this success against alternative methods in Appendix \ref{Testing the kinematic model}.  

We primarily use S\'ersic profile moment-0 maps from the NSA catalogue during PSF convolution and Voronoi binning since they model the intrinsic brightness profiles. However, there are a number of high S\'ersic index galaxies (typically $\rm n_{S\acute{e}rsic}>4$) whose kinematic fits are visually improved by adopting the \textsc{dap} moment-0 maps (see Fig. \ref{fig:PVSquiggle_2Dfit_pvdiagram}, for example). We choose to adopt the \textsc{dap} moment-0 maps for these galaxies, but we recognise that they are PSF convolved and are shallower than the true intrinsic brightness profiles. We therefore repeat the analysis in this paper, adopting the simplistic kinematic model introduced in the next section for the $\rm \sim 400$ galaxies for which we adopt \textsc{dap} moment-0 maps in the fiducial analysis, and confirm that the key results are stable to this test in Appendix \ref{Testing the kinematic model}.

\subsubsection{Corrected velocity dispersions}
\begin{figure*}
    \includegraphics[width=\textwidth]{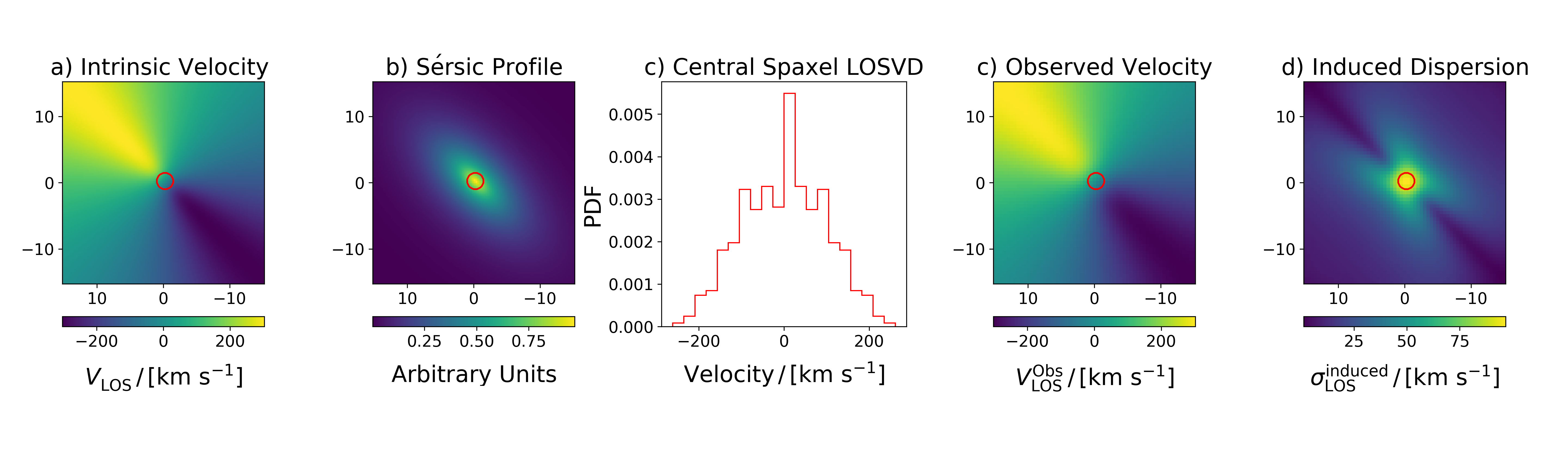}
    \caption{Inducing velocity dispersion from differential disc rotation. \textit{From left to right, panel a:)} The LOS velocity map for a mock galaxy with a large kinematic inclination ($\rm 60^{\circ}$), a short kinematic lengthscale relative to the MaNGA PSF ($r_c=2.5\, \mathrm{arcsec}$), and a relatively large Vmax ($\rm 350\, km~s^{-1}$) . The mock galaxy properties are deliberately chosen to induce a large velocity dispersion. \textit{Panel b:} The assumed light profile, with $\rm n_{S\acute{e}rsic}=1$. \textit{Panel c:} The central spaxel LOS velocity distribution (LOSVD). This is the distribution of the LOS velocities of all stars, weighted by their surface brightness and PSF response, with the PSF centred on the central spaxel. The LOSVD corresponds to the emission/absorption line profile expected for spaxels with zero intrinsic velocity dispersion. \textit{Panel d:} The observed LOS velocity map, found by calculating the mean of the LOSVD in each spaxel. This map is blurry relative to \textit{panel a}, which is a well-known consequence of beam smearing. \textit{Panel e:} The induced velocity dispersion map, found by calculating the standard deviation of the LOSVD in each spaxel. The induced dispersion is largest in regions with large velocity gradients, as expected. The red ellipse in all maps represents
    the MaNGA PSF centred on the central spaxel, shown for scale.}
        \label{fig:InducedSigma_inc_kin=60.0}
\end{figure*}

Artificial velocity dispersion is induced when a galaxy exhibiting differential disc rotation is observed with a finite PSF. We simulate this effect for a mock galaxy in Fig. \ref{fig:InducedSigma_inc_kin=60.0}. In \textit{panel a}  we show the intrinsic LOS velocity map, which is the inclined rotating disc model by design, and in \textit{panel b}  we show the galaxy's surface brightness map, which is arbitrarily chosen to have a S\'ersic profile. We demonstrate the effect of differential disc rotation by considering the central spaxel, but note that the following discussion applies equally to every spaxel in the map. The central spaxel receives flux not only from stars at the galaxy centre, which have LOS velocity $V_\mathrm{LOS}=0$, but also from stars offset from the galaxy centre, which have non-zero $ V_\mathrm{LOS}$.  In \textit{panel c} of Fig. \ref{fig:InducedSigma_inc_kin=60.0}  we show the  distribution of stellar velocities in this set up, with the stellar velocity of each region weighted by the product of the brightness and the PSF amplitude. This distribution is known as the line of sight velocity distribution (LOSVD). The observed LOS velocity, $\rm V^\mathrm{obs}_\mathrm{LOS}$, is given by the mean of the LOSVD, which is $\rm \sim$zero, as expected. In \textit{panel d}  we show the $\rm V^\mathrm{obs}_\mathrm{LOS}$ map, and note the blurring of this map relative to the intrinsic $ V_\mathrm{LOS}$ map. This blurring is commonly referred to as beam smearing.

Our focus here is on the non-zero width of the LOSVD. The ordered rotation of the galaxy has thus induced a non-zero velocity dispersion. For MaNGA observations, this effect will broaden the gas emission lines and stellar absorption features, leading to overestimates of the intrinsic velocity dispersion. In the right panel of Fig. \ref{fig:InducedSigma_inc_kin=60.0}, we  parameterise this broadening via the standard deviation of the LOSVD in each spaxel, $\rm \sigma_\mathrm{LOS}^\mathrm{induced}$ - i.e. the induced LOS velocity dispersion.

As expected, $\rm \sigma_\mathrm{LOS}^\mathrm{induced}$ is large in regions that have a significant $ V_\mathrm{LOS}$ gradient, such as the galaxy centre. Similarly, galaxies with large $V_\mathrm{max}$, steep rotation curves (i.e. small $r_c$), and small inclination angles have steep $ V_\mathrm{LOS}$ gradients and consequently have $\rm \sigma_\mathrm{LOS}^\mathrm{induced}$ as large as $\rm  150\,km~s^{-1}$ in the central regions. This demonstrates the need to correct the observed $\sigma_\mathrm{LOS}^\mathrm{obs}$ estimates from the \textsc{dap}, but we note that these extreme galaxies are rare in practice. Indeed, we find that the induced velocity dispersion typically results in only a small ($\sim 10$ per cent) overestimation of $\rm \sigma_\mathrm{LOS}^\mathrm{induced}$.

Nonetheless, we rigorously correct for the effect of differential disc rotation.  We use the kinematic model to estimate  $\rm \sigma_\mathrm{LOS}^\mathrm{induced}$, and we calculate $\sigma_\mathrm{LOS}^\mathrm{intrinsic}$ as follows:
\begin{equation}
\label{EQN:SigmaQuadratureCorrection}
    \sigma_\mathrm{LOS}^\mathrm{intrinsic}=\sqrt{{\left(\sigma_\mathrm{LOS}^\mathrm{obs}\right)}^2-{\left(\sigma_\mathrm{LOS}^\mathrm{induced}\right)}^2}
\end{equation} 
where $\sigma_\mathrm{LOS}^\mathrm{obs}$ is the LOS velocity dispersion estimate reported by the \textsc{dap}. 

Estimating $\rm \sigma_\mathrm{LOS}^\mathrm{induced}$ requires a reliable kinematic fit of the ordered rotation, so equation \ref{EQN:SigmaQuadratureCorrection} cannot be used to correct the moment-2 maps of galaxies that are inconsistent with the inclined rotating disc model. We note that $\rm \sigma_\mathrm{LOS}^\mathrm{induced}$ is likely to be small in these galaxies since they generally lack strong velocity gradients (i.e. they do not appear to be rotating). Furthermore, they tend to be spheroidal galaxies with large $\sigma_\mathrm{LOS}^\mathrm{obs}$, and the difference between the $\sigma_\mathrm{LOS}^\mathrm{intrinsic}$ and $\sigma_\mathrm{LOS}^\mathrm{obs}$ at fixed $\rm \sigma_\mathrm{LOS}^\mathrm{induced}$ decreases with increasing $\sigma_\mathrm{LOS}^\mathrm{obs}$, as shown in equation \ref{EQN:SigmaQuadratureCorrection}. These compounding effects ensure that any overestimation of the velocity dispersion in galaxies that we fail to fit is likely to be small and to have little influence on our key results.  

The velocity dispersion correction completes our detailed kinematic model, so we take a moment to summarise the methodology as follows:
\begin{enumerate}
    \item First, we take  preexisting estimates of moment-0. 
    \item Second, we use moment-0, the observed PSF and the observed LOS velocity map to determine moment-1 via an inclined rotating disc model.
    \item Last, we use moment-0, moment-1 and the observed PSF  to correct moment-2 for the induced effect of differential disc rotation.
\end{enumerate} 
This approach ignores any backward steps in which higher order moments are used to constrain lower order moments, such as the simultaneous use of moment-1 and moment-2 to constrain the kinematic centre. However, we note that these steps are second order effects, and emphasise that our model includes the dominant, first order dependencies between moment-0, moment-1 and moment-2, as outlined above. 

\subsection{Simplistic kinematic model}\label{Simplistic kinematic model}
We find that $\rm \sim30$ per cent of galaxies passing the data quality cuts have kinematics that are inconsistent with inclined disc rotation. In this section, we present an alternative, simplistic method for estimating their kinematics. 

The goal of the simple method is to achieve approximate constraints on the rotation of galaxies that are inconsistent with the inclined rotating disc model. The majority of these galaxies are slowly rotating, and knowledge of this alone is sufficient for our study of galaxy quenching. Thus, the simple method is designed not to give a precise estimate of $V_\mathrm{max}$, but to constrain $V_\mathrm{max}$ sufficiently such that it can be compared with  the velocity dispersion to identify a galaxy as a slow rotator or fast rotator. Indeed, we will show that different formulations of the simple method achieve relatively tight bounds on the rotational state of galaxies, even in the absence of full kinematic fitting.

Previous works have adopted the `histogram technique', typically used to determine kinematics from HI linewidths \citep{Catinella2012}, to measure the maximum rotational velocity in IFU data (e.g. \citealt{Cortese2014, Barat2019, Oh2020}). In this approach, $ V_\mathrm{max}^\mathrm{Simple}$ is given by
\begin{equation}
\label{EQN:VmaxSimple}
    V_\mathrm{max}^\mathrm{Simple} = \frac{V_{95}-V_{5}}{2\sin(\mathrm{inc})}
\end{equation}
where $\rm V_{95}$ and $\rm V_{5}$ are the 95th and 5th percentiles of the histogram of voxel LOS stellar velocities within $\rm 1.5\,R_e$. Note, we use the 95th and 5th percentiles, rather than the 90th and 10th that are typically used, since we consider the histogram of voxel LOS velocities, which are less noisy than their spaxel counterparts. 

The denominator in equation \ref{EQN:VmaxSimple} requires an estimate of the kinematic inclination angle. Previous works have approximated this angle using the photometric axis ratio (i.e. taking ${\mathrm{inc}=\arccos((b/a)_\mathrm{phot})}$), but the finite thickness of galaxies ensures that $\arccos((b/a)_\mathrm{phot})$ is in fact a lower limit of the true inclination. Adopting this estimate in equation \ref{EQN:VmaxSimple} therefore achieves an effective upper limit on $ V_\mathrm{max}^\mathrm{Simple}$. This effect is more significant for galaxies with large intrinsic axial ratios, so we treat photometric discs and photometric spheroids separately in the simple method. 

Disc galaxies have low intrinsic axial ratios \citep{Catinella2012, Bluck2014, Cortese2014, Oh2020}, so $\arccos((b/a)_\mathrm{phot})$ is a reasonable proxy of their inclination angles. We compare the simple method and the full kinematic model for galaxies that have a disc ($\rm n_{S\acute{e}rsic}<3$) and are well modelled as inclined disc rotators. This corresponds to 85 per cent of the discs in our sample.  We find that $ V_\mathrm{max}^\mathrm{Simple}$ found using $\arccos((b/a)_\mathrm{phot})$ to approximate inc underestimates $\rm V_{max}^{Fit}$ taken from the full kinematic model, with $\mathrm{Bias}=-25.3\mathrm{\, km~s^{-1}}$. 

\begin{figure}
    \includegraphics[width=\columnwidth]{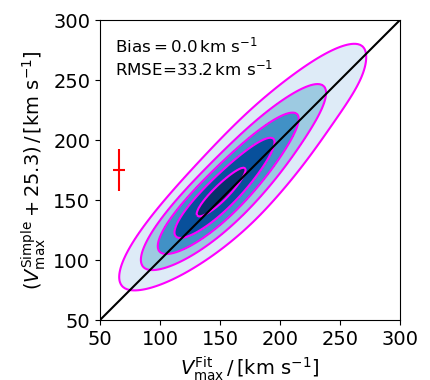}
    \caption{A direct comparison of $V_\mathrm{max}$ estimated using the full kinematic model on the x-axis and the bias-corrected simple method on the y-axis for discs that have good kinematic fits. Linearly spaced density contours are shown with blue shading and magenta lines, and we display the bias, which is zero by design, and the root mean square error (RMSE) of residuals from the black 1-1 line. The estimates are highly correlated with low scatter. The red error bars represent the $1\sigma_\mathrm{err}$ uncertainties. Note, the uncertainty on $V_\mathrm{max}^\mathrm{Fit}$ is estimated from our tests of the inclined rotating disc on mock galaxy data. The uncertainties on all other kinematic parameters presented in this paper are estimated by propagating the uncertainties on the the LOS stellar velocity and LOS stellar velocity dispersion taken from the DAP. The uncertainty on $V_\mathrm{max}$  Note, the best fit line does not perfectly intersect the peak of the density contours since the distribution is not symmetric.} 
        \label{fig:Vmax_FitVsSimple}
\end{figure}

We correct for this small bias and in the left panel of Fig. \ref{fig:Vmax_FitVsSimple}  we show the bias-corrected estimate of $ V_\mathrm{max}^\mathrm{Simple}$ on the y-axis and  $\rm V_{max}^{Fit}$ on the x-axis. The consistency between the two estimates is striking; they are highly correlated with low scatter. We emphasise that the inclined rotating disc model is far more complex than the simple method, most notably in its accounting for the effect of beam smearing and explicitly fitting the kinematic inclination (rather than merely assuming it via photometry). This good agreement between the two methodologies thus acts as an important check on our kinematic modelling and builds confidence in our approach. 

We adopt the bias-corrected simple method for the 15 per cent of discs that do no have a good fit, on the grounds that they are not systematically different to the discs shown in Fig. \ref{fig:Vmax_FitVsSimple}. This assumption may be overly simplistic, but it impacts only 15 per cent of discs and hence it is unlikely to have a significant influence on our results. We also test restricting our analysis to galaxies that are well fit by the inclined rotating disc model, and we confirm in Appendix \ref{Testsonthestabilityoftheresults} that our key results are robust. 

The treatment of photometric spheroids ($\rm n_{S\acute{e}rsic}>3$) is more challenging since they have large intrinsic axial ratios. Adopting ${\mathrm{inc}=\arccos((b/a)_\mathrm{phot})}$ in equation \ref{EQN:VmaxSimple}  for these galaxies could therefore return a significant overestimate of $V_\mathrm{max}$. Instead, we explore four different estimates of inc and examine their influence on our key results. 
\begin{enumerate}
    \item First, we use $\arccos((b/a)_\mathrm{phot})$, which provides an upper limit on $ V_\mathrm{max}^\mathrm{Simple}$.
    \item Second, we assume the galaxies are perfectly edge-on with $\rm inc=90^{\circ}$, which provides a lower limit on $ V_\mathrm{max}^\mathrm{Simple}$. Given our restricted focus to galaxies with $\arccos((b/a)_\mathrm{phot})>25^{\circ}$,  these first two approaches bound $ V_\mathrm{max}^\mathrm{Simple}$ to within a factor of $\sim2.5$.
    \item Third, we adopt  $\rm {inc=60^{\circ}}$, which is expectation value for the viewing angle of galaxies when distributed isotropically in 3D space. If the galaxy has $\rm {\arccos((b/a)_{phot})>60^{\circ}}$, however, we employ ${\mathrm{inc}=\arccos((b/a)_\mathrm{phot})}$ as a known lower limit.
    \item Finally, and most precisely, we estimate inc in a Bayesian fashion. We model the distribution of the intrinsic axial ratios ($(b/a)_\mathrm{int}^\mathrm{model}$) of photometric spheroids as a Gaussian with mean value $\mu$ and standard deviation $\sigma$. We then transform this distribution into a model distribution of observed photometric axial ratios ($(b/a)_\mathrm{phot}^\mathrm{model}$) by randomly drawing objects from the distribution of $(b/a)_\mathrm{int}^\mathrm{model}$, viewing each object with random angles drawn form the isotropic distribution of viewing angles in 3D space, and calculating the corresponding observed photometric axis ratios using the following equation equation taken from \citet{Cappellari2016}:
    \begin{equation}
    \label{eqn:IntrinsicObservedAxisRatio}
    \sin(\mathrm{inc}) = \sqrt{\frac{1-\left(\frac{b}{a}\right)_\mathrm{phot}^2}{1-\left(\frac{b}{a}\right)_\mathrm{int}^2}}\,.
    \end{equation}
    We compare the model distribution of $(b/a)_\mathrm{phot}^\mathrm{model}$ with the observed distribution of $(b/a)_\mathrm{phot}$  for SDSS spheroids, and use \textsc{lmfit} and $\rm \chi^2$ minimisation to find the following best-fitting parameters:  $\mu=0.71$ and  $\sigma=0.16$. 
    
    We then calculate inc for each spheroid in our sample by comparing the mean $(b/a)_\mathrm{int}^\mathrm{model}$ with $(b/a)_\mathrm{phot}$ (using equation \ref{eqn:IntrinsicObservedAxisRatio}). We assume the galaxy is viewed edge-on (i.e. inc=$90^{\circ}$) if $(b/a)_\mathrm{phot}$ is less than the mean $(b/a)_\mathrm{int}^\mathrm{model}$. Of course, not all galaxies have intrinsic axial ratios equal to the mean of the distribution, but this approach gives the correct inclination angle on average across the sample. 
    
    We estimate the typical error, from the Bayesian approach,  on inc by experimenting with three values of $(b/a)_\mathrm{int}$: first, we adopt the value  $\mu-\sigma$ and label the corresponding value of the inclination $\mathrm{inc}_{\mu-\sigma}$; second, we adopt the value $\mu$ and label the corresponding value of the inclination $\mathrm{inc}_{\mu}$; and last, we adopt the value $\mu+\sigma$ and label the corresponding value of the inclination, $\mathrm{inc}_{\mu+\sigma}$. We note that the correct value of the inclination, $\rm inc_{true}$, will lie in the range ${[\mathrm{inc}_{\mu-\sigma},\mathrm{inc}_{\mu+\sigma}]}$ 68 per cent of the time. By considering equation \ref{eqn:IntrinsicObservedAxisRatio} with fixed $(b/a)_\mathrm{phot}$, it is straightforward to show that $\sin(\mathrm{inc}_{\mu+\sigma})\sim 0.85\sin(\mathrm{inc}_{\mu})$ and $\sin(\mathrm{inc}_{\mu-\sigma})\sim1.37 \sin(\mathrm{inc}_{\mu})$, where we have used $\rm \mu=0.71$ and $\sigma=0.16$. The maximum velocity in the simple method is inversely related to the sine of the inclination angle (see equation \ref{EQN:VmaxSimple}), and hence we estimate a typical uncertainty on $ V_\mathrm{max}^\mathrm{Simple}$ of only 20-30 per cent.

\end{enumerate}  
We adopt the Bayesian approach in our fiducial sample since it gives the most accurate constraints. Nonetheless, we have also rigorously tested using the other three approaches, just to see how sensitive our results are to this issue. Happily, all methods yield identical final conclusions and so our results are incredibly stable to our ignorance of kinematic inclination angle in spheroids.

We extend the methodology of equation \ref{EQN:VmaxSimple} to derive spatially resolved circular velocity estimates by rearranging equations \ref{EQN:LOSVelocity} as follows:
\begin{equation}
\label{EQN:LOS_to_Circular}
    V_\mathrm{c}^\mathrm{Simple}(r) = \frac{V_\mathrm{LOS}(r,\theta)-V_\mathrm{sys}}{\cos(\theta)\sin(\mathrm{inc})}
\end{equation}
where $\theta$  is the angle measured anticlockwise from $\rm PA_{phot}$ in the plane with inclination angle, inc, and we estimate $V_\mathrm{sys}$ as the median LOS stellar velocity within $\rm 1\,R_e$. Voxels close to the photometric minor axis have $\rm \theta\sim90^{\circ}$. The function $\cos(\theta)^{-1}$ is steep in this regime and tends asymptotically to infinity, so even  small errors on $\theta$ can cause very large errors on $V_\mathrm{c}^\mathrm{Simple}$. We therefore only calculate $V_\mathrm{c}^\mathrm{Simple}$ in voxels that are more than $\rm 30^{\circ}$ offset from the minor axes, and rely on our assumption of axisymmetry when deriving global parameters in the next section. As in equation \ref{EQN:VmaxSimple}, we adopt $\mathrm{inc}=\arccos((b/a)_\mathrm{phot})$ for discs. We adopt the Bayesian approach for spheroids and have also tested against the following three alternatives: $\mathrm{inc}=\arccos((b/a)_\mathrm{phot})$, edge-on ($\rm inc=80^{\circ}$), $\rm inc=60^{\circ}$. We define edge-on here as $\rm inc=80^{\circ}$, rather than $\rm inc=90^{\circ}$, since $\rm \theta$ is not defined for true edge-on systems\footnote{The values of $\sin(90^{\circ})$ and $\sin(80^{\circ})$ differ by only $\sim 1$ per cent, so we make this pragamatic choice to achieve a well defined lower limit for $V_\mathrm{c}^\mathrm{Simple}$. Though somewhat arbitrary, $\rm 80^{\circ}$ is chosen for consistency with our removal of $\arccos((b/a)_\mathrm{phot})>80^{\circ}$ galaxies in the previous section.}.

\subsection{Kinematic parameters}\label{Kinematic parameters}
We study the global quenching of galaxies in this work. We do not examine the relationship between spatially resolved kinematics and quenching, since previous works have found that the shutdown of star formation is governed primarily by processes that affect galaxies as a whole, rather than processes that operate on local scales within galaxies \citep{Bluck2020a, Bluck2020b}. Moreover, it is natural and sensible to start with the more simple problem of global quenching before examining quenching on spatially resolved scales. In this section, we define and estimate seven global kinematic parameters that we use in later sections to study quenching. We also include an eighth parameter, the global stellar mass ($M_{\star}$), given its prominence in the quenching literature (e.g. \cite{Baldry2006, Peng2010, Peng2012}).

Although the MaNGA survey is designed to map galaxies out to $\rm 1.5\,R_e$, we find a number of galaxies whose annuli beyond $\rm 1\,Re$ are only partially mapped. We therefore measure the global parameters over spaxels within $\rm 1\,R_{e,kin}$ to ensure that all galaxies have data measured on the same spatial scales, where $\rm 1\, R_{e,kin}$ is the locus of points separated by $\rm 1\,R_e$ from the kinematic centre, measured in the plane of the kinematic disc. We adopt the $\rm 1\, R_{e,kin}$ radius rather than its photometric counterpart to measure the global parameters, since it reflects the assumed axisymmetry of galaxy kinematics.

The average circular velocity is defined as follows:
\begin{equation}
\label{EQN:AvgCircularVelocity}
\overline{V}=\frac{\sum (\Sigma_{\star}\cdot V_\mathrm{c})}{\sum (\Sigma_{\star})}
\end{equation} 
where $\rm \Sigma_{\star}$ is the stellar mass surface density taken from \textsc{pipe3d}, and $ V_\mathrm{c}$ is the estimated circular velocity of a given spaxel. Unless otherwise stated, the sums in this section are defined over all spaxels within $\rm 1\,R_{e,kin}$. Large values of $ \overline{V}$ relate to galaxies in which the stars orbit at high speed. This occurs in galaxy discs, where young stars form, so it is reasonable to expect a relationship  between $ \overline{V}$ and the level of star formation within galaxies, and perhaps with galaxy quenching. 

The average velocity dispersion is similarly defined:
\begin{equation}
\label{EQN:AvgDispersion}
\overline{\sigma}
=\frac{\sum (\Sigma_{\star}\cdot\sqrt{3}\,\sigma_\mathrm{LOS})}{\sum(\Sigma_{\star})}
\end{equation} 
As discussed, we use $\sigma_\mathrm{LOS}=\sigma_\mathrm{LOS}^\mathrm{intrinsic}$ when we have a good kinematic fit, and $\sigma_\mathrm{LOS}=\sigma_\mathrm{LOS}^\mathrm{obs}$ otherwise. The factor of $\rm \sqrt{3}$ converts the LOS velocity dispersion to the total dispersion in 3D space, with the implicit assumption that the velocity dispersion vector is  isotropic. This assumption is invalid for individual galaxies \citep{Cappellari2016}, but it does not introduce a systematic bias and is reasonable on average since our sample contains $\rm \sim 2000$ galaxies with a wide range of orientations. Previous works have identified a strong relationship between galaxy quenching and velocity dispersion measured in the central kpc \citep{Wake2012, Bluck2016, Bluck2020a, Bluck2020b}. We define $\overline{\sigma}$ within $\rm 1\, R_{e,kin}$ for consistency with the other parameters in our set, but we have confirmed that our results hold for both definitions. 

We quantify the ratio of ordered to disordered velocity, $\overline{V}\,/\,\overline{\sigma}$, which is the kinematic analogue of the disc to bulge mass ratio ($D/B$). Prominent spheroidal structures have frequently been associated with galaxy quenching \citep{Wuyts2011, Bell2012,  Bluck2014, Omand2014, Morselli2016, Pandya2017}, not least because of the common understanding that discs are mostly blue whilst spheroids are mostly red (e.g. \citealt{CameronDriver2009,  Gadotti2009, Cappellari2011a, Bell2012, Lang2014, Omand2014, Bluck2014, Bluck2016}). Nonetheless, $D/B$ is a crude descriptor of a how spheroidal a particular galaxy is. Indeed, \citet{Lilly2016} show that a `bulge' can be reproduced in a model of pure disc galaxies in which the disc scale length increases with time. The kinematic ratio $\overline{V}\,/\,\overline{\sigma}$, on the other hand, cleanly separates spheroidal and disc galaxies through their fundamental difference: discs are rotation-dominated, whilst spheroids are dispersion-dominated.

We define two parameters that are fundamentally connected to the physics of stellar orbits: the average specific kinetic energy and the average specific angular momentum of the stars. The average specific kinetic energy is defined as follows: 
\begin{equation}
\label{EQN:KineticEnergy}
\overline{\mathcal{E}_k}=\frac{\mathrm{KE}}{M_{\star}}=\frac{1}{2}\cdot V_\mathrm{rms}^2
\end{equation}
where KE is the total kinetic energy of the stars within $\rm 1\,R_{e,kin}$, and $V_\mathrm{rms}$ is the root mean square total velocity of the stars within $\rm 1\,R_{e,kin}$ \citep{Cappellari2016}, defined as follows:
\begin{equation}
\label{EQN:TotalVelocity}
V_\mathrm{rms}=\sqrt{\frac{\sum \left(\Sigma_{\star} \cdot \left(V_\mathrm{c}^2 +3\sigma_\mathrm{LOS}^2\right)\right)}{\sum \left(\Sigma_{\star}\right)}}.
\end{equation} 
The galaxies in our sample do not show signs of a recent merger or interaction with a companion galaxy, so we assume that they are virialised. Invoking the virial theorem, the specific gravitational potential energy is given by
\begin{equation}
\label{EQN:VirialTHeorem}
\phi_\mathrm{G}=-2\overline{\mathcal{E}_k}.
\end{equation} 
The specific gravitational potential energy is technically  the mean mass-weighted gravitational potential experienced by the stellar system, and it  depends on a galaxy's dynamical mass. We choose not to estimate the dynamical mass this way, since the gravitational radius and virial parameter are largely unknown.

The average specific angular momentum is defined as follows:
\begin{equation}
\label{EQN:AngMom}
\overline{j}=\frac{\sum (\Sigma_{\star} \cdot r \cdot V_\mathrm{c})}{\sum(\Sigma_{\star})}
\end{equation} 
where $r$ is the distance of a spaxel (in kpc) from the kinematic centre in the galaxy plane. The specific angular momentum builds on $ \overline{V}$ by accounting for the spatial distribution of the rotation, such that it distinguishes between galaxies with rotation on large and small spatial scales.

We include the dimensionless spin parameter, $\lambda$, since it is commonly used in the literature as a crude classifier of a galaxy's kinematic state. The advantage of this parameter is that it does not rely on a parametric model of the ordered rotation, and is measured directly using the observed LOS velocity as follows:
\begin{equation}
\label{EQN:SpinParam}
\lambda=\frac{\sum  (F\cdot r \cdot |V^\mathrm{obs,voxel}_\mathrm{LOS}|)}{\sum \left(F\cdot r \cdot \sqrt{{V_\mathrm{LOS}}^2+\sigma_\mathrm{LOS}^2}\right)}
\end{equation} 
where we take F as the $g$-band flux of a particular voxel from the \textsc{dap}. Unlike equations \ref{EQN:AvgCircularVelocity}-\ref{EQN:AngMom},  the sums in equation \ref{EQN:SpinParam} are defined over voxels (rather than spaxels) within 1$\rm R_{e}$ (rather than $\rm 1\, R_{e,kin}$) for consistency with the literature \citep{Emsellem2007,Cappellari2016, Graham2018}. We have implicitly corrected for the effect of beam smearing on velocity dispersion by using $\sigma_\mathrm{LOS}=\sigma_\mathrm{LOS}^\mathrm{intrinsic}$ in the denominator of equation \ref{EQN:SpinParam} where possible. Similar to the relationship between $\overline{j}$ and $ \overline{V}$,  $\lambda$ builds on $\overline{V}\,/\,\overline{\sigma}$ by accounting for the spatial distribution of the kinematics.

We briefly mentioned the challenges of measuring the dynamical mass via the virial theorem. We circumvent these issues by considering galactic dynamics, rather than energetics. Stellar orbits trace the total mass within the orbital radius under the assumption of a simple spherical geometry, by Newton's first Theorem. We balance the gravitational force at $\rm 1\,R_{e,kin}$ with the centrifugal force and estimate the dynamical mass within $\rm 1\, R_{e,kin}$  as follows:
\begin{equation}
\label{EQN:DynamicalMass}
M_\mathrm{D}=V_\mathrm{rms,1Re}^{2}\cdot\frac{R_{e}}{G}=\overline{\left(V_\mathrm{c, 1R_e}^2+3\sigma_\mathrm{LOS,1Re}^2\right)}\cdot\frac{R_{e}}{G}
\end{equation} 
where G is the gravitational constant, and the subscript $\rm 1\,R_e$ makes explicit that the average is taken over spaxels in a thin elliptical annulus at $\rm 1\,R_{e,kin}$ with total width $\rm 0.25\,R_{e,kin}$, rather than over all the spaxels within  $\rm 1\,R_{e,kin}$. 

Unlike mass estimates from  spectroscopy or photometry which often track only a single phase of mass, $M_\mathrm{D}$ tracks all components of mass, including baryonic (stellar as well as gas in all phases) and dark matter. This highlights one of the key advantages of a kinematic study of quenching. Nonetheless, we also include in our parameter set an estimate of the total stellar mass, $M_{\star}$, taken from the NSA catalogue. We adopt the total mass, rather than the mass within $\rm 1\,R_{e,kin}$, given its frequent use in the literature. In the next section, however, we do briefly consider the total stellar  mass within $\rm 1\, R_{e,kin}$, which is simply found by summing $\rm \Sigma_{\star}$ estimates from \textsc{pipe3d} within an aperture of the same size.

\subsection{Tests: Kinematic scaling relations and connection to morphology}\label{Tests: Kinematic scaling relations and connection to morphology}
\begin{figure*}
    \includegraphics[width=\textwidth]{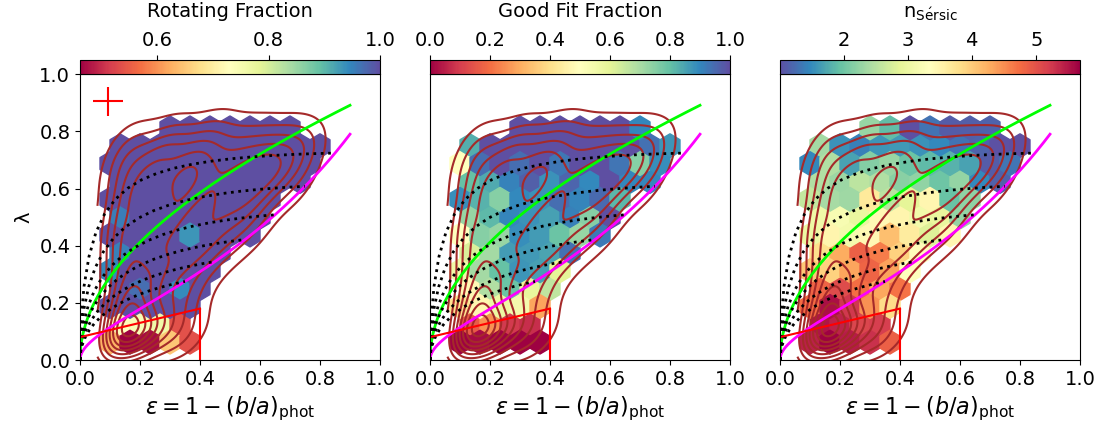}
    \caption{The ($\lambda,\epsilon$) plane. For all panels, the green line is the prediction for an edge-on isotropic rotator \citep{Binney2005}, and the magenta line is the prediction for an edge-on rotator with anisotropy parameter $\rm \delta = 0.7\epsilon_\mathrm{intr}$ \citep{Cappellari2007}. The black dotted lines show how galaxies lying on the magenta line with a fixed intrinsic ellipticity appear as they are viewed at decreasing inclination angle - i.e. from edge-on to face-on. The red lines in the lower-left corner of each panel mark the slow rotator region \citep{Emsellem2011}, and linearly spaced brown contours depict the density of galaxies in the plane.  \textit{Left panel:} The hexagons are colour coded by the fraction of galaxies that are visually determined to show ordered rotation. Slow rotator-classified galaxies often don't appear to be rotating, whilst almost all fast rotator-classified galaxies show visual evidence of rotation. The red error bars represent the $1\sigma_\mathrm{err}$ uncertainties. \textit{Central panel:} The hexagons are colour coded by the fraction of galaxies that are well fit by our kinematic model. We are able to fit fast rotators but not slow rotators, and our ability to fit fast rotators increases with $\lambda$. This good agreement between our kinematic model and the $(\lambda, \epsilon)$ plane is a clear success of our method.  \textit{Right panel:} The hexagons are colour coded by the mean S\'ersic index. This panel highlights the connection between kinematics and morphology: high S\'ersic index galaxies are generally slow rotators and low Sersic index are fast rotators. Yet there is a considerable population of high S\'ersic index fast rotators, rendering the S\'ersic index parameter an imperfect proxy of galaxy kinematics.}
        \label{fig:LambdaEps}
\end{figure*} 

The ultimate goal of this work is to study the relationship between the parameters derived in the previous section and galaxy quenching. Before deploying them in this novel context, we first validate the parameter set against well established scaling relations, as well as traditional estimates of galaxy kinematics and morphology. 

\subsubsection{The $(\lambda, \epsilon)$ plane}\label{The (lambda, epsilon) plane}

In Fig. \ref{fig:LambdaEps}  we show the $(\lambda, \epsilon)$ plane, which has traditionally been used to separate galaxies by kinematic type \citep{Emsellem2007, Emsellem2011, Fogarty2015, Cappellari2016, Graham2018, Wang2020}. In each panel  we show $\lambda$ on the y-axis and ${\rm \epsilon=1-(b/a)_{phot}}$ on the x-axis. The colour coding, which varies from panel-to-panel, will be discussed later in this section. We first focus on the distribution of our galaxy sample in the plane shown with brown density contours, and examine its relation to theoretical predictions and previous observations. 

The $(\lambda, \epsilon)$ plane reveals two kinematic populations. First, we identify the population of fast rotators. These galaxies have large $\lambda$ and the brown density contours show that they are distributed consistently with the theoretical prediction for rotators with anisotropy parameter $\rm \delta<0.7\epsilon_\mathrm{intr}$, where $\epsilon_\mathrm{intr}$ is the intrinsic ellipticity \citep{Cappellari2007}. To see this, we include the magenta line which is the prediction for a $\rm \delta=0.7\epsilon_\mathrm{intr}$ rotator viewed edge-on, and the black dotted lines which are the tracks of these same galaxies with fixed $\epsilon_\mathrm{intr}$ as they are viewed at decreasing inclination angle, reaching $\rm (\lambda, \epsilon)=(0,0)$ for face-on systems. For comparison, we show in lime green the theoretical prediction for edge-on isotropic rotators \citep{Binney2005} and note that many galaxies have $\lambda$ below this line, which shows that their flattening cannot be entirely explained by rotation and that it must be due partially to velocity anisotropy. Second, we identify the population of slow rotators. These galaxies have small $\lambda$ and $\rm \epsilon$ and lie in the slow rotator region parameterised by \citet{Emsellem2011}, which is bounded by two red lines in the lower left corner of each panel in Fig. \ref{fig:LambdaEps}. The key point is that our sample includes both fast rotators and slow rotators, thus spanning the full range of kinematic states. We note the lack of galaxies with $\rm \epsilon<0.1$, which is a direct consequence of our removing all galaxies with $\arccos((b/a)_\mathrm{phot})<25^{\circ}$ in Section \ref{2D Kinematic Model}. 
 
\begin{figure*}
    \includegraphics[width=\textwidth]{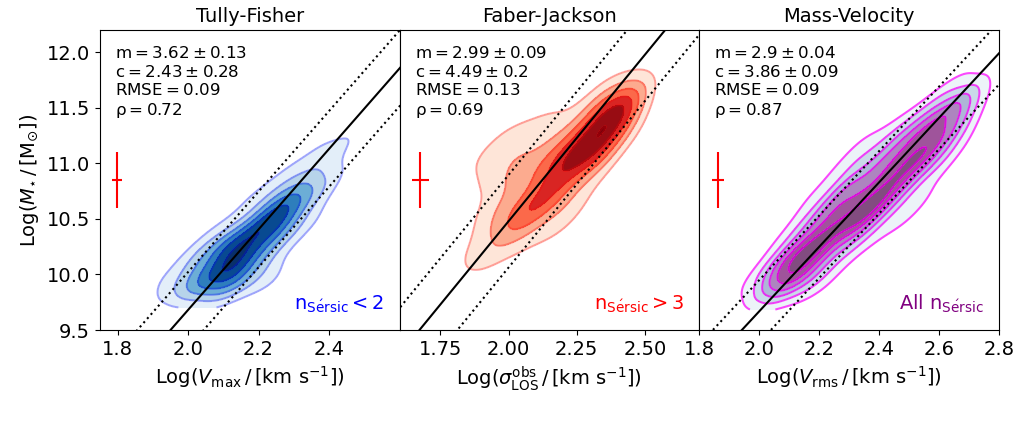}
    \caption{Kinematic scaling relations. \textit{Left panel:} The TF relation for disc galaxies (i.e. S\'ersic index less than two). \textit{Middle panel:} The FJ relation for spheroidal galaxies (i.e. S\'ersic index greater than three). Fast rotator spheroidal galaxies lie above the FJ relation and skew the contours, further highlighting the limitations of S\'ersic index as a proxy of galaxy kinematics.  \textit{Right panel:} The new Mass-Velocity (MV) relation, which is shown for all morphological galaxy types (i.e. all S\'ersic indices).  In each panel, we display the  gradient (m), intercept (c) and root mean square error (RMSE) of linear regression fits (ODR) to the relations, and show the best fit and  $\rm \pm 1~\sigma$ scatter lines with solid and dashed black lines respectively. We also report the Pearson correlation coefficient ($\rm \rho$). The red error bars represent the $1\sigma_\mathrm{err}$ uncertainties. Linearly spaced density contours are shown in all panels. The MV relation is tighter (i.e. it has lower RMSE) and has a higher Pearson correlation strength than either of the TF and FJ relations.  
    }
        \label{fig:StellarMassVsKinematicParams}
\end{figure*} 

In the left and central panels of Fig. \ref{fig:LambdaEps} we compare the dimensionless spin parameter with two alternative kinematic classifiers of galaxy type. In the left panel, we colour code the $(\lambda, \epsilon)$ plane by the fraction of galaxies with velocity maps that are visually suggestive of rotation. This classification is different to our inspection of the fits in Section \ref{2D Kinematic Model}. Here, we are not concerned with the quality of the fit per se, but with answering the following question: which of the \textsc{dap} LOS velocity maps exhibit velocity gradients that are typical of galaxy rotation? Answering this question for a specific galaxy is somewhat subjective, but the general distribution of these galaxies in the $(\lambda, \epsilon)$ plane is striking. Almost all galaxies in the fast rotator region show clear visual evidence of rotation, whilst those in the slow rotator region often do not appear to be rotating. This consistency check is not surprising, since the dimensionless spin parameter and our visual inspection are both measurements made directly on the observed data, but it does demonstrate the dimensionless spin parameter's success as a scalar,  non-parametric quantity capable of classifying galaxies. 

In the central panel, we colour code the $(\lambda, \epsilon)$ plane by the good fit fraction, which is the fraction of galaxies whose kinematics are consistent with the inclined rotating disc model. This parameter varies significantly in the plane, such that galaxies in the slow rotator region are inconsistent with the inclined disc rotation model, whilst those in the fast rotator region show good consistency, with the fraction of well fit fast rotators increasing with $\lambda$. In other words, our model is able to describe the kinematics of fast rotators, but not of slow rotators. Consistency (or lack thereof) with the inclined rotating disc model is therefore a powerful way to constrain a galaxy's kinematic type. We stress that the $(\lambda, \epsilon)$ method for classifying galaxies is independent of our kinematic modelling. The good agreement between the two approaches is thus a clear success of our method.

In the right panel, we colour code the $(\lambda, \epsilon)$ plane by $\rm n_{S\acute{e}rsic}$ to directly compare galaxy kinematics and morphology. The S\'ersic index probes the concentration of a galaxy's brightness profile, such that galaxies with a prominent bulge typically have $\rm n_{S\acute{e}rsic}>3$, whilst disc galaxies typically have $\rm n_{S\acute{e}rsic}<2$, with intermediate bulge plus disc systems occupying the range $\rm 2<n_{S\acute{e}rsic}<3$. To first order, Fig. \ref{fig:LambdaEps} shows good agreement between $\rm n_{S\acute{e}rsic}$ and galaxy kinematics. Galaxies with high $\rm n_{S\acute{e}rsic}$ are mostly located in the slow rotator region, whilst those with low $\rm n_{S\acute{e}rsic}$ are mostly located in the fast rotator region. In other words, photometric discs tend to be fast rotators and photometric spheroids tend to be slow rotators. This result supports the extensive use of $\rm n_{S\acute{e}rsic}$ to separate the two galaxy types. The relationship is imperfect, however, and we find a significant population of high $\rm n_{S\acute{e}rsic}$ photometric spheroids in the fast rotator region of the $(\lambda, \epsilon)$ plane. This second order effect is one of the key motivations of this work, where we attempt to study galaxy evolution through direct probes of galaxy kinematics, without relying on imperfect morphological proxies such as $\rm n_{S\acute{e}rsic}$.  

Overall, we stress the good consistency between crude non-parametric kinematics ($\lambda$), detailed kinematic modelling, and morphology ($\rm n_{S\acute{e}rsic}$), where slow rotators generally have low $\lambda$, low $\overline{V}\,/\,\overline{\sigma}$ and large $\rm n_{S\acute{e}rsic}$, whilst fast rotators generally have large $\lambda$, large $\overline{V}\,/\,\overline{\sigma}$  and low $\rm n_{S\acute{e}rsic}$. S\'ersic index and $\lambda$  are well established classifiers of galaxy type, so we highlight this result as a major success of our kinematic modelling.

\subsubsection{Stellar mass-kinematics scaling relations}\label{Stellar mass-kinematics scaling relations}
There are a number of kinematic scaling relations that have a long precedent in the literature \citep{Fisher1977, Faber1976} and it is important to test that they are consistent with our kinematic estimates. In Fig. \ref{fig:StellarMassVsKinematicParams} we show galaxy kinematics as a function of stellar mass. In each panel we report gradients and intercepts of the scaling relations, as well as the scatter about the best fit lines, with all of the best fit lines determined using orthogonal distance regression (ODR). We compare our gradients to results from \citet{Aquino-Ortiz2020}, which were also calibrated using IFU data from the MaNGA and CALIFA surveys \citep{Aquino-Ortiz2018}, though we do not compare intercepts, since these are highly dependent on a number of assumptions, such as the IMF, the assumed templates used in SED fitting, and the spatial scale over which $M_{\star}$ is determined.

In the left panel we show the $\rm M_{\star}-V_{max}$ relation, commonly known as the Tully-Fisher (TF) relation \citep{Fisher1977}, with $M_{\star}$ on the y-axis and $V_\mathrm{max}$ on the x-axis. The TF relation is generally associated with disc galaxies, so we  only consider those with $\rm n_\mathrm{S\acute{e}rsic}<2$. There is a strong positive correlation between $V_\mathrm{max}$ and  $M_{\star}$, with low scatter. The gradient of the best fit line is in relatively good agreement with those of \citet{Aquino-Ortiz2018} ($\rm m=3.3$) and \citet{Aquino-Ortiz2020} ($\rm m=3.2$), and is consistent with \citet{Avila-Reese2008} within the scatter ($\rm m=3.7$), with whom \citet{Aquino-Ortiz2020} compare.  In the middle panel, we examine the $\rm M_{\star}-\overline{\sigma^\mathrm{obs}_\mathrm{LOS}}$ relation for $\rm n_\mathrm{S\acute{e}rsic}>3$ photometric spheroids, with $M_{\star}$ on the y-axis and $\rm \overline{\sigma^\mathrm{obs}_\mathrm{LOS}}$ on the x-axis, where $\rm \overline{\sigma^\mathrm{obs}_\mathrm{LOS}}$  is the linear average of the observed LOS velocity dispersion. This relationship is referred to as the Faber-Jackson (FJ) relation \citep{Faber1976}.  Note, we do not mass-weight the average or correct the velocity dispersion for the effect of differential disc rotation to enable a fair comparison with \citet{Aquino-Ortiz2018, Aquino-Ortiz2020}. We observe a strong positive correlation between $M_{\star}$ and  $\rm \overline{\sigma^\mathrm{obs}_\mathrm{LOS}}$, with scatter similar to the TF. The slight offset between the peak of the density contours and the best fit line is caused by a population of high $\overline{V}\,/\,\overline{\sigma}$ galaxies lying above the FJ relationship. Once again, we find good consistency with \citet{Aquino-Ortiz2018} ($\rm m=3.2$) and \citep{Aquino-Ortiz2020} ($\rm m=3.1$). We emphasise that the good agreement between our kinematic estimates and these well established TF and FJ scaling relations is an important check on our method, and builds confidence in our kinematic parameter set.

We seek an extension of the TF and FJ relations that includes both photometric discs and spheroids. The underlying physics of the TF and FJ is the virial theorem, so a natural progression is the relationship between the  total velocity (both disordered and ordered, i.e. $V_\mathrm{rms}$, recall its definition in equation \ref{EQN:TotalVelocity}) and stellar mass, which we call  the \textit{Mass-Velocity} (MV) relation. In the right panel of Fig. \ref{fig:StellarMassVsKinematicParams}, we compare stellar mass on the y-axis with $V_\mathrm{rms}$ on the x-axis. In this diagram, photometric spheroids and photometric discs form a single population, in which $M_{\star}$ is highly dependent on $V_\mathrm{rms}$, with low scatter. 

We note that there is a slight offset between spheroids and discs in the MV relation, such that discs have larger $M_{\star}\,/\,V_\mathrm{rms}$ than spheroids, which is likely a consequence of spheroids being more compact than discs at fixed mass \citep{vanderWel2014}. We have neglected this effect in our first order application of the virial theorem, and do not consider the gravitational radii or the virial parameters of individual galaxies. Nonetheless, this offset is small, and we find that the MV relation is tighter, and has a higher Pearson correlation strength, than either the TF and FJ, which is particularly impressive given its application to the \textit{full range of galaxy types}.

Summarising, the analysis in this section confirm that our kinematic measurements are reliable and indeed even more effective than the simpler alternatives from the literature. 

\begin{figure}
    \includegraphics[width=\columnwidth]{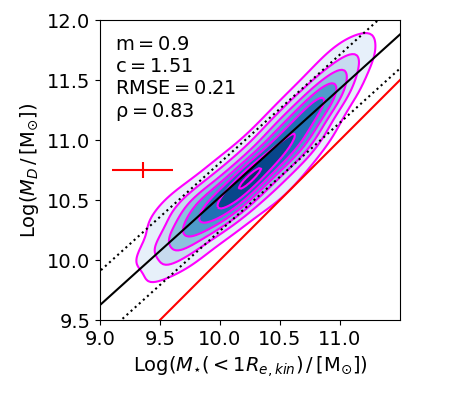}
    \caption{A comparison of our dynamical mass estimates, calculated within $\rm 1\, R_{e,kin}$, and stellar mass estimates within $\rm 1\, R_{e,kin}$, calculated  by integrating the Pipe3D stellar mass surface density maps.  The two mass estimates are therefore compared on the same spatial scales. The estimates are strongly correlated with low scatter, and the dynamical mass consistently exceeds the stellar mass, as expected. We display the gradient (m), intercept (c) and root mean square error (RMSE) of linear regression fits (ODR) to the relations. We also report the Pearson correlation coefficient ($\rm \rho$). We show the best fit and  $\rm \pm 1~\sigma$ scatter lines with solid and dashed black lines respectively, and we show the 1:1 relationship in solid red. The red error bars represent the $1\sigma_\mathrm{err}$ uncertainties. }
        \label{fig:StellarMassDynamicalMass}
\end{figure} 
\subsubsection{Dynamical mass vs stellar mass}\label{Dynamical mass vs stellar mass}

In Fig. \ref{fig:StellarMassDynamicalMass} we show the relationship between dynamical mass derived via equation \ref{EQN:DynamicalMass} and stellar mass measured within $\rm 1\, R_{e,kin}$ by integrating $\rm \Sigma_{\star}$ from \textsc{pipe3d}, $\rm M_{\star}(<R_{e,kin})$. As expected, we find a strong positive correlation between the two mass estimates ($\rho=0.83$) and low scatter about the linear best fit line ($\rm RMSE=0.21\,dex$). Dynamical mass traces all components of mass, including stellar, gas (molecular, neutral and ionised) and dark matter, so we expect  $M_\mathrm{D}$ to exceed $\rm M_{\star}(<R_{e,kin})$. Indeed, $M_\mathrm{D}$ exceeds $\rm M_{\star}(<R_{e,kin})$ by more than $\rm 0.5\,dex$, which is an important success of our kinematic modelling.

\section{Results}\label{Results}
In this section, we explore the connection between galaxy kinematics and quenching using the global kinematic parameter set derived and validated for 1862 MaNGA galaxies in the previous section. 

We comment at the outset that our galaxy sample contains both centrals ($\sim75$ per cent) and satellites ($\sim25$ per cent), as classified by \citet{Yang2007}. It is commonly thought that internal processes quench centrals and environmental processes quench satellites \citep{Peng2010, Peng2012}, and hence it would be best to analyse the two galaxy types separately. Consequently, we have applied the analysis in this section to a pure sample of centrals, and we confirm that our key results are robust to this test. However, there are too few galaxies to perform a statistically robust analysis of a pure sample of satellites.  We therefore analyse  centrals and satellites together and leave a full separate analysis of both galaxy types for later work. This decision is supported by \citet{Bluck2020b}, who
show that environmental effects dominate galaxy quenching only in low mass satellites. In the high mass regime analysed in this work, they find that centrals and satellites quench similarly, which implies that they can be analysed together.

\subsection{Star forming and quenched classification}\label{Star forming and quenched classification}
\begin{figure*}
    \includegraphics[width=\textwidth]{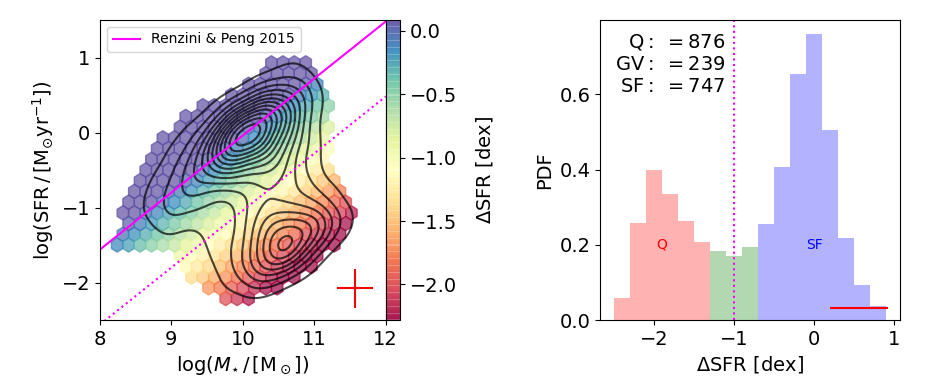}
    \caption{The star forming/quenched classification scheme. \textit{Left panel:} The global star forming main sequence for SDSS galaxies ($z<0.085$), with linearly spaced density contours shown in black. The solid magenta line is the \citet{Renzini2015} fit to the main sequence, and the dotted magenta line marks the minimum of the $\rm \Delta SFR$ histogram, $\rm 1~dex$ below the main sequence (see right hand panel). The hexagons are colour coded by their logarithmic offset from the main sequence, $\rm \Delta SFR$, which cleanly separates the star forming galaxies in the upper density peak from the passive galaxies in the lower density peak. Note, SFR of many passive galaxies is strictly an upper limit set at $\rm log(sSFR\,/\,[yr^{-1}])=-12$. \textit{Right panel:} The 1D distribution of $\rm \Delta SFR$ for SDSS galaxies. The dotted magenta line marks the minimum of the bimodal distribution at $\rm \Delta SFR=-1~dex$. The star forming,  green valley and quenched regions are shown in blue, green and red, respectively. In the legend, we quote the number of star forming, green valley and quenched MaNGA (i.e. rather than SDSS) galaxies analysed in this study. In both panels the red error bars represent the $1\sigma_\mathrm{err}$ uncertainties. 
    }  
        \label{fig:MS_HistDeltaSFR}
\end{figure*} 
We must first define the star forming and quenched populations. The most common approach is to consider the distribution of galaxies in the local $\mathrm{SFR}-M_{\star}$ plane, and to define galaxies that lie on the star forming main sequence (SFMS) as star forming and those that have SFR considerably offset below the SFMS as quenched or passive. 

We construct the classification scheme using a statistically robust and representative sample of SFR and $M_{\star}$ estimates for galaxies drawn from the SDSS, which is the parent sample of MaNGA. In the left panel of Fig. \ref{fig:MS_HistDeltaSFR} we show the distribution in the $\mathrm{SFR}-M_{\star}$ plane of $\rm \sim250,000$ SDSS galaxies with redshift $z<0.085$. We apply the redshift cut for consistency with \citet{Renzini2015}, who used these data to characterise the SFMS, but we note that this cut is approximately the same as the redshift range in our sample. The black linearly spaced density contours are strongly bimodal; star forming galaxies reside in the upper density peak, whilst quenched galaxies reside in the lower density peak. We show the best fit to the SFMS from \citet{Renzini2015} in solid magenta and note its explicit form:
\begin{equation}
    \label{eqn:mainsequnce}
    \mathrm{log\left(\frac{SFR_{MS}}{M_{\odot}~yr^{-1}}\right) = 0.76\times log}\left(\frac{M_{\star}}{\mathrm{M_{\odot}}}\right)-7.64 .
\end{equation}
The uncertainty on the linear coefficients is $\sim$1-2 per cent \citep{Renzini2015}. As expected, the best fit line tracks the ridge of the star forming density contours.

We define a galaxy's logarithmic offset from the SFMS as follows, as in \citet{Bluck2014, Bluck2016}:
\begin{equation}
    \label{eqn:deltasfr}
    \rm
    \Delta SFR = log(SFR) - log(SFR_{MS}).
\end{equation}
We colour code hexagonal bins in the $\mathrm{SFR}-M_{\star}$ plane by $\rm \Delta SFR$ and note that all star forming galaxies in the upper density peak have $\rm \Delta SFR \sim 0$ (coloured blue), whilst all quenched galaxies in the lower density peak have $\rm \Delta SFR<-1~dex$ (coloured red). The offset from the main sequence thus unambiguously separates star forming and quenched galaxies.

The 1D distribution of $\rm \Delta SFR$ shown in the right panel of Fig. \ref{fig:MS_HistDeltaSFR} emphasises the bimodalidy of star forming and quenched galaxy properties. We observe  peaks at $\rm \Delta SFR=0~dex$ and $\rm \Delta SFR=-1.8~dex$ corresponding to star forming and quenched galaxies, respectively.  It is important to note that \citet{Brinchmann2004} calculate SFR using a combination of emission line diagnostics and D4000 where possible, but that they fix sSFR of quenched galaxies with low SNR emission lines at $\rm log(sSFR\,/\,[yr^{-1}])=-12$, which is strictly an upper limit. The peak at $\rm \Delta SFR=-1.8~dex$ is therefore slightly misleading, and the true distribution is more accurately thought of as a long tail tending to SFR=0, or $\rm \Delta SFR=-\infty~dex$.  

We use the distribution of $\rm \Delta SFR$ to build a quantitative classification of star forming and quenched systems. The dotted magenta line in the right panel of Fig. \ref{fig:MS_HistDeltaSFR} marks the minimum of the 1D density distribution, at $\rm \Delta SFR=-1~dex$. This boundary effectively separates the star forming and quenched peaks. Galaxies with intermediate $\rm \Delta SFR\sim -1~dex$ are commonly referred to as green valley galaxies and are thought to be in the process of quenching \citep{Wyder2007, Martin2007,Schawinski2014}. Green valley galaxies cannot be uniquely associated with either of the star forming or quenched populations, so we introduce a $\rm 0.6~dex$ buffer region and partition the $\rm \Delta SFR$ distribution into the following three regimes:
\begin{enumerate}
    \item Quenched/Passive (Q/Pa): $\rm \Delta SFR<-1.3~dex$
    \item Green Valley (GV): $\rm  -1.3~dex<\Delta SFR<-0.7~dex$
    \item Star forming (SF): $\rm \Delta SFR>-0.7~dex$
\end{enumerate}
We have checked that our results are not strongly dependent on the precise values of these cuts. We mark the star forming, green valley and quenched regions of the $\rm \Delta SFR$ distribution in blue, green and red, respectively. 

\subsection{General relationship between star formation and kinematics}\label{General relationship between star formation and kinematics}
In this sub-section, we visualise the data and present simple quantitative analyses, before adopting rigorous statistical techniques in Sections \ref{Random forest analysis} and \ref{Correlation analysis and quenching angle}.

\begin{figure*}
    \includegraphics[width=\textwidth]{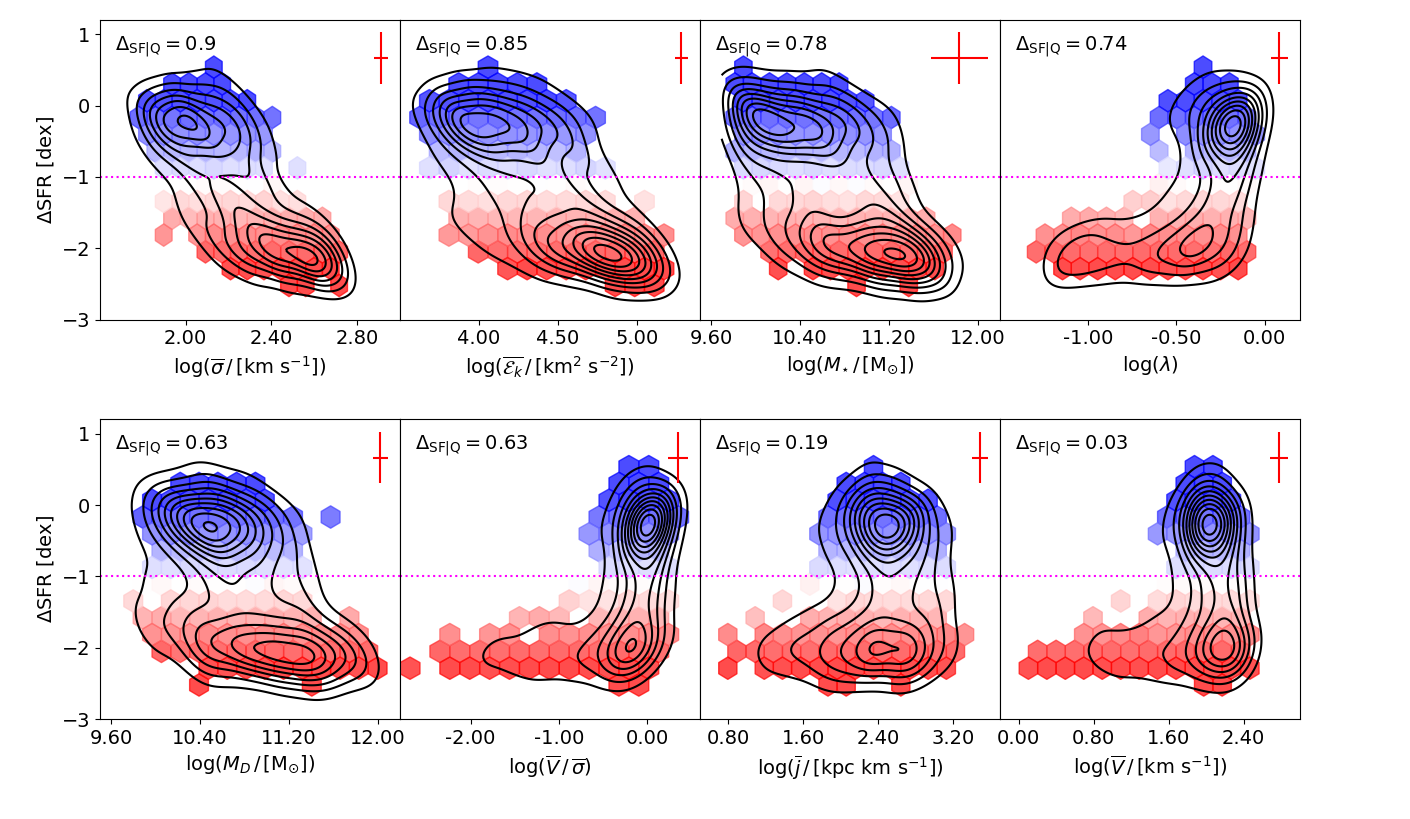}
    \caption{The relationship between star forming state, as expressed by $\rm \Delta SFR$, and each of our kinematic parameters. In each panel, linearly spaced density contours are shown in black, and the boundary between the star forming (coloured blue) and quenched (coloured red) populations at $\rm \Delta SFR=-1~dex$ is shown with a dotted magenta line. The red error bars represent the $1\sigma_\mathrm{err}$ uncertainties. The panels are arranged from left to right and top to bottom in order of decreasing $\rm \Delta_{SF|Q}$. Parameters with large $\rm \Delta_{SF|Q}$ separate the star forming and passive populations along the x-axis and are likely associated with quenching, whilst parameters with low $\rm \Delta_{SF|Q}$ show degeneracies across the two populations (i.e. both star forming and passive galaxies existing at a given fixed parameter) and are hence not associated with quenching. 
    }
        \label{fig:FullCorrelations_KinematicParamsDeltaSFR}
\end{figure*}

In Fig. \ref{fig:FullCorrelations_KinematicParamsDeltaSFR}, we collapse the $\mathrm{SFR}-M_{\star}$ plane into a 1D problem, and compare $\rm \Delta SFR$ (i.e. star forming state) on the y-axis with the parameter being investigated on the x-axis. The bimodal distribution of galaxies is visible in each panel, where we observe a star forming (coloured blue) and quenched (coloured red) density peak at high and low $\rm \Delta SFR$, respectively.  We are looking for parameters that are effective at separating these two peaks along the x-axis. Of course, no single parameter is able to perfectly predict a galaxy's star forming state, but Fig. \ref{fig:FullCorrelations_KinematicParamsDeltaSFR} does show a range of behaviour that can be used to rank the parameters. 

In the upper left panel, for example, we observe the star forming density peak at low $ \overline{\sigma}$ and the quenched density peak at high $\overline{\sigma}$, and the two density contours show little overlap when compared with other panels. Contrast this with behaviour of $\overline{V}$  in the lower right panel, where the star forming and quenched density contours overlap significantly. This suggests that $\overline{\sigma}$ is  more closely related than $ \overline{V}$ to galaxy quenching. 

Visually comparing the absolute separation along the x-axis of the density contours in each panel is flawed since the parameters have different dynamic ranges.  We must first normalise the absolute separation of each parameter by its variability to make a fair comparison, which we achieve by introducing the following parameter (similar to \citet{Bluck2020a}):
\begin{equation}
    \label{eqn:DeltaSFQ}
    \rm
    \Delta_{SF|Q} = \frac{med(\mathit{X})_{Q} - med(\mathit{X})_{SF}}{IQR(\mathit{X})}
\end{equation}
where $\rm med(\mathit{X})_Q$ and $\rm med(\mathit{X})_{SF}$ are the median values of parameter $X$ for the quenched and star forming populations, respectively,  and IQR($X$) is the interquartile range of parameter $X$ for the full population, including both star forming and quenched galaxies. 

We report $\rm \Delta_{SF|Q}$ in each panel of Fig. \ref{fig:FullCorrelations_KinematicParamsDeltaSFR} and arrange the panels in order of decreasing $\rm \Delta_{SF|Q}$, from left to right and top to bottom.  Of all the parameters, the star forming and quenched systems differ most in terms of their average velocity dispersion. This is our first quantitative evidence of $\overline{\sigma}$'s significant role in galaxy quenching. The $\rm \Delta_{SF|Q}$ statistic shows that other parameters are also effective at predicting quenching. Indeed, $\overline{\mathcal{E}_k}$ has $\rm \Delta_{SF|Q}$ similar to that of $\overline{\sigma}$, and six of the eight parameters have $\rm \Delta_{SF|Q}>0.5$. Although these six parameters separate the star forming and quenched populations, we stress that both $\overline{\sigma}$ and  $\overline{\mathcal{E}_k}$ have larger $\rm \Delta_{SF|Q}$ than $M_{\star}$. \textit{This shows that galaxy kinematics are more effective at predicting quenching than a photometric measurement of stellar mass.} 

There are two parameters, however, with significantly lower $\rm \Delta_{SF|Q}$ that seem totaly unrelated to quenching: $\overline{V}$ and $\overline{j}$. Interestingly, these are the only two parameters that are completely independent of velocity dispersion; the stellar mass is related to $\overline{\sigma}$ by the viral theorem, and all other parameters have an explicit dependence on velocity dispersion, as outlined in Section \ref{Kinematic parameters}. One may naively expect galaxies that have massive discs (i.e. large $\overline{V}$ and $\overline{j}$) to be star forming since discs are commonly the site of molecular gas and active star formation. Nonetheless, we find a number of these galaxies in the quenched population, which shows that quenching is governed not by properties of the disc, but rather by properties of the bulge/spheroidal component, as quantified by the  average velocity dispersion. Thus, star formation and quenching appear to be distinct physical phenomenon (see \citet{Bluck2020a} for further evidence).

\subsection{Random forest analysis}\label{Random forest analysis}
The distribution of $\rm \Delta SFR$ in Fig. \ref{fig:FullCorrelations_KinematicParamsDeltaSFR} demonstrates that the star forming state of galaxies is bimodal. Moreover, the $\rm \Delta SFR$ values in the quenched population are mostly upper limits, and so there is no information in their specific numerical values.  Hence the study of galaxy quenching is more concerned with the classification of a galaxy (i.e. star forming or quenched) rather than its specific value of $\rm \Delta SFR$. In this section, we use a random forest classifier, which is a machine learning algorithm, to quantify the importance of each parameter in the quenching process. 

We repeat the analysis in Appendix \ref{Testsonthestabilityoftheresults}, where we conduct the following tests: only analysing galaxies that have good fits, without using the simple method at all; using the simple method for all galaxies, even for those that have a good kinematic fit; only trusting fits that are successful when assuming the S\'ersic flux map; and finally, testing the role of differential measurement uncertainty, where we find that no feasible level of uncertainty on our measurement of $\overline{V}$ could have wrongly led to our conclusion that $\overline{\sigma}$ is the most important parameter. We confirm in advance that the key results of this section are completely stable to all of these tests.

\subsubsection{Random forest method}
A random forest is a machine learning algorithm that is able to  identify non-linear features in multidimensional data that are useful for classification. The algorithm is trained using data that has previously been classified with `truth' labels (i.e. `training data'), and in most applications, the trained random forest is then used to predict the classification state of new, unseen data. In our work, we train a random forest to identify features in our parameter set ($\overline{V},~ \overline{\sigma},~ \overline{V}\,/\,\overline{\sigma},~ \overline{\mathcal{E}_{k}},~\overline{j},~\lambda,~ M_\mathrm{D},~\mathrm{and}~M_{\star}$) that are useful for predicting the star forming state of a galaxy, i.e. star forming or quenched. Our goal is not to use the trained algorithm to predict the star formation state of new galaxies that do no yet have $\rm \Delta SFR$ estimates, but rather to compare the relative importance of each parameter in the classification scheme.

We briefly review the methodology of the random forest algorithm, and we refer the interested reader to \citet{Bluck2020b} for  a full discussion. A random forest consists of a number of individual decision trees that ask a series of binary questions to categorise data. The input data for each decision tree is found by bootstrapped  random sampling with return, so each decision tree in the random forest produces subtly different results. This encourages the algorithm to learn general features of the data, rather than those that are unique to a particular sample. 

Each fork in a decision tree tests the validity of a simple inequality statement relating to a single parameter. As an example, a decision fork in our work may ask `does the galaxy have $ \overline{\sigma}>220\mathrm{\, km~s^{-1}}$?' This fork would split the galaxies into two groups, or nodes: those that have $ \overline{\sigma}>220\mathrm{\, km~s^{-1}}$ and those that do not. Starting from the top of the tree, at the parent node with the full sample of training data, the random forest algorithm identifies the feature and criterion that maximises the reduction in impurity, which is quantified using the Gini coefficient. 

The Gini coefficient of a particular node $n$ is given by
\begin{equation}
    I_\mathrm{G}(n)=1-\sum^{i=2}_{i=1}\left(p_i(n)^2\right),
    \label{eqn:Gini}
\end{equation}
where $p_i$ is the probability of randomly selecting an object with classification state $i$. The summation in equation \ref{eqn:Gini} spans the two classification states in this work, star forming and quenched. In words, the Gini coefficient of a node gives the probability of a random sample being labelled inaccurately if it is labelled in accordance with the distribution of samples in the node.

Each decision fork splits the sample into two nodes, where the process is repeated. This could continue indefinitely until the final nodes, leaf nodes, contain pure samples of the two classification states (either all quenched or all star forming in our case), but we terminate the random forest before it achieves these pure leaf nodes to prevent overfitting. Impure leaf nodes, containing at least one object of each classification state, are classified by the modal truth value of their training data. We can then use the trained algorithm to classify new data, where the classification of a particular object is given by the mean classification of its leaf node modal values across all of the trees in the forest. 

There are alternative machine learning algorithms that could be used for this work, but we choose the random forest algorithm since it has a transparent and simple methodology for mapping multidimensional features of the data into a classification state. At each node, the random forest selects the most effective parameter for separating the two classification states. We can quantify the reduction in Gini coefficient that is associated with each parameter in a tree and average across all trees in the forest to find the relative importance of each parameter in the overall classification scheme. The parameter with the largest relative importance is responsible for the greatest reduction of impurity in the data. The random forest algorithm is thus not only an effective predictive tool, but also a powerful system for objectively ranking the importance of features in multidimensional data.

We perform the random forest classification using the \textsc{randomforest-classifier} from the \textsc{scikit-learn}\footnote{\href{https://www.scikit-learn.org}{https://www.scikit-learn.org}} python package. We remove green valley galaxies from our sample in this section due to their ambiguous classification, and randomly select two equally sized samples of star forming and quenched objects. We note that keeping the green valley galaxies in the sample, and simply classifying those with $\rm \Delta SFR<-1~dex$ as quenched and those with $\rm \Delta SFR>-1~dex$ as star forming, does not have a significant impact on our results. It is important that the data have equal numbers of star forming and quenched galaxies to ensure that the algorithm is not biased towards learning the features of only one galaxy type. We construct the largest possible balanced sample of 1494 galaxies, containing 747 star forming and 747 quenched objects, and split it into two equally sized training and testing samples. Our random forest has 250 trees per forest, and we allow each tree to have a maximum of 250 nodes.

We perform the random forest analysis with max-features set to `All'\footnote{Formally, this is achieved by setting max-features to `None' within the \textsc{randomforest-classifier} function.}, which allows each fork in the decision tree to consider  all of the parameters when choosing the optimal split to reduce the Gini coefficient. We fine tune the min-samples-leaf parameter, reducing it as far as possible to improve the algorithm performance without inducing overfitting. We use the area under the receiver operating curve parameter, AUC, to quantify the performance of the algorithm, and compare the relative performance of the algorithm in classifying the training and testing data sets via the difference in their respective AUC values, $\rm \Delta AUC=AUC_{Training}-AUC_{Testing}$. A large value of $\rm \Delta AUC$ confirms that the model is significantly better at describing the training data, which it has seen, than the testing data, which it has not seen. This suggests that the algorithm is learning pathologies unique to the training data rather than true features of the galaxy population, an effect known as overfitting. The likelihood of overfitting increases as we allow the algorithm to form increasingly small leaf nodes. We therefore decrease min-samples-leaf as far as possible whilst ensuring that the algorithm performs similarly for the training and testing data, as quantified by our constraint $\rm \Delta AUC<0.02$. We find an optimal min-samples-leaf value of 40.   

The random forest returns a single scalar value quantifying the relative importance of each parameter. We repeat the analysis 10 times to estimate its uncertainty. In each run, we choose a new random sample of 1494 galaxies containing equal numbers of star forming and quenched objects, and redefine the training and testing sample. We take the mean relative importance of each parameter across the 10 runs as the typical value, and the standard deviation of the relative importance as the typical uncertainty. We similarly use the 10 runs to calculate the typical performance of the model and its uncertainty. The final trained model has a testing performance $\rm AUC_{Testing}=0.92\pm 0.01$, which is commonly considered to be outstanding \citep{Teimoorinia2016}. This impressive performance suggests that the parameters investigated in this paper contain the vast majority of the information required to predict the star forming state of a galaxy. In other words,  our parameter set is not missing any measurements that are crucial to predict quenching to a very high level of accuracy.. 
 
\subsubsection{Random forest results}
\begin{figure}
    \includegraphics[width=\columnwidth]{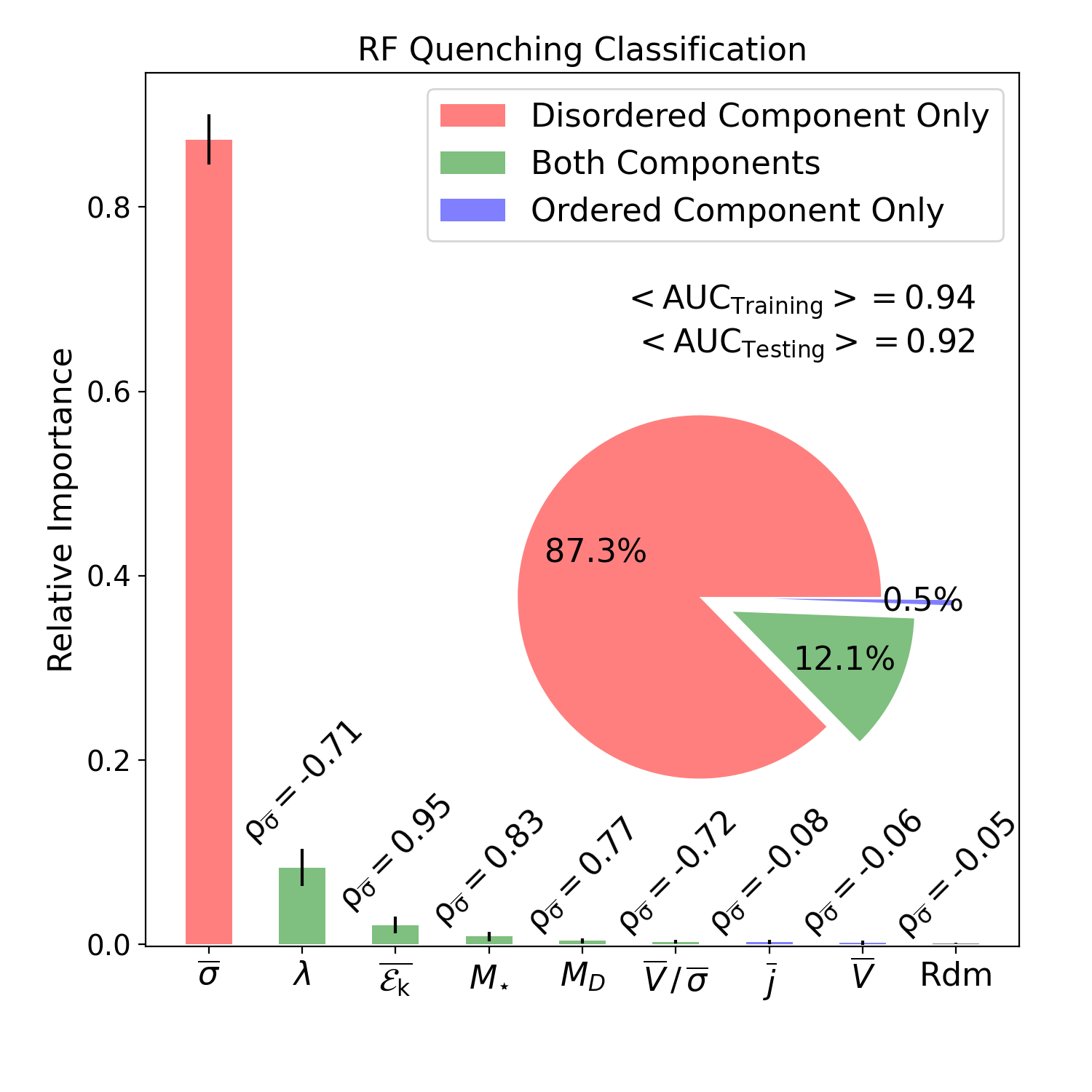}
    \caption{A random forest classification to quantify the relative importance of kinematic parameters for predicting the star forming state of galaxies. We arrange the parameters along the x-axis in order of decreasing relative importance, and we include the relative importance of a random variable, Rdm, for scale. The black error bars indicate the scatter in the relative importances across ten bootstrapped random training sets. The Spearman rank correlation strength between each parameter and the most important parameter, the average velocity dispersion, is printed above each bar to highlight the strong correlations within the parameter set. We report the AUC values for the training and testing sets, both of which are greater than 0.9, which is commonly considered to be outstanding. The colour of each bar indicates the category of the parameter under investigation, as shown in the legend. The `Ordered Component Only' category contains $\overline{\sigma}$, the `Both Components' category includes $\lambda$, $\overline{V}\,/\,\overline{\sigma}$, $M_{\star}$, $\overline{\mathcal{E}_{k}}$ and  $M_\mathrm{D}$, and the `Disordered Component Only' includes $\overline{V}$ and $\overline{j}$ . We present a pie chart chart showing the combined relative importance of the parameters in each category. The random forest analysis clearly shows that $\overline{\sigma}$ is overwhelmingly the most important parameter for predicting galaxy quenching, and all other parameters are relatively insignificant.} 
        \label{fig:AllGalaxiesRF_RelativeImportance}
\end{figure} 

In Fig. \ref{fig:AllGalaxiesRF_RelativeImportance} we show the relative importance of each parameter for predicting the star forming state of galaxies. We organise the parameters along the x-axis in order of decreasing relative importance and include a random variable for scale. The random variable, which is completely disconnected from galaxy quenching by design, is the clear loser with relative importance consistent with zero, as expected.  

Once again, we find that $\overline{\sigma}$ is the most important parameter for predicting quenching.  Indeed, $\overline{\sigma}$ is at least 10 times more important than any other parameter in the set, and  $\overline{\sigma}$ is classified as the most  important parameter with confidence level greater than $20\,\sigma_\mathrm{err}$. 

The remaining parameters only play a marginal role in predicting the star forming state of galaxies. One might naively expect parameters that consider both the ordered and disordered components of the velocity to have the best predictive power. After all, they include the same information as $\overline{\sigma}$ as well as extra information about the ordered rotation. The reduced predictive power of  $\overline{\mathcal{E}_{k}}$, $\lambda$ and $\overline{V}\,/\,\overline{\sigma}$ relative to $\overline{\sigma}$, however, shows that adding information about the ordered stellar motion actually $decreases$ the predictive power, and is akin to adding noise. This is supported by the poor performance of $\overline{V}$ and $\overline{j}$ in the random forest, which have relative importances only marginally better than the random parameter. This once again demonstrates that parameters that relate exclusively to the disc properties have little predictive power over quenching.   

The added insight of Fig. \ref{fig:AllGalaxiesRF_RelativeImportance} is that galaxy quenching is not strongly dependent on the ratio of ordered to disordered motion either, as expressed by both $\lambda$ and $\overline{V}\,/\,\overline{\sigma}$. We remind the reader that the dimensionless spin parameter is measured independently of our kinematic modelling. Thus even if our kinematic modelling is unsuccessful, the enhanced predictive power of $\overline{\sigma}$ relative to $\lambda$ conclusively demonstrates that it is the absolute level of velocity dispersion that is important for predicting quenching, not the relative levels of ordered and disordered velocity.

We recognise that we have allowed the random forest to consider all of the parameters at each decision fork and that this approach is generally more susceptible to bias than versions of the random forest algorithm that consider only a subset of the parameters at each decision fork. The tests of Appendix \ref{Testsonthestabilityoftheresults} are thus an important check, and they demonstrate that the  result in Fig. \ref{fig:AllGalaxiesRF_RelativeImportance} is not caused by the random forest interpreting pathologies in our kinematic parameter set. We have also repeated the random forest analyses with max-features set to `Sqrt', which allows each fork in the decision trees to consider only a random sample of the square root of the number of parameters when choosing the optimal split for reducing the Gini coefficient. Unlike the analysis in Fig. \ref{fig:AllGalaxiesRF_RelativeImportance}, $\overline{\sigma}$ is often unavailable to the decision forks, so the random forest gives increased importance to secondary parameters that are correlated with $\overline{\sigma}$ \citep{Piotrowska2021}. However, the key result is unchanged and  $\overline{\sigma}$ is once again identified as the most important parameter for quenching.

Our kinematic parameters display strong inter-correlations. To see this, we report the Spearman rank correlation coefficient of each parameter with $\overline{\sigma}$, the most important parameter in our set, above each bar in Fig. \ref{fig:AllGalaxiesRF_RelativeImportance}. The power of our random forest analysis with max-features set to `All' lies in the way it simultaneously compares all of our kinematic parameters and evaluates their importance for predicting quenching in a competitive framework. Thus, although a parameter such as the specific kinetic energy is highly correlated with the velocity dispersion and therefore is a good predictor of quenching, the random forest recognises  that $\overline{\sigma}$ is the fundamental parameter and that the additional knowledge of $\overline{\mathcal{E}_{k}}$ does little to improve the predictive power.

\subsection{Correlation analysis and quenching angle}\label{Correlation analysis and quenching angle}

\begin{figure*}
    \includegraphics[width=\textwidth]{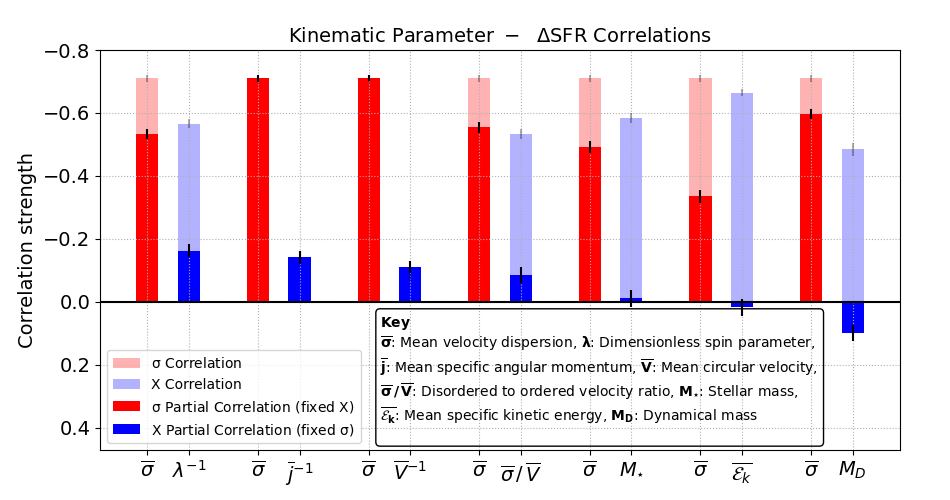}
    \caption{Full and partial correlation analysis of galaxy quenching. Correlation strengths are shown on the y-axis, and the  parameter set is arranged along the x-axis in groups of two, comparing the performance of each parameter to that of the average velocity dispersion. The y-axis is inverted such that taller bars represent stronger positive correlations with quenching. The black error bars indicate the scatter in correlation strengths across 100 bootstrapped random samples. We show the full correlation strength between each parameter and $\rm \Delta SFR$ using light shaded bars. The solid bars show the partial correlation strengths between the parameter in question and $\rm \Delta SFR$, whilst holding the second parameter in the pair fixed. The pairings are arranged in order of decreasing partial correlation strength with quenching, at fixed velocity dispersion. Quenching is more correlated with velocity dispersion (both full and partial correlation strengths) than with any other parameter. Furthermore, the correlation strengths between $\rm \Delta SFR$ and the secondary parameters are significantly reduced when $\overline{\sigma}$ is held constant, showing that the correlations are mostly spurious and do not reflect fundamental relationships. Note, the light shaded bars (i.e. full correlation) are sometimes hidden (i.e. for $\overline{V}$ and $\overline{j}$). This occurs when the partial correlation is equal to or larger than the full correlation.}
        \label{fig:AllGalaxiesDeltaSFRCorretionAnalysisBarPlots}
\end{figure*} 
In this section, we perform a correlation analysis to identify the most important parameter for quenching. This approach tests the random forest results, using a simpler (and hence more familiar) technique. It also contains the potential for a simple visual presentation, which we take advantage of. 

We quantify the degree of correlation using the Spearman rank correlation coefficient. The Spearman rank correlation coefficient between two variables, X and Y, is given by the Pearson correlation coefficient between the rank order statistics of X and Y. We adopt the Spearman rank correlation rather than the Pearson correlation, since the Pearson correlation is a measure of the linear relationship between two variables, which is not appropriate for our highly non-linear data.

A strong correlation coefficient suggests an association between two variables, such that Y changes as X changes, but it does not necessarily imply a fundamental relationship. This is particularly true in  multidimensional data with high degrees of inter-correlations, such as the data in this work. For example, parameters X and Y may appear to be highly correlated simply due to their both having strong correlations with a third, confounding parameter, Z.

We introduce the partial correlation coefficient to assess the influence of confounding variables. The partial correlation coefficient is defined as follows:
\begin{equation}
    \rho_{X,Y|Z} = \frac{\rho_{X,Y}-\rho_{X,Z}\,\rho_{Y,Z}}{\sqrt{1-\rho_{X,Z}^2}\sqrt{1-\rho_{Y,Z}^2}}
    \label{eqn:partialcorrelation}
\end{equation}
where $\rho_{X,Y}$ is the Spearman rank correlation coefficient between parameters X and Y. The partial correlation coefficient, $\rho_{X,Y|Z}$, is the degree of correlation between X and Y whilst a third parameter, Z, is held constant. A significant $\rho_{X,Y|Z}$ thus rules out the possibility that the confounding parameter Z is the cause of the correlation between X and Y. 

One may worry that there might be a confounding variable that is not part of our parameter set. Of course, we cannot use the partial correlation analysis to test for an unknown confounding parameter, but we note that this issue holds for any parameter set that we may choose to define, regardless of its size and breadth of quantities. We thus do not consider it further, and we remind the reader that we have chosen a broad set of physically motivated kinematic parameters that have a high AUC value in the random forest, which suggests that there are no crucial parameters missing from our set.

In Fig. \ref{fig:AllGalaxiesDeltaSFRCorretionAnalysisBarPlots} we show the correlation strength between $\overline{\sigma}$ and $\rm \Delta SFR$ in light red shaded bars, and the correlation strength between the remaining parameters and $\rm \Delta SFR$ with light blue shaded bars. The light red shaded bar is repeated for each of the remaining parameters, which will be useful for the partial correlation analysis. The uncertainty on each bar is given by the standard deviation of 100 estimates of the correlation using bootstrapped random sampling with return, and the y-axis is oriented with increasingly negative correlation strength from bottom to top, such that the height of a bar above the x-axis correlates with the degree of quenching (i.e. negative $\rm \Delta SFR$). 
We  use $\lambda^{-1}$, $\big(\overline{V}\big)^{-1}$, $\big(\overline{j}\big)^{-1}$ and $ \overline{\sigma}\,/\,\overline{V}$, rather than $\lambda$, $ \overline{V}$, $\overline{j}$ and $\overline{V}\,/\,\overline{\sigma}$, so that all of the light shaded bars have the same orientation. We note that this does not impact the magnitude of the correlations. 

Focusing only on the light shaded bars, it is clear that $\overline{\sigma}$ is the parameter that is most correlated with quenching. This is consistent with our earlier result that velocity dispersion is the most important parameter for classifying quenched objects, but it is important to note that the correlation strength between $\rm \Delta SFR$ and $\overline{\sigma}$ is only $\rm \sim30$ per cent larger than the correlation strengths between $\rm \Delta SFR$ and the following five parameters: $\lambda$, $\overline{V}\,/\,\overline{\sigma}$, $M_{\star}$, $\overline{\mathcal{E}_{k}}$ and  $M_\mathrm{D}$.

We therefore perform a partial correlation analysis to check for the influence of confounding variables. In Fig. \ref{fig:AllGalaxiesDeltaSFRCorretionAnalysisBarPlots}, we compare the correlations of each parameter with $\rm \Delta SFR $ at fixed $\overline{\sigma}$. We choose to fix the velocity dispersion since it has the strongest correlation with $\rm \Delta SFR$ and is the most important parameter for quenching in the random forest analysis. For each parameter, $X$, we show the partial correlation strength between  $\rm \Delta SFR $ and $\overline{\sigma}$ at fixed $X$ in solid red bars, and the partial correlation strength between  $\rm \Delta SFR $ and $X$ at fixed $\overline{\sigma}$ in solid blue bars. We order the parameters along the x-axis in order of decreasing height of the solid blue bars - i.e. in order of decreasing positive correlation strength with quenching at fixed $\overline{\sigma}$. 

The difference between the full and partial correlations is striking. Holding a parameter, $X$, fixed has little effect on the correlation between $\rm \Delta SFR $ and $\overline{\sigma}$. This suggests that the strong relationship between velocity dispersion and $\rm \Delta SFR$ is genuine and is not driven by a confounding variable in our parameter set. The strength of the correlations between the remaining parameters and $\rm \Delta SFR $, on the other hand, is significantly reduced when  $\overline{\sigma}$ is fixed. This shows that the bulk of the relationship between these parameters and quenching is not genuine, but is caused by the confounding variable $\overline{\sigma}$. Comparing the heights of the solid red and solid blue bars, the partial correlation between $\overline{\sigma}$ and $\rm \Delta SFR $ at fixed $X$ exceeds the partial correlation between $X$ and $\rm \Delta SFR$ at fixed $\overline{\sigma}$ by at least $\rm \sim 10$ times the typical uncertainty on the correlation strengths. This highly statistically significant result is clear evidence that the velocity dispersion is the parameter in our set that is most connected with galaxy quenching, and that the relationship between the other parameters and quenching is incidental rather than fundamental, which confirms the random forest result. 

\subsubsection{Visualising the statistical results}
\begin{figure*}
    \includegraphics[width=\textwidth]{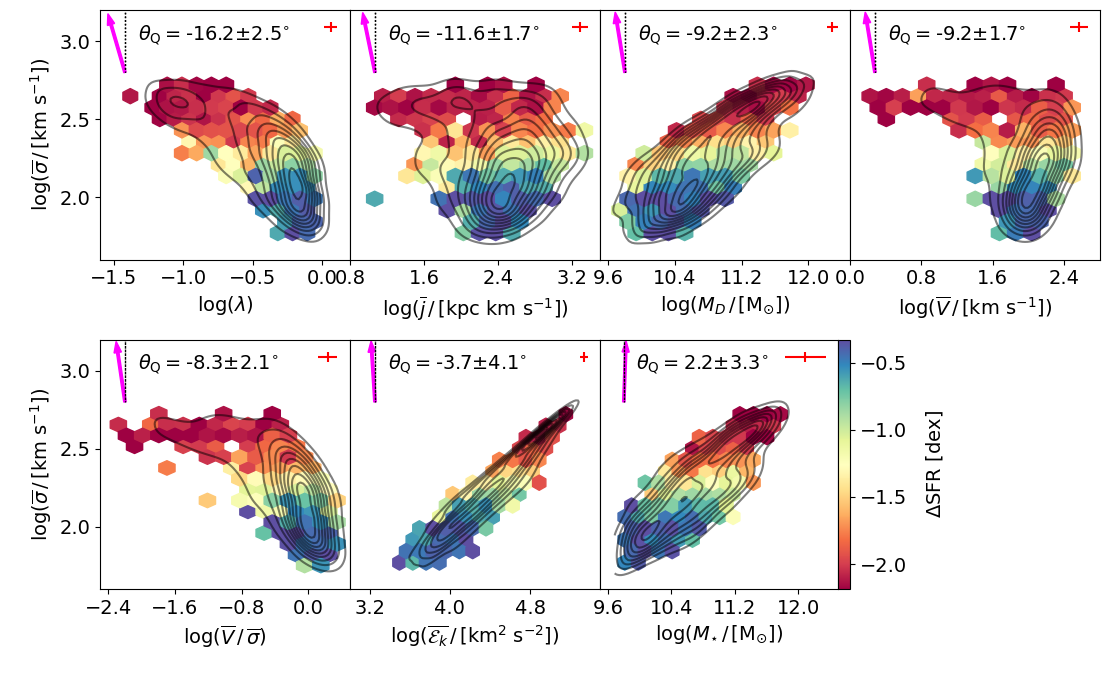}
    \caption{Visual representation of the correlation analysis. In each panel, the average velocity dispersion is shown on the y-axes and a second variable from the parameter set is shown on the x-axis. The hexagonal bins are colour coded by $\rm \Delta SFR$, and linearly spaced density contours are shown in black. Colour gradients generally run parallel to the y-axis, suggesting that this is the optimal direction for quenching galaxies. The quenching angle, $\theta_{\mathrm{Q}}$, quantifies this effect, and the magenta arrows point in the optimal direction. The uncertainty is calculated as the standard deviation of 100 estimates of $\theta_{\mathrm{Q}}$ using bootstrapped random sampling with return. All quenching angles are significantly smaller in magnitude than $\rm 45^{\circ}$, demonstrating that the average velocity dispersion is the best predictor of galaxy quenching, and most quenching angles are $\rm \sim 0^{\circ} $, demonstrating that the secondary parameters are related to quenching only due to their correlation with average velocity dispersion. The red error bars represent the $1\sigma_\mathrm{err}$ uncertainties.}
        \label{fig:PartialCorrelationPlanes_KinematicParamsDeltaSFR}
\end{figure*} 
We visualise the partial correlation analysis in Fig. \ref{fig:PartialCorrelationPlanes_KinematicParamsDeltaSFR}. In each panel we show the distribution of galaxies in the ($X$, $\overline{\sigma}$) plane with black contours, and colour code hexagonal bins by their mean $\rm \Delta SFR$. The colour gradients run parallel to the y-axis, such that the optimal path for transitioning from the star forming to the quenched population of galaxies is in the direction of increasing $\overline{\sigma}$. We formalise this visual assessment using the `quenching angle' \citep{Bluck2020a}, which is given by:
\begin{equation}
    \theta_{\mathrm{Q}} = \arctan\left(\frac{\rho_{\Delta \mathrm{SFR},X|\overline{\sigma}}}{\rho_{\Delta \mathrm{SFR},\overline{\sigma}|X}}\right).
    \label{eqn:quenchingangle}
\end{equation}
$\theta_{\mathrm{Q}}$ is the optimal angle to traverse the ($X$, $\overline{\sigma}$) plane for  decreasing $\rm \Delta SFR$, or in other words, for quenching galaxies. A value of $\rm \theta_{\mathrm{Q}}=0^{\circ}$ suggests that quenching is best achieved by changing $\overline{\sigma}$ and that quenching is independent of $X$, whilst a value of $\rm \theta_{\mathrm{Q}}=90^{\circ}$ suggests that quenching is best achieved by changing $X$ and that quenching is independent of $\overline{\sigma}$. An intermediate value of $\rm \theta_{\mathrm{Q}}=45^{\circ}$ suggests that the optimal path changes $X$ and $\overline{\sigma}$ in equal measure, and hence that both parameters are equally important for quenching galaxies. We report $\rm \theta_{\mathrm{Q}}$ in each panel, and organise the panels from left to right and top to bottom in order of decreasing $\rm |\theta_{\mathrm{Q}}|$. The magenta arrow in each panel points in the optimal direction for quenching. All of the quenching angles are small and the quenching arrows align closely with the y-axis direction. This is yet further evidence that the average velocity dispersion is the most important parameter for quenching.

The largest quenching angle occurs in the top left panel comparing $\overline{\sigma}$ and $\lambda$. Nonetheless, the quenching angle is significantly smaller than $\rm 45^{\circ}$. The ($\lambda$, $\overline{\sigma}$) plane shows that whilst $\lambda$ is good at separating slow rotators (density peak at small $\lambda$) from fast rotators (density peak at large $\lambda$), it is ineffective at separating star forming and quenched galaxies. Indeed, there exists both star forming and quenched fast rotators that have similar values of $\lambda$. These galaxies, however, differ greatly in terms of  $\overline{\sigma}$, where the quenched fast rotators have large $\overline{\sigma}$ and the star forming fast rotators have small $\overline{\sigma}$. The average velocity dispersion is thus more effective than  the dimensionless spin parameter at predicting a galaxy's star forming state. This result is consistent with \citet{Wang2020} who found that there exists both fast rotators and slow rotators below the SFMS, and that one needs to combine the kinematic classification (i.e. fast rotator or slow rotator) with $M_{\star}$ to constrain the star forming state of a galaxy.

The correlation and quenching angle analysis offers further confirmation that $\overline{\sigma}$ is the fundamental parameter in the random forest. We demonstrate this by considering the dependence of quenching on the specific kinetic energy, the parameter that is most correlated with $\overline{\sigma}$, but the argument applies to all of the panels in Fig. \ref{fig:PartialCorrelationPlanes_KinematicParamsDeltaSFR}. In the ($\overline{\mathcal{E}_{k}}$, $\overline{\sigma}$) panel, we see that quenched galaxies have both large $\overline{\mathcal{E}_{k}}$ and large $\overline{\sigma}$, and the black contours emphasise the high degree of correlation between the two parameters. Nonetheless, the quenching angle is able to break this degeneracy. Indeed, the quenching angle is $2^{\circ}$, which confirms that increasing $\overline{\mathcal{E}_{k}}$ at fixed $\overline{\sigma}$ has  no effect on the star forming state of a galaxy. In other words, increasing the specific kinetic energy does increase the likelihood of a galaxy being quenched, but only because of the corresponding increase in $\overline{\sigma}$. The average velocity dispersion is thus the fundamental  parameter for quenching. We highlight that under the assumption of virialisation, $\overline{\mathcal{E}_{k}}$ is related to the specific gravitational potential energy and total specific energy of a galaxy's stellar system. The success of $\overline{\sigma}$ over $\overline{\mathcal{E}_{k}}$ thus shows that \textit{it is not the total energy of the system that is important for quenching, but the fraction of this energy that is contained in a disordered state}.

In the bottom row of Fig. \ref{fig:PartialCorrelationPlanes_KinematicParamsDeltaSFR}, we compare the role of $\overline{\sigma}$ and galaxy mass (as expressed by $ M_\mathrm{\star}~\mathrm{and}~M_\mathrm{D}$) in determining quenching. We find quenching angles that are close to zero, showing that at fixed $\overline{\sigma}$, galaxy quenching is almost independent of mass. This result is striking given the prevalence of the view that mass is responsible for quenching, in the so called `mass-quenching' paradigm \citep{Baldry2006, Peng2010, Peng2012}. We note that mass is highly related to quenching when $\overline{\sigma}$ is not held fixed, as shown by the full correlations in Fig. \ref{fig:AllGalaxiesDeltaSFRCorretionAnalysisBarPlots}. Thus, massive galaxies do tend to be more quenched, but this result is merely a consequence of the correlation between galaxy mass and the true predictor of quenching, average velocity dispersion (see also \citealt{Wake2012, Bluck2016, Bluck2020a, Bluck2020b}). We point out that $M_\star$ is estimated from SDSS photometry \citep{Blanton2011}, so the poor predictive power of stellar mass for quenching cannot be attributed to any possible flaws in our kinematic modelling. Indeed, it is inconceivable that we could have accidentally and erroneously modelled the kinematics in a way that artificially returns estimates of  $\overline{\sigma}$ that are more predictive of galaxy quenching than stellar mass. The significantly superior predictive power of $\overline{\sigma}$ over $M_\star$ is thus genuine and points to a new `velocity dispersion quenching' paradigm (see also \citealt{Wake2012, Bluck2016}).

The ($\overline{V}$, $\overline{\sigma}$) plane explicitly compares the importance of ordered and disordered stellar orbits in galaxy quenching. We consider three regimes for the quenching angle: $\theta_{\mathrm{Q}}\sim0^{\circ}$ would suggest that quenching is most correlated with the average velocity dispersion, which relates to the galaxy bulge/spheroidal component; $\theta_{\mathrm{Q}}\sim45^{\circ}$ would suggest that quenching is determined by $\overline{V}\,/\,\overline{\sigma}$, which quantifies how discy or spheroidal a galaxy is; $\theta_{\mathrm{Q}}\sim0^{\circ}$ would suggest that quenching is most related to the ordered rotation of stars, which relates primarily to the galaxy disc. The small quenching angle thus confirms that quenching is largely independent of the disc properties or the size of the disc relative to the bulge. In fact, it is the absolute value of the velocity dispersion, or properties of the bulge/spheroidal component, that is important for determining the star forming state of a galaxy.

\section{Discussion - How do galaxies quench?}\label{Discussion - How do galaxies quench?}

In this section, we discuss the strong performance of $\overline{\sigma}$ and weak performance of $\overline{V},~ \overline{\sigma},~ \overline{V}\,/\,\overline{\sigma},~ \overline{\mathcal{E}_{k}},~\overline{j},~\lambda,~ M_\mathrm{D},~\mathrm{and}~M_{\star}$ in the context of different quenching mechanisms and previous results. It is important to note at the outset that the strong connection between $\overline{\sigma}$ and quenching does not imply that the velocity dispersion is somehow the cause of galaxy quenching, but rather that any viable quenching mechanism must be able to explain the dominance of $\overline{\sigma}$ in our analysis. We also remind the reader of our choice to define quenching with reference to maintenance mode - i.e. as the processes which prevent the reaccretion of hot halo gas and the subsequent rejuvenation of star formation (see the introduction).  

Our analysis is consistent with previous morphological studies of galaxy quenching which show that the mass of the galaxy bulge is more effective at separating star forming and quenched galaxies than either the total mass of the galaxy, the mass of the galaxy disc, or the bulge-to-total mass ratio ($B/T$) \citep{Lang2014, Bluck2014,  Bluck2021}. We remind the reader of the following advantages of this kinematic study: the kinematic parameters relate to the fundamental physics of bulges (i.e. velocity dispersion) and discs (i.e. circular velocity); the kinematic parameters which describe the
stellar orbits trace all components of mass in a galaxy, unlike the
bulge mass and disc mass which correspond exclusively to the mass
of the stellar system; and the kinematic estimates are free from
a number of assumptions, such as the IMF, stellar templates and
star formation history assumed in SED fitting. Nonetheless, the good
agreement between kinematic and morphological studies clearly shows that quenching
is not connected to the galaxy disc.

Together, the weak performance of $B/T$, $\overline{V}\,/\,\overline{\sigma}$ and $\lambda$ challenges the existence of a deep connection between galaxy morphology and quenching (e.g. \citet{CameronDriver2009,  Gadotti2009, Cappellari2011a, Bell2012,  Omand2014}), and favours a `dispersion-colour' or `bulge-colour' relationship rather than a `morphology-colour' relationship. In other words, the common notion that `elliptical galaxies are red and spiral galaxies are blue' is a little misleading. A more accurate summary is that `galaxies with a prominent bulge are red and galaxies without a prominent bulge are blue', regardless of the existence or extent of any surrounding disc structure. 

The poor predictive power of parameters relating to galaxy discs (i.e. $\overline{V}$ and $\overline{j}$) may seem surprising since the galaxy disc is where stars form. It demonstrates that galaxy quenching is not governed by the same processes that regulate star formation. Quenching is an entirely different process that is regulated by distinct physical mechanisms that are largely independent of the galaxy disc.  This is consistent with \citet{Bluck2020a} who demonstrate that star formation is governed by local processes whilst quenching is fundamentally a global process, which implies that star formation and quenching are (counter-intuitively) distinct phenomena.

We begin our focus on the average velocity dispersion by noting that the virial theorem relates  $\overline{\sigma}$ to the mass density of galaxies, which has long been associated with galaxy quenching (e.g. \citealt{Kauffmann2003b, Brinchmann2004, Franx2008, Wuyts2011, Wake2012, Cheung2012, Fang2013, Bluck2014, Bluck2021}). \citet{Lilly2016} argue that this close connection between galaxy quenching may be an artefact of `progenitor bias'. In this hypothesis, the connection between mass density and quenching is caused by the fact that the star forming progenitors of today's quenched galaxies stopped forming stars at earlier times, when galaxies were smaller at fixed mass and therefore more dense \citep{Trujillo2007, Buitrago2008, Newville2014, vanderWel2014}.  In this framework, galaxies are not quenched \textit{because} of their high densities in any mechanistic sense, but rather the connection between galaxy density and quenching is incidental, i.e. arising out of an \textit{a causal} link to an independent factor. 

The progenitor bias argument is a strong reminder of the potential risks of extrapolating from a correlation to a causal relationship, but we do not consider it further as an explanation here for four reasons. Firstly, galaxies in the green valley region of the $\mathrm{SFR}-M_\star$ plane are thought to  be currently transitioning between the star forming and quenched populations \citep{Wyder2007, Martin2007, Schawinski2014}, so the mass-size relation will have had little time to evolve since the onset of quenching. The progenitor bias effect therefore seems incapable of explaining the larger densities of galaxies in the green valley \citep{Bluck2016}. Secondly, \citet{Bluck2021} show that the strong relationship between galaxy density and quenching is stable since at least $z\sim2$. The evolution of the mass-size relation prior to cosmic noon is unknown, but it is likely to be less significant in the $\rm 4\,Gyr$ between the Big Bang and cosmic noon than the $\rm 10\,Gyr$ between cosmic noon and the current epoch. The progenitor bias argument is thus less credible for explaining the observed importance of galaxy density for the quenching of galaxies at $z\sim2$, which importantly remains invariant to $z\sim0$. 

Thirdly, and most importantly, the virial theorem relates the galaxy density to the \textit{total} velocity of the stellar orbits, and hence the progenitor bias argument cannot by itself explain the observed overwhelming importance of the \textit{disordered} component (i.e. $\overline{\sigma}$) over the ordered component (i.e. $\overline{V}$) of the stellar orbits found in this work. Finally, the progenitor bias argument does not suggest a physical cause of galaxy quenching and offers no physical solution to the cooling catastrophe.

\subsection{Viable Quenching Mechanisms}
Morphological quenching is one possible causal quenching mechanism that is related to $\overline{\sigma}$, where the presence of a central bulge stabilises the galaxy disc against gravitational collapse and thereby reduces the star formation efficiency and quenches the galaxy \citep{Martig2009}. The strong prediction of the morphological quenching scenario is that galaxies with larger bulges (or larger $\overline{\sigma}$) are more likely to be quenched, which our results clearly support. Nonetheless, there are two independent challenges to the morphological quenching framework. Firstly, galaxies do not exist in isolation, and it is difficult to see how the morphological quenching scheme could be stable to galaxy mergers. Galaxy interactions will likely disturb the stabilising influence of the galaxy bulge and hence rejuvenate star formation, at least temporarily or periodically (which is not observed in quenched systems). 

Secondly, the morphological quenching mechanism does not reduce the amount of molecular gas in a galaxy. One would therefore expect quenched galaxies to have large gas reservoirs, but this is not observed and conversely we find that quenched galaxies have gas fractions significantly lower than galaxies on the main sequence (e.g. \citealt{Saintonge2016, Saintonge2017, Piotrowska2020, Brownson2020, Ellison2019a, Ellison2021}). Thus, morphological quenching may be important for reducing the star formation efficiency within galaxies, but we require an additional mechanism that solves the cooling catastrophe and reduces the gas fraction. 

We now consider causal galaxy quenching mechanisms in the context of the cooling catastrophe. In particular, we examine the consistency of our analysis with supernovae feedback (e.g. \citealt{Cole2000, Henriques2019}), virial heating (e.g. \citealt{Dekel2006, Woo2013}), and AGN feedback (e.g. \citealt{Croton2006, Bower2008, Hopkins2008}).

Supernovae are unlikely to emit sufficient energy to keep massive halos hot (e.g. \citealt{Cole2000, Croton2006, Bower2008}). This is particularly true in quenched systems, which lack Type-II supernovae. The total stellar mass records the star formation rate integrated over the lifetime of a galaxy, and it is therefore related to the number of, and energy released by, supernovae \citep{Bluck2020a}. Under the supernovae feedback solution to the cooling catastrophe, therefore, one would naively expect $M_\star$ to be the strongest predictor of galaxy quenching. Yet, we find that  at fixed velocity dispersion, stellar mass is not correlated with quenching at all within the uncertainties, and hence we rule out supernovae feedback as a viable solution to the cooling catastrophe. This result is remarkable given the numerous studies focusing on the correlation between $M_\star$ and quenching \citep{Baldry2006, Peng2010, Peng2012}. 

Alternatively, quenching by virial shock heating depends strongly on the mass of the halo \citep{Dekel2006,Woo2013, Bluck2020a}. Previous studies show that at fixed stellar mass quenched galaxies live in more massive halos than star forming galaxies \citep{Woo2013, Mandelbaum2016}, and in fact that the correlation  between halo mass and quenching is stronger than the correlation between stellar mass and quenching \citep{Woo2013}. The  measurements of the central density of galaxies, however, prove even more effective than halo mass \citep{Bluck2014, Woo2015}. We have not directly considered halo mass in this work, but we note that $M_\mathrm{D}$ is implicitly related to the halo mass for galaxies that are not in dense groups or clusters, and we  highlight the  dominance of $\overline{\sigma}$ over $M_\mathrm{D}$ for predicting quenching. More directly, we recall the dominance of $\overline{\sigma}$ over halo mass in our previous work, where halo mass is only weakly related to galaxy quenching at fixed stellar velocity dispersion \citep{Bluck2016, Bluck2020a, Piotrowska2021}. Thus, we conclude that quenching is caused not by halo shock heating, but rather by some alternative mechanism connected with the velocity dispersion.

The success of $\overline{\sigma}$ in predicting galaxy quenching begs for a physical mechanism that is related to the velocity dispersion. The $M_\mathrm{BH}-\overline{\sigma}$ relationship is a tight relationship between black hole mass  and stellar velocity dispersion (e.g. \citealt{Ferrarese2000, McConnell2011,McConnell2013, Kormendy2013, Saglia2016}). We note that different versions of $\overline{\sigma}$ are used in the literature for the $M_\mathrm{BH}-\overline{\sigma}$ relation (e.g. the average velocity dispersion measured within the central kpc or within $1\,\mathrm{R_e}$), but we find that our key result is not dependent on the precise definition. Combined with our key result, the $M_\mathrm{BH}-\overline{\sigma}$ relationship shows that galaxies with larger black hole masses are more likely to be quenched than galaxies with smaller black hole masses. This interpretation is consistent with modern theoretical simulations of galaxy evolution, which require feedback from AGN to quench galaxies \citep{Scahye2015, Vogelsberger2014a, Vogelsberger2014b, Weinberger2017, Henden2018, Henriques2019, Zinger2020, Piotrowska2021}.

Galaxy mergers are a possible cause  of the $M_\mathrm{BH}-\overline{\sigma}$ relationship. In this scenario,  mergers create spheroids from discs (e.g. \citealt{Toomre1972}), thereby increasing $\overline{\sigma}$,  and simultaneously drive gas inflows towards the galaxy centre which can trigger nuclear star bursts and energetic AGN feedback, thereby increasing $M_\mathrm{BH}$ \citep{Hopkins2008}. This is $one$ physical interpretation of the $M_\mathrm{BH}-\overline{\sigma}$ relationship, but we remain agnostic about the true origin of the connection between  $M_\mathrm{BH}$ and $\overline{\sigma}$ in this work. We simply note the empirical fact that galaxies with larger $\overline{\sigma}$ host more massive dynamically measured black holes, and look for quenching mechanisms related to $M_\mathrm{BH}$.

The black hole mass traces the total energy released during the black hole growth (e.g. \citealt{Sotan1982, Silk2012, Bluck2011, Fabian2012}) and is therefore related to AGN feedback. As mentioned in the introduction, there are two modes of AGN feedback: the high Eddington ratio `quasar mode' (e.g. \citealt{DiMatteo2005, Hopkins2008, Maiolino2012, Bischetti2019}) and the low Eddington ratio `preventative mode' (e.g. \citealt{Croton2006, Bower2008, Fabian2006, Sijacki2007, Zinger2020}). We now provide three reasons for favouring preventative-mode AGN feedback over quasar-mode AGN feedback as the mechanism  responsible for quenching galaxies. 

Firstly, quasar-mode feedback is a violent event that is more closely related to the rate of accretion of gas onto the black hole (i.e. the rate of growth of the black hole, e.g. \citealt{DiMatteo2005, Hopkins2008, Maiolino2012}). The energy released during preventative-mode feedback, on the other hand, is related to the integrated black hole growth rate over time (e.g. \citealt{Croton2006, Bower2008, Fabian2006}), or in other words, the black hole mass \citep{Bluck2020a}. The prominence of $\overline{\sigma}$ in our random forest analysis and the existence of the $M_\mathrm{BH}-\overline{\sigma}$ relationship therefore points to preventative-mode feedback rather than quasar-mode feedback as the mechanism driving galaxy quenching. We also note that violent quasar-mode feedback is a rare event. Preventative-mode feedback, on the other hand, regularly deposits smaller levels of energy into the halo. It could therefore keep the halo hot and prevent the reaccretion of gas onto the galaxy over long timescales.

Secondly, \citet{Bluck2020a, Bluck2021} show that galaxy quenching depends on global galaxy properties rather than their local counterparts, which appears most consistent with preventative-mode feedback. This is because quasar-mode feedback drives massive galactic outflows (e.g. \citealt{Maiolino2012, Cicone2014, Cicone2015, Fluetsch2019}) which are more likely to influence the central bulge region rather than the outer disc. The accretion rate is lower in the preventative-mode feedback paradigm, however, and the energy released does not significantly influence  the galaxy directly. Instead, it simply heats the halo and prevents the accretion of pristine gas, which  simultaneously starves the bulge and disc components of gas in equal measure. 

Finally, the ejection of gas in the quasar mode is expected to halt star formation on short timescales, whilst the halo heating from preventative-mode feedback should not affect extant gas in the galaxy disc and should therefore allow star formation to continue temporarily, even after the accretion of pristine gas has stopped. It is possible to differentiate between these rapid and delayed/preventative quenching scenarios by comparing the stellar metallicity of star forming and quenched galaxies \citep{Peng2015, Trussler2020, Bluck2020b}. These studies find that star formation continues whilst the galaxy is starved of pristine gas, which is in good agreement with the preventative-mode feedback paradigm. 

There is weak direct evidence of preventative feedback in Figs. \ref{fig:AllGalaxiesDeltaSFRCorretionAnalysisBarPlots} and \ref{fig:PartialCorrelationPlanes_KinematicParamsDeltaSFR}. At fixed $\overline{\sigma}$, galaxies with more prominent discs (i.e. larger $\lambda$, $\overline{j}$, $\overline{V}$) have less negative $\rm \Delta SFR$ and are less quenched. The positive correlation between the prominence of the disc and $\rm \Delta SFR$ could be because galaxies with a significant disc have a larger extant gas supply and can therefore continue forming stars after the onset of galaxy quenching. Galaxies without a disc, on the other hand, which are likely to have undergone a recent  merger, probably do not have a large extant gas supply and will stop forming stars soon after the cessation of gas accretion. It is important to recognise that these secondary partial correlations with $\lambda$, $\overline{j}$, $\overline{V}$ are weak. We cautiously offer an explanation of their origin, but we reiterate the key result that these correlations are significantly weaker than the partial correlations with $\overline{\sigma}$ at fixed $\lambda$, $\overline{j}$, or $\overline{V}$. The true quenching mechanism is thus overwhelmingly related to $\overline{\sigma}$, for which we have suggested an explanation via (most probably preventative) AGN feedback.

\section{Summary}\label{Summary}
In this paper we study the connection between galaxy kinematics and quenching for 1862 galaxies taken from the MaNGA survey. The galaxies in our sample have $\rm log(\mathit{M_{\star}}\,/\,M_{\odot})>9.8$, show no evidence of a recent merger or interaction with a companion galaxy, and pass the data quality cuts established in Section \ref{2D Kinematic Model} as necessary for effective kinematic modelling. 

First, we model the moment-1 maps of 70 per cent of galaxies, using an inclined rotating disc model, carefully accounting for the effect of beam smearing where possible. We use an alternative simplistic method, the `histogram technique', for the remaining 30 per cent of galaxies that are inconsistent with the inclined rotating disc model. Second, we use the  moment-1 model to correct the moment-2 maps for the effect of differential disc rotation. The estimates of the intrinsic rotational velocity and velocity dispersion are then used to define the  following kinematic parameters: the mean circular velocity ($\overline{V}$), the mean velocity dispersion ($\overline{\sigma}$), the ratio of ordered to disordered stellar orbital velocity  ($ \overline{V}\,/\,\overline{\sigma}$), the mean specific kinetic energy ($\overline{\mathcal{E}_k}$), which is also an accurate estimate of the gravitational potential and the total specific energy of the system (under the assumption of virialisation), the mean specific angular momentum ($\overline{j}$), the dimensionless spin parameter  ($\lambda$), and the dynamical mass ($ M_\mathrm{D}$). All parameters are calculated within $\rm 1\, R_{e,kin}$  except $\lambda$ which is calculated within 1$\rm R_{e}$. We also add the global stellar mass ($ M_{\star}$) to our sample.   

We rigorously validate our kinematic model by testing its performance on synthetic galaxy data, and by comparing its outputs to more traditional galaxy properties and scaling relations. The key performance metrics/tests are as follows:
\begin{enumerate}
    \item The inclined rotating disc model is able to recover the maximum rotation velocity with a typical accuracy of 3 per cent for the data considered in this work, as determined through the analysis of simulated mock data.
    \item The success of our inclined rotating disc model is consistent with a galaxy's classification as a slow rotator or fast rotator, as prescribed by its position in the $(\lambda, \epsilon)$ plane. Thus, our model is able to accurately characterise the kinematic state of a galaxy.
    \item The kinematic estimates are consistent with the Faber-Jackson and Tully-Fisher relations. We also define a new scaling relation, the Mass-Velocity relation, which  compares a galaxy's stellar mass with the total velocity of the stellar orbits (i.e. considering both ordered and disordered motion). The Mass-Velocity relation is tighter than both the FJ and TF relations, clearly indicating the improvement in information content of our generalised approach.
\end{enumerate}
These  tests on the accuracy of our kinematic parameters encourage their use in the  study of galaxy quenching. 

In Section \ref{Results}, we study the relationship between the kinematic parameters and the star forming state of a galaxy, parameterised by its logarithmic offset from the SFMS, $\rm \Delta SFR$. We perform a random forest analysis to identify the parameters that are best at separating star forming and quenched galaxies. Our key findings are as follows:
\begin{enumerate}
    \item The average velocity dispersion is the most important parameter for determining whether a galaxy is star forming or quenched. Galaxies with $\overline{\sigma}>220\mathrm{\, km~s^{-1}}$ are mostly quenched, whilst galaxies with $ \overline{\sigma}<220\mathrm{\, km~s^{-1}}$ are mostly star forming. Note, that these values of $\overline{\sigma}$ are in 3D space - i.e. $\sqrt{3}$ times the LOS dispersion. 
    \item Parameters that relate exclusively to the disc, i.e. $\overline{j}$ and $\overline{V}$, are not important in the random forest. Thus, although disc properties are related to the SFR of star forming galaxies, we find that quenching is governed by properties of the bulge/spheroidal component, as parameterised by $\overline{\sigma}$. This shows that quenching is not simply the inverse of the process of forming stars. It is an altogether different phenomenon regulated through radically different galaxy properties.
    \item Quenching is not constrained by properties that quantify whether a galaxy is rotation- or dispersion-dominated, such as $ \overline{V}\,/\,\overline{\sigma}$ and $\lambda$. Thus, the commonly held view that disc galaxies are star forming and spheroids are quenched is misleading. In fact, the absolute level of the velocity dispersion is most important for quenching, not relative levels of velocity dispersion and ordered rotation.
    \item We complement the random forest analysis with a partial correlation analysis. This confirms that when $\overline{\sigma}$ is held constant, the other parameters are only marginally related to quenching. 
    \item We emphasise that parameters that related to the total mass and morphology of the system  (i.e. $\overline{\mathcal{E}_k}$, $M_{\star}$, $ M_\mathrm{D}$, $ \overline{V}\,/\,\overline{\sigma}$ and $\lambda$) show strong correlations with quenching, but these  correlations are almost entirely removed when $\overline{\sigma}$ is held constant. Thus, we agree with all prior `mass-quenching' and `morphology-colour' works that parameters related to mass and morphology are phenomenological related to quenching, but we point out that this connection  is merely a proxy of the true quenching mechanism. What really matters for galaxy quenching is the absolute level of disordered motion. 
\end{enumerate}

We construct an argument for the physical origin of the clear connection between $\overline{\sigma}$ and quenching via the paradigm of AGN feedback. The average velocity dispersion is well correlated with  black hole mass, $M_{\mathrm{BH}}$ (e.g. \citealt{Ferrarese2000, McConnell2011,McConnell2013,Saglia2016}), which traces the total energy released by preventative-mode feedback during the growth of a black hole. In other words, our analysis supports a scenario in which galaxies quench due to significant preventative feedback. This is consistent with theoretical predictions from numerical simulations (e.g. \citealt{Bluck2016, Davies2019, Terrazas2020, Zinger2020, Bluck2020a, Piotrowska2021}), and is explained physically by AGN injecting energy into galaxy halos, keeping them hot and thereby preventing further accretion of gas and  halting star formation.

The data disfavours the following alternative quenching scenarios: quenching via supernovae feedback, since $M_{\star}$ is less predictive of quenching than $\overline{\sigma}$; morphological quenching, since $\overline{V}\,/\,\overline{\sigma}$ and $\lambda$ are less predictive of quenching than $\overline{\sigma}$; and halo quenching, since $M_\mathrm{D}$ and $M_\mathrm{H}$ \citep{Bluck2016, Bluck2020a} are less predictive of quenching than $\overline{\sigma}$.

\section*{Acknowledgements}
SB, AFLB, RM, and GJ acknowledge ERC Advanced Grant
695671 QUENCH, and support from the UK Science and Technology Facilities Council (STFC). We also thank the referee for their careful reading of our manuscript and many helpful suggestions.

Funding for the Sloan Digital Sky Survey IV has been provided by the Alfred P. Sloan Foundation, the U.S. Department of Energy Office of  Science, and the Participating Institutions. 

SDSS-IV acknowledges support and  resources from the Center for High Performance Computing  at the University of Utah. The SDSS website is \href{https://www.sdss.org}{https://www.sdss.org}.

SDSS-IV is managed by the Astrophysical Research Consortium for the Participating Institutions  of the SDSS Collaboration including the Brazilian Participation Group, the Carnegie Institution for Science, Carnegie Mellon University, Center for Astrophysics | Harvard \&  Smithsonian, the Chilean Participation Group, the French Participation Group,  Instituto de Astrof\'isica de  Canarias, The Johns Hopkins  University, Kavli Institute for the  Physics and Mathematics of the  Universe (IPMU) / University of  Tokyo, the Korean Participation Group,  Lawrence Berkeley National Laboratory,  Leibniz Institut f\"ur Astrophysik  Potsdam (AIP),  Max-Planck-Institut  f\"ur Astronomie (MPIA Heidelberg), Max-Planck-Institut f\"ur  Astrophysik (MPA Garching),  Max-Planck-Institut f\"ur  Extraterrestrische Physik (MPE),  National Astronomical Observatories of  China, New Mexico State University,  New York University, University of  Notre Dame, Observat\'ario  Nacional / MCTI, The Ohio State  University, Pennsylvania State  University, Shanghai  Astronomical Observatory, United  Kingdom Participation Group,  Universidad Nacional Aut\'onoma  de M\'exico, University of Arizona,  University of Colorado Boulder,  University of Oxford, University of  Portsmouth, University of Utah,  University of Virginia, University  of Washington, University of Wisconsin, Vanderbilt University,  and Yale University.

\section*{Data Availability}
The MaNGA data underlying this article are publicly available and can be accessed at \href{https://www.sdss.org/dr15/manga/manga-data/data-access/}{https://www.sdss.org/dr15/manga/manga-data/data-access/}.
The kinematic data products produced in this article are publicly available and can be accessed at \href{https://www.kicc.cam.ac.uk/files/brownson22_publicdata.zip}{https://www.kicc.cam.ac.uk/files/brownson22\_publicdata.zip}. Researchers interested in using these data are encouraged to first consult the README file. Questions can be directed to SB and AFLB.



\bibliographystyle{mnras}
\bibliography{Kinematics}




\appendix

\section{Example Fits}\label{Example Fits}
\begin{figure*}
    \includegraphics[width=\textwidth]{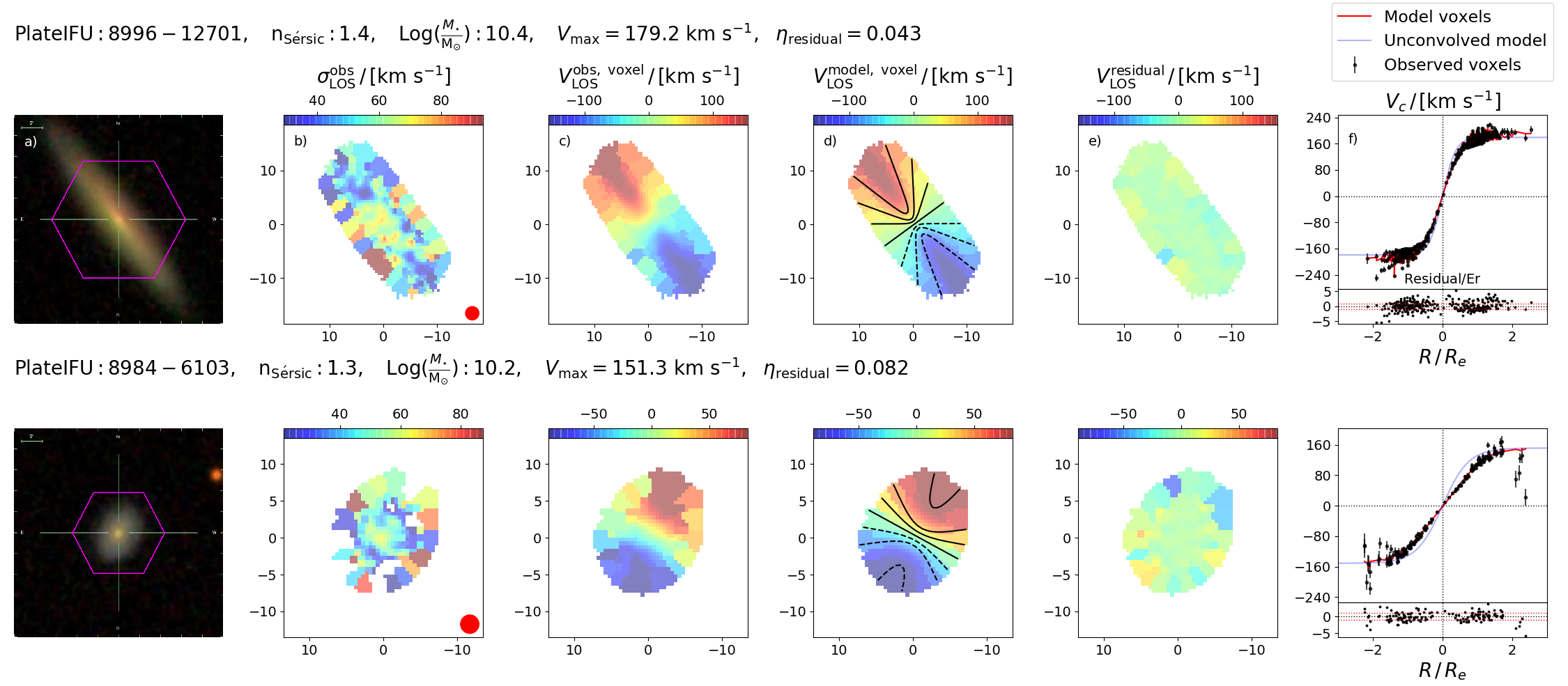}
    \caption{Examples of two successful fits of low $\rm n_{S\acute{e}rsic}$ galaxies. This figure has the same structure as Fig. \ref{fig:GoodFits_2Dfit_pvdiagram}. Both galaxies show strong velocity gradients, and the model residuals are small. 8996-12701 has $\rm \arccos((b/a)_\mathrm{phot})\sim70^{\circ}$, demonstrating our model's ability to fit the kinematics of relatively inclined systems.}
        \label{fig:LowNGoodFits_2Dfit_pvdiagram}
\end{figure*} 
\begin{figure*}
    \includegraphics[width=\textwidth]{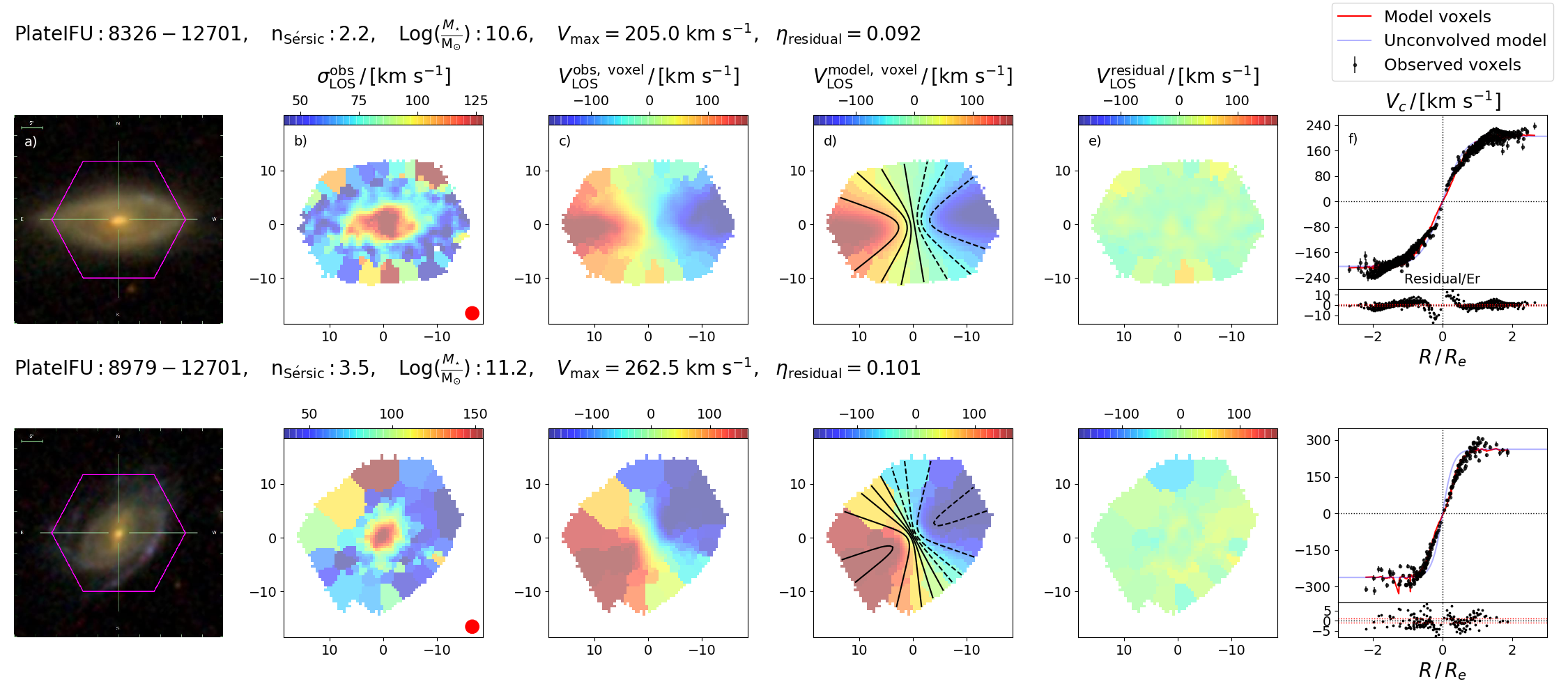}
    \caption{Examples of two successful fits of intermediate $\rm n_{S\acute{e}rsic}$ galaxies. This figure has the same structure as Fig. \ref{fig:GoodFits_2Dfit_pvdiagram}. The SDSS $g,r,i$ composite images show that these galaxies are bulge plus disc systems. Nonetheless, the moment-1 maps still exhibit gradients that are consistent with inclined disc rotation.}
        \label{fig:MidNGoodFits_2Dfit_pvdiagram}
\end{figure*} 
In this appendix we show a number of example fits to demonstrate the performance of the inclined disc rotation model. All figures in this section have the same structure as Fig. \ref{fig:GoodFits_2Dfit_pvdiagram}, which we advise the reader to re-examine prior to reading this section. 

In Fig. \ref{fig:LowNGoodFits_2Dfit_pvdiagram} we show two examples of good fits at low S\'ersic index ($\rm n_{S\acute{e}rsic}<2$), which are generally considered to be rotation-dominated in the local Universe \citep{Wisnioski2015, Ubler2019}. Indeed, both galaxies exhibit clear velocity gradients consistent with simple ordered rotation, and our model is able to recover their kinematics, with low residuals in both the sky plane and the PV diagram. We also note the high inclination of galaxy 8996-12701, as shown in the SDSS $g,r,i$ composite image. This galaxy has $\rm \arccos((b/a)_\mathrm{phot})\sim70^{\circ}$, yet the model is still able to accurately fit its kinematics. This is consistent with our tests on simulated data, where the model is able to accurately constrain the kinematics of galaxies with inclination lower than $\rm 80^{\circ}$ with accuracy better than 25 per cent.

In Fig. \ref{fig:MidNGoodFits_2Dfit_pvdiagram} we show two good fits of galaxies with intermediate S\'ersic index ($\rm 2<n_{S\acute{e}rsic}<4$), which are generally bulge plus disc systems. Indeed, both galaxies in Fig. \ref{fig:MidNGoodFits_2Dfit_pvdiagram} show a disc structure and a bright central peak in their SDSS $g,r,i$ composite images, as well as a clear increase in $\sigma_\mathrm{LOS}^\mathrm{obs}$ towards the galaxy centre. Nonetheless, both galaxies have stellar velocity maps with strong gradients that are well modelled by inclined disc rotation.  We note that there is a noticeable increase in the residuals in the PV diagrams for the $\rm n_{S\acute{e}rsic}>2$ systems. This likely reflects the fact that these systems contain a bulge and have more complex kinematics, which we make no attempt to model. However, the increase in the PV residuals is small and the overall characterisation of the rotation curve in the PV diagram is clearly still excellent. 

\begin{figure*}
    \includegraphics[width=\textwidth]{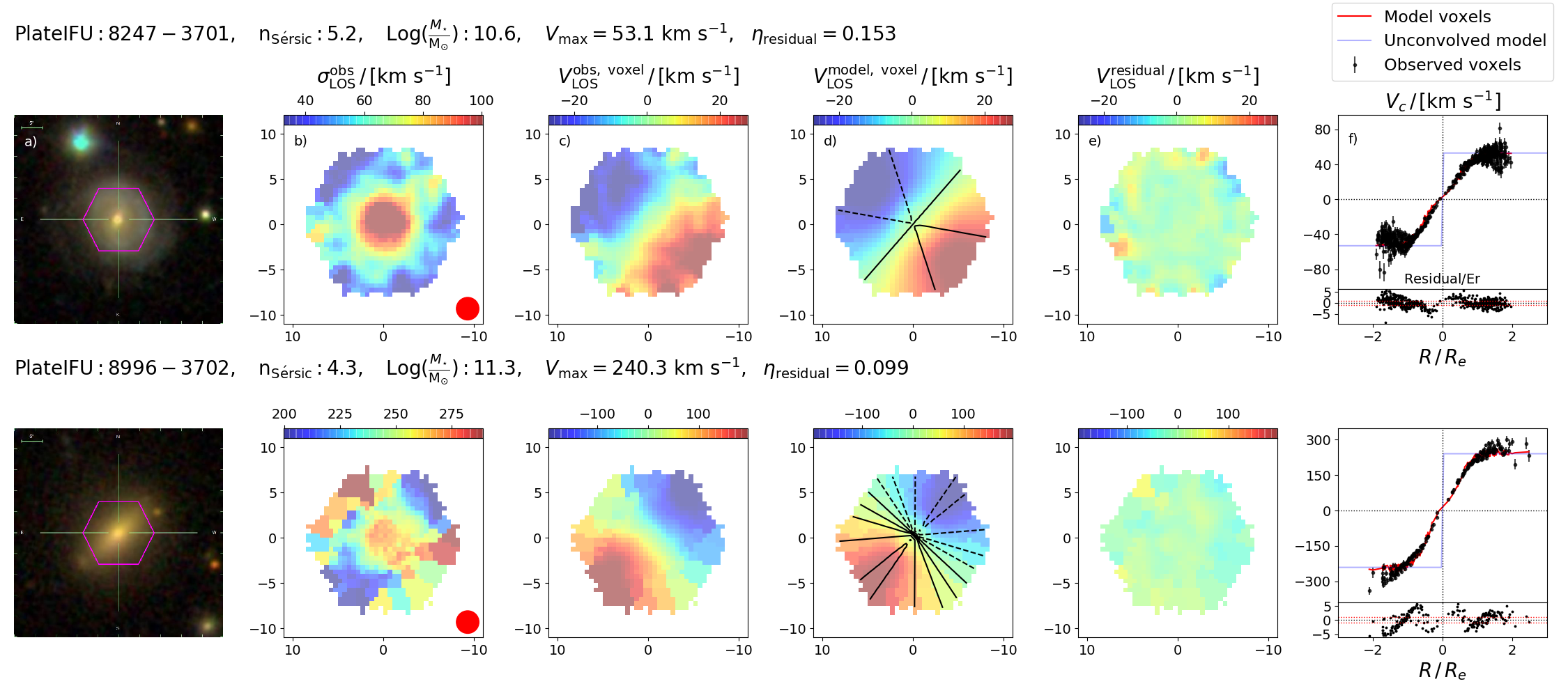}
    \caption{Examples of two successful fits of high $\rm n_{S\acute{e}rsic}$ galaxies. This figure has the same structure as Fig. \ref{fig:GoodFits_2Dfit_pvdiagram}. The model is able to fit the ordered rotation even in these dispersion-dominated systems.}
        \label{fig:HighNGoodFits_2Dfit_pvdiagram}
\end{figure*} 

In Fig. \ref{fig:HighNGoodFits_2Dfit_pvdiagram}, we show two examples of good fits of high S\'ersic index galaxies ($\rm n_{S\acute{e}rsic}>4$), which are spheroidal and are generally dispersion-dominated. Indeed, both galaxies have large velocity dispersion when contrasted with their maximum circular velocities. Nonetheless, they  still  have velocity maps that demonstrate ordered rotation, and remarkably, our model is able to fit the rotational kinematics. This result validates our approach of attempting to fit all galaxies in our sample with the inclined rotating disc model, including dispersion-dominated spheroids. Both galaxies display steeply rising unconvolved rotation curves shown in blue, which probably reflect the presence of a significant bulge component \citep{Noordermeer2007, Lelli2016, Lelli2021}.

\begin{figure*}
    \includegraphics[width=\textwidth]{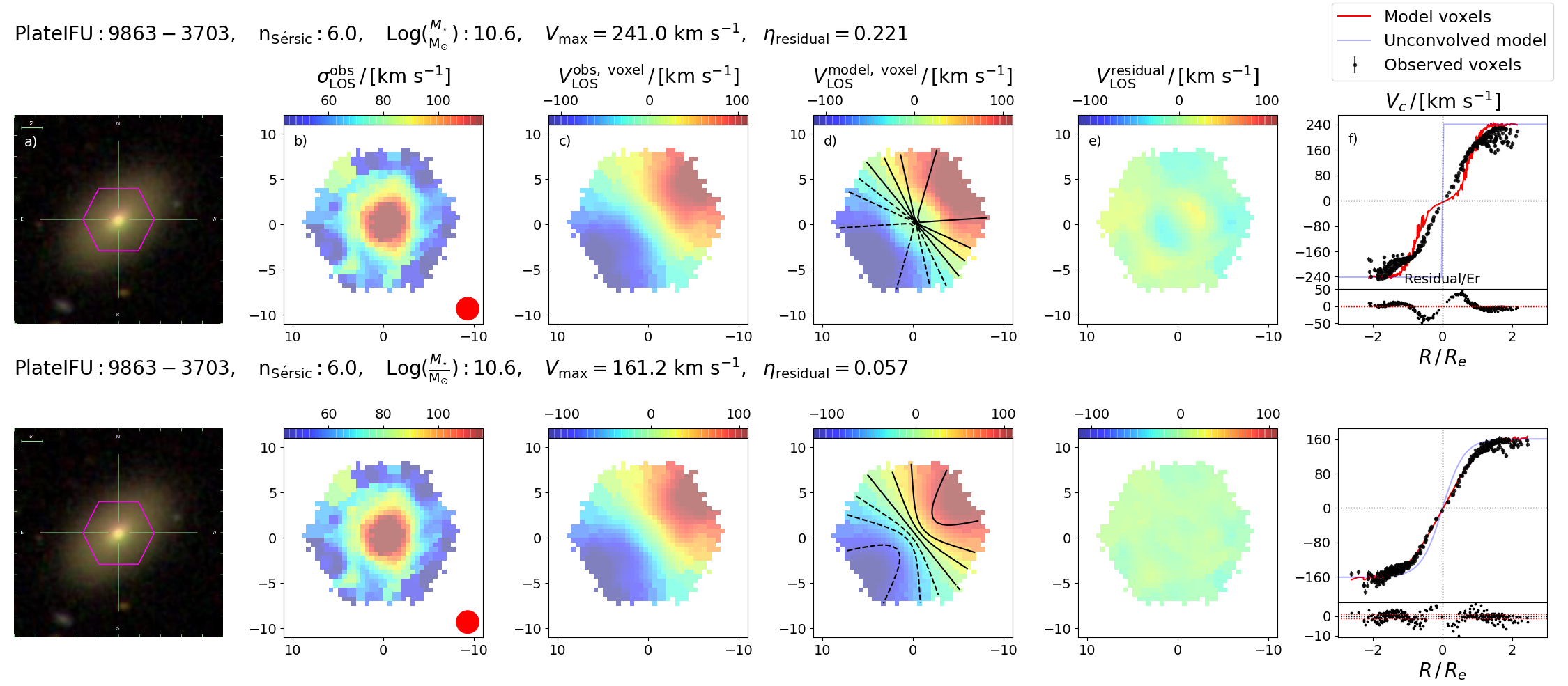}
    \caption{Example of a fit improved by adopting the \textsc{dap} flux map. The figure has the same structure as Fig. \ref{fig:GoodFits_2Dfit_pvdiagram}. The top row shows the best fitting model when adopting the S\'ersic brightness profile, whilst the bottom row shows the best fitting model (of the same galaxy) when adopting the flux map from the \textsc{dap}. The \textsc{dap} flux map fit removes the suppression of $\rm |V^{model, voxel}_\mathrm{LOS}|$ caused by the steep, centrally-concentrated S\'ersic profile.}
        \label{fig:PVSquiggle_2Dfit_pvdiagram}
\end{figure*} 

In Fig. \ref{fig:PVSquiggle_2Dfit_pvdiagram} we show an example galaxy (9863-3703) whose fit is improved by using the \textsc{dap} $g$-band flux map during PSF convolution and Voronoi binning, as first mentioned in Section \ref{2D Kinematic Model}. In the top row  we show the fit returned when assuming the S\'ersic light profile, and in the bottom row we show the fit returned when assuming the \textsc{dap} $g$-band flux map. We highlight the discrepancy between the data and the model when the S\'ersic light profile is assumed, where  the model underestimates $\rm |V^{obs, voxel}_\mathrm{LOS}|$ in the central regions and a `squiggle' is observed in the PV diagram. As with most galaxies showing this effect, 9863-3703 is a high $\rm n_{S\acute{e}rsic}$ galaxy, with a centrally peaked surface brightness profile. Any slight error in the S\'ersic model will therefore have a strong influence on the fit, particularly in the central regions. The \textsc{dap} flux map improves the fit since it is shallower than the S\'ersic profile, which prevents the central region from dominating the PSF convolution. The \textsc{dap} flux map improves the fit for $\rm \sim 400$ galaxies, though we note that the PV squiggle is generally smaller than that of 9868-3703, which has been chosen to highlight the effect.  

We have investigated the possible degeneracy between the S\'ersic and \textsc{dap} flux maps in fitting the inclined rotating disc model. Galaxies that do not show a squiggle in the PV diagram have $V_\mathrm{max}$ estimates from both models that are consistent within $\rm \sim10\,km~s^{-1}$ on average. This suggests that the \textsc{dap} flux map can be taken as a good proxy of the S\'ersic profile for our purposes, even though it is known to be shallower and less centrally concentrated than the true brightness profile since it is PSF convolved. However, we recognise that the PV squiggle could be evidence of genuine inconsistency with the inclined rotating disc model, which the \textsc{dap} flux map fits wrongly conceal. We therefore repeat the analysis in this work using the simple method for all galaxies such as 9868-3703 that show the PV squiggle, rather than adopting the \textsc{dap} flux map fit, and confirm that the results are stable to this test in Appendix \ref{Testing the kinematic model}.

\section{Tests on the stability of the results}\label{Testsonthestabilityoftheresults}
In this appendix, we make subtle changes to the random forest analysis to test the stability of our key result that the average velocity dispersion is the most important parameter for predicting galaxy quenching. In Section \ref{Testing the kinematic model}, we test different implementations of the kinematic model; and in Section \ref{Testing the effect of measurement uncertainty}, we test the effect of differential measurement uncertainty.

\subsection{Testing the kinematic model}\label{Testing the kinematic model}

\begin{figure}
    \includegraphics[width=\columnwidth]{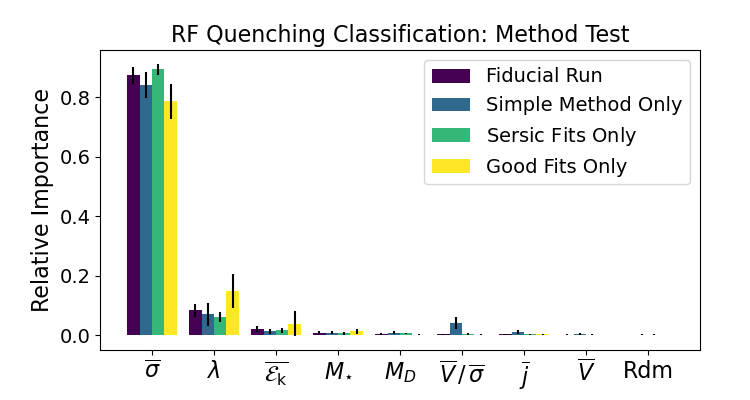}
    \caption{The random forest analysis for  different versions of the kinematic model. The `Fiducial Run'  results are identical to those presented in Fig. \ref{fig:AllGalaxiesRF_RelativeImportance}, and are included for comparison. The `Good Fits Only' test only considers  galaxies that have a successful kinematic fit, the `Simple Method Only' test assumes the simple method for all galaxies and does not use the kinematic model at all, and the `S\'ersic Fits Only' test does not use the \textsc{dap} $g$-band flux map in the kinematic modelling. The average velocity dispersion is consistently ranked the most important parameter for quenching, which demonstrates that our key result is not dependent on the precise prescription of the kinematic model.}
        \label{fig:RFTests_GoodFitsOnl_SimpleMethodOnly_SersicFitsOnly}
\end{figure}

In Fig. \ref{fig:RFTests_GoodFitsOnl_SimpleMethodOnly_SersicFitsOnly}  we show the random forest  for three alternative formulations of the kinematic model. It is important to note that it is not possible to know which, if any, of these formulations is truly correct. The spirit of this section is to try a range of reasonable approaches. The key point is that the ordering of the parameters is the same for all three tests, and the average velocity dispersion is consistently found to be the most important parameter for predicting quenching.  Thus, taken together, these tests provide  strong evidence that our key result is robust and not strongly dependent on the choices outlined in Section \ref{Galaxykinematics}. We discuss the three tests in order.

First, in the `Simple Method Only' test, we do not use our kinematic model at all, but rather we model the kinematics of all galaxies using the simple method. We remind the reader that the simple method is entirely independent of the kinematic model estimates, and indeed it is a far more simplistic, non-parametric method for constraining kinematics. Nonetheless, the random forest identifies $\overline{\sigma}$ as the most predictive parameter, thus providing independent support for our key result. The consistency of this test with the fiducial run demonstrates that the simple method is sufficiently accurate for studying galaxy quenching in kinematic parameters.

We note that we do not apply the bias-correction to the simple method for $\rm n_{S\acute{e}rsic}<3$ galaxies (see Section \ref{Simplistic kinematic model}) here since it could introduce an additional artificial distinction between $\rm n_{S\acute{e}rsic}<3$ and $\rm n_{S\acute{e}rsic}>3$ galaxies. Instead, we treat all galaxies equally and use the simple method without any bias-correction. We are not as concerned about introducing an artificial distinction between $\rm n_{S\acute{e}rsic}<3$ and $\rm n_{S\acute{e}rsic}>3$ galaxies in the fiducial run, since the bias-correction is only applied to the 15 per cent of discs that are not well fit by the inclined rotating disc model and it is therefore unlikely to have a significant impact on the results.

Second, in the `S\'ersic Fits Only' analysis, we estimate the kinematics using either the kinematic model with the S\'ersic flux map or the simplistic method. In other words, we do not include any galaxies whose kinematics are determined using the kinematic model and adopting the \textsc{dap} $g$-band flux as the moment-1 map. Instead, we resort to the simple method when the kinematic model together with the S\'ersic flux map returns a fit that exhibits a squiggle in the PV diagram, as described in Section \ref{2D Kinematic Model} and Appendix \ref{Example Fits}. We recognise that the \textsc{dap}  $g$-band flux map is not an accurate representation of the true surface brightness since it is PSF convolved. Nonetheless, this test demonstrates that the $g$-band flux is sufficiently representative of the true surface brightness for the purposes of studying galaxy quenching.

Finally, in the `Good Fits Only' test, we only consider galaxies that have a good kinematic fit, and we remove the 30 per cent of galaxies that are inconsistent with the kinematic model rather than resorting to the simple method. The `Good Fits Only' test thus constitutes our sample with the most reliable kinematic estimates. It is important to recognise that the high accuracy of this sample comes at the cost of lacking genuine slow rotators, which are fundamentally inconsistent with inclined disc rotation and consequently have failed fits. Indeed, the distribution of galaxies for this sample in the ($\overline{V}$, $\overline{\sigma}$) plane is missing the large $\overline{\sigma}$, small $\overline{V}$ galaxies seen in Fig. \ref{fig:PartialCorrelationPlanes_KinematicParamsDeltaSFR}. Nonetheless, the random forest analysis of this sample is consistent with the fiducial run. This demonstrates that it is not only slow rotators whose quenching is dominated by $\overline{\sigma}$, but the quenching of fast rotators is also predicted best by velocity dispersion. The dimensionless spin parameter, by contrast, is effective at identifying quenched slow rotators, but it is not particularly effective at separating star forming fast rotators and quenched fast rotators, as shown in Fig. 
\ref{fig:FullCorrelations_KinematicParamsDeltaSFR}. This is the key advantage of $\overline{\sigma}$ over $\lambda$ for predicting quenching. 

\subsection{Testing the effect of measurement uncertainty}\label{Testing the effect of measurement uncertainty}
In this section of the appendix, we investigate the effect of measurement error on the random forest analysis. The predictive power of a parameter will decrease as its measurement uncertainty increases, so one may wonder whether the success of velocity dispersion in predicting quenching is caused by it being more precisely measured than parameters that relate to the ordered velocity. We test this possibility by  adding increasingly significant random noise to our estimates of $\overline{\sigma}$ in the random forest, essentially to mimic the possibility of significant differential measurement uncertainty. It is important to note that the random forest's  tolerance of differential measurement uncertainty decreases  with increased correlation between variables in the random forest (see \citealt{Bluck2021} for a discussion).  Indeed when the parameters exhibit no inter-correlations, there is no  level of differential measurement uncertainty that could result in a secondary parameter (i.e. not the fundamental predictor) being crowned as the most important parameter, since the secondary parameters are completely independent of the fundamental parameter and are therefore akin to random noise.

\begin{figure}
    \includegraphics[width=\columnwidth]{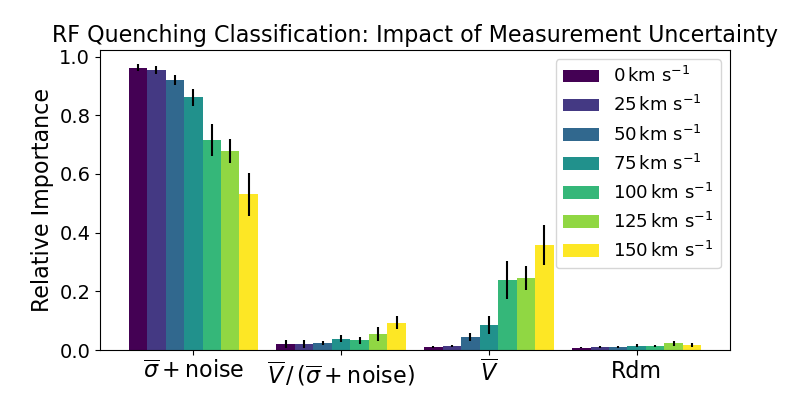}
    \caption{Testing the effect of measurement uncertainty on the random forest analyses. We add Gaussian noise to the average velocity dispersion, both in the $\overline{\sigma}$ term and $\overline{V}\,/\,\overline{\sigma}$ term, with standard deviation of the Gaussian noise distribution specified by the colour coding in the legend. Our result that average velocity dispersion is the most important parameter for predicting quenching is robust even when we add measurement uncertainty as large as $\rm 150 \, km~s^{-1}$ to $\overline{\sigma}$, which is six times the average error on $\overline{V}$. We thus confirm that our key results cannot be attributed to the possibility that $\overline{\sigma}$  is measured with greater precision than the other parameters in this study (which we also note is unlikely in any case).}
        \label{fig:RFTestsAddingNoiseToSigma}
\end{figure}

In Fig. \ref{fig:RFTestsAddingNoiseToSigma}  we show the results of a random forest analysis that considers only $\overline{\sigma}$; $(\overline{V}\,/\,\overline{\sigma})$; $\overline{V}$; and a random parameter. The analysis thus directly compares the ordered and disordered velocity for their effectiveness at predicting quenching. The colour coding reflects the standard deviation of Gaussian random noise that has been added to $\overline{\sigma}$, both in the $\overline{\sigma}$ term and $\overline{V}\,/\,\overline{\sigma}$ term. As the noise increases, the relative importance of $\overline{\sigma}$ decreases and the relative importance of $\overline{V}$ increases. This is expected, since the addition of noise washes out some of the information within $\overline{\sigma}$ that is useful for predicting quenching. Nonetheless, we find that the average velocity dispersion is the most important parameter even when its measurement uncertainty is increased by $\rm 150 \, km~s^{-1}$, which is approximately six times larger than the typical total error on $\overline{V}$ ($\rm \sim 25 \, km~s^{-1}$), shown for example in Fig. \ref{fig:FullCorrelations_KinematicParamsDeltaSFR}. We thus rule out the scenario in which $\overline{V}$ is the most important parameter for quenching with $6\sigma_\mathrm{err}$ confidence.

It is important to stress that the test in Fig. \ref{fig:RFTestsAddingNoiseToSigma} should not be interpreted as evidence that the estimates of $\overline{V}$ have measurement uncertainty $\rm 150 \, km~s^{-1}$
larger than that of $\overline{\sigma}$. This narrative is entirely inconsistent with our estimates of the uncertainties presented in figures throughout this paper, where we estimate a typical error on $\overline{V}$ of  $\rm \sim 25 \, km~s^{-1}$. 

One may worry that we have underestimated the typical error on $\overline{V}$. However, we can use the tightness of the Tully-Fisher and Mass-Velocity scaling relations shown in Fig. \ref{fig:StellarMassVsKinematicParams} as an independent indicator on the maximum allowed error on $\overline{V}$, which places a robust upper limit of $\rm \sim 50 \, km~s^{-1}$ and still leads to a confidence of  $>3\sigma_\mathrm{err}$. We note that the true confidence using this approach is likely much grater since the considerable error on $M_\star$ ($\sim0.2~\rm dex$) contributes significantly to the scatter about the Tully-Fisher and Mass-Velocity scaling relations. In other words, we cannot attribute all of the scatter to  uncertainty on $\overline{V}$. Thus, the two estimates of our confidence  ($6\sigma_\mathrm{err}$ inferred directly and $>3\sigma_\mathrm{err}$ inferred indirectly) are in good agreement.

The test in Fig. \ref{fig:RFTestsAddingNoiseToSigma} shows that $\overline{V}$ becomes the most important parameter only when the measurement uncertainty on $\overline{\sigma}$ is increased by $\rm \sim150 \, km~s^{-1}$. We note, however, that no amount of measurement uncertainty results in  $\overline{V}\,/\,\overline{\sigma}$ being the most important parameter. Indeed, $\overline{\sigma}$ has the largest relative importance when low levels of measurement uncertainty are added to $\overline{\sigma}$, and $\overline{V}$ has the largest relative importance when very high levels of measurement uncertainty are added to $\overline{\sigma}$. These data thus conclusively rule out the scenario in which $\overline{V}\,/\,\overline{\sigma}$ is the most important parameter for predicting galaxy quenching. Finally, we note that although we have chosen $\overline{V}\,/\,\overline{\sigma}$ as a specific example in Fig. \ref{fig:RFTestsAddingNoiseToSigma}, we have found similar results for the other parameters that are a function of both $\overline{V}$ and $\overline{\sigma}$. This makes sense in a scenario in which $\overline{V}$ is unimportant for quenching, in which it would not be possible for a combination of $\overline{V}$ and $(\overline{\sigma}+\mathrm{noise})$ to outperform $(\overline{\sigma}+\mathrm{noise})$ alone.

\label{lastpage}
\end{document}